\documentclass[rmp,twocolumn,aps,nofootinbib,floatfix]{revtex4}

\usepackage{amsmath,amssymb,bm}
\usepackage{times}
\usepackage{txfonts}
\usepackage{graphicx}

\mathchardef\ogon="012C%
\newcommand{\as}{a\kern-0.22em\lower.40ex\hbox{$_{\ogon}$}}

\begin{document}

  \title{Production of cold molecules via magnetically
    tunable Fesh\-bach resonances}
  \author{Thorsten K\"{o}hler}
  \author{Krzysztof G{\'o}ral}
  \affiliation{Clarendon Laboratory, Department of Physics,  
    University of Oxford, Parks Road, Oxford, OX1 3PU, United Kingdom}
  \author{Paul S.\ Julienne}
  \affiliation{Atomic Physics Division, National Institute of Standards and 
    Technology, 100 Bureau Drive Stop 8423, Gaithers\-burg, 
    Maryland 20899-8423}

\begin{abstract}
Magnetically tunable Fesh\-bach resonances were employed to associate cold 
diatomic molecules in a series of experiments involving both atomic Bose as 
well as two spin component Fermi gases. This review illustrates theoretical 
concepts of both the particular nature of the highly excited Fesh\-bach 
molecules produced and the techniques for their association from unbound atom 
pairs. Coupled channels theory provides the rigorous formulation of the 
microscopic physics of Fesh\-bach resonances in cold gases. Concepts of 
dressed versus bare energy states, universal properties of Fesh\-bach 
molecules, as well as the classification in terms of entrance- and 
closed-channel dominated resonances are introduced on the basis of practical 
two-channel approaches. Their significance is illustrated for several 
experimental observations, such as binding energies and lifetimes with respect 
to collisional relaxation. Molecular association and dissociation are 
discussed in the context of techniques involving linear magnetic field sweeps 
in cold Bose and Fermi gases as well as pulse sequences leading to Ramsey-type 
interference fringes. Their descriptions in terms of Landau-Zener, two-level 
mean field as well as beyond mean field approaches are reviewed in detail, 
including the associated ranges of validity. 

\end{abstract}
\date{\today}
\maketitle
\tableofcontents

\section{Introduction}
Resonances in general refer to the energy dependent enhancement of 
inter-particle collision cross sections due to the existence of a meta-stable 
state. Since the beginning of quantum mechanics such phenomena have been the 
subject of numerous studies in nuclear \cite{BlattWeisskopf52} as well as 
atomic and molecular physics \cite{BransdenJoachain03,Child74}. The 
meta-stable state may be described in terms of tunnelling across a barrier of 
the potential energy or coupling of a bound level of a subsystem to its 
environment
\cite{RiceJCP33,FanoNuovoCimento35,FanoPR61,FeshbachAnnPhys58,FeshbachAnnPhys62}. These scenarios are respectively referred to as shape and Fesh\-bach 
resonances. 

In the context of cold atomic gases, collision phenomena associated with 
Fesh\-bach resonances were predicted first for systems of spin polarised 
hydrogen and deuterium \cite{StwalleyPRL76,UangPRL80} as well as lithium 
\cite{UangJCPLi81} in the presence of magnetic fields. The associated 
resonance energies depend on the field strength via the Zeeman effect in the 
hyperfine levels. This research has gained substantially new experimental 
perspectives since the achievement of Bose-Einstein condensation 
\cite{BoseZPhys24,EinsteinSBPreuss24,EinsteinSBPreuss25} of dilute vapours of 
alkali atoms
\cite{AndersonScience95,DavisPRL95,BradleyPRL95,BradleyPRL97}. Contrary to
conventional gases, such atomic clouds with densities five orders of magnitude 
less than air and sub $\mu$K temperatures give rise to binary collision 
energies close to the threshold between scattering and molecular binding. 
In this extraordinary regime, magnetically tunable Fesh\-bach resonances can
be employed to manipulate the inter-atomic forces determined by the scattering 
length \cite{TiesingaPRA93}, as well as for the production of diatomic 
molecules at rest \cite{TimmermansPhysRep99}.  

\begin{figure}[htbp]
  \includegraphics[width=\columnwidth,clip]{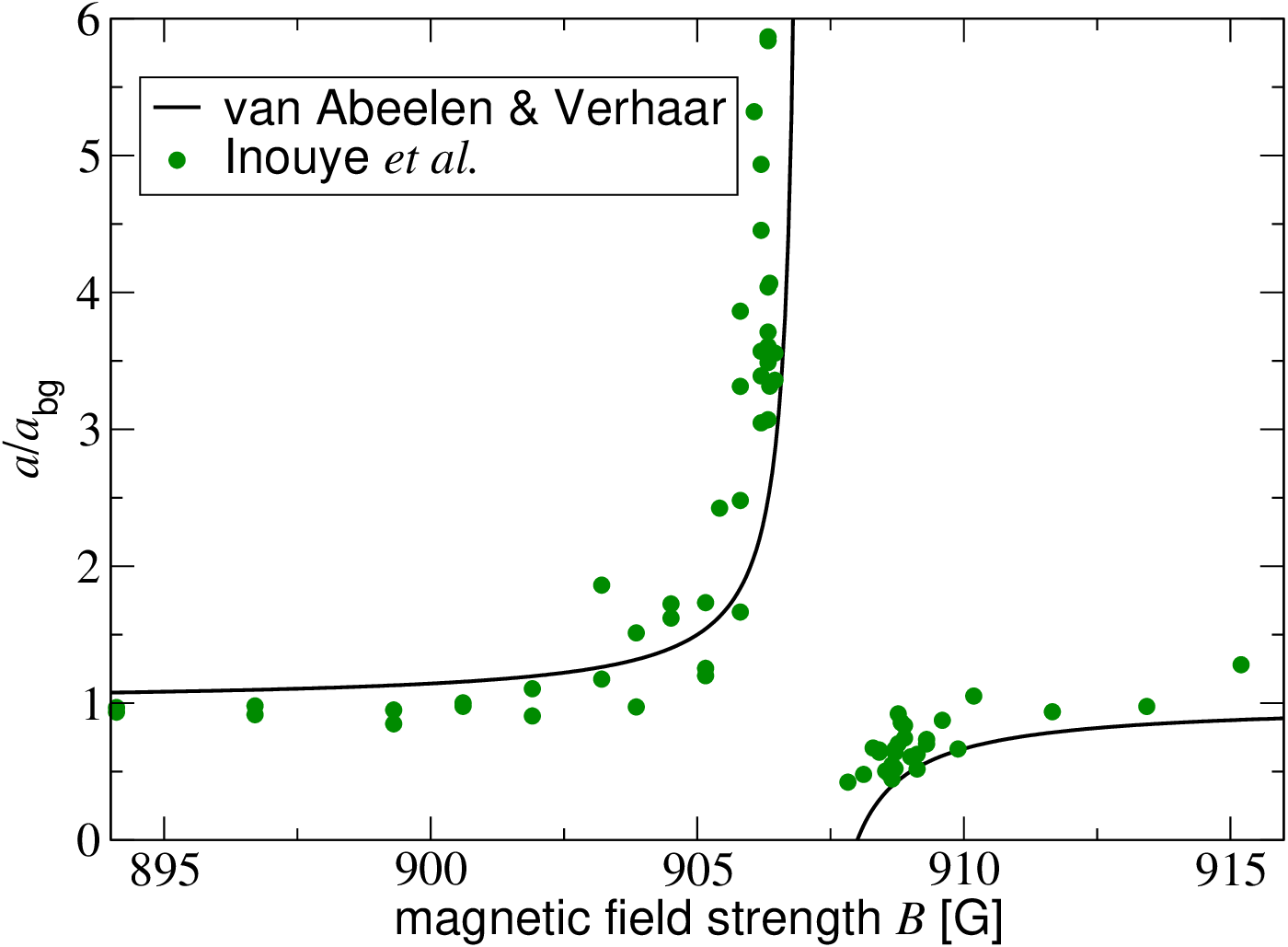}
  \caption{(Colour in online edition) 
    Order of magnitude variation of the scattering length, $a$, as a 
    function of the magnetic field strength, $B$, in the vicinity of the 
    907\,G (90.7\,mT) zero energy resonance of $^{23}$Na. The circles refer to 
    measurements using sodium Bose-Einstein condensates \cite{InouyeNature98}, 
    while the solid curve indicates associated theoretical predictions 
    \cite{Abeelenscatteringlengths}. The scattering length is normalised to 
    its asymptotic value $a_\mathrm{bg}$ away from the singularity at 
    $B_0=907\,$G. Adapted by permission from Macmillan Publishers Ltd:
    Nature (London), \cite{InouyeNature98}, copyright (1998).}
  \label{fig:Inouyeabyabg}
\end{figure}

Both techniques were demonstrated in several series of experiments with 
widespread applications throughout the physics of cold gases. 
Figure~\ref{fig:Inouyeabyabg} illustrates the manipulation of the scattering 
length, $a$, in a $^{23}$Na Bose-Einstein condensate exposed to a spatially
homogeneous magnetic field of variable strength, $B$ \cite{InouyeNature98}. 
The pole in Fig.~\ref{fig:Inouyeabyabg} at about 907\,G\footnote{The Standard 
International unit for magnetic field is Tesla, whereas most of the papers 
quoted in this review use Gauss as the unit. Consequently, we use G here, 
where $1\,$G=$10^{-4}\,$T.} is due to the near degeneracy of the energy 
associated with a Fesh\-bach resonance and the threshold. Such a singularity, 
usually referred to as a zero energy resonance \cite{Taylor72}\footnote{In the 
context of cold gases, a singularity of the scattering length is also often 
referred to simply as a Fesh\-bach resonance.}, allows the scattering length, 
in principle, to assume all values between $-\infty$ and $\infty$. Cold gases 
with such widely tunable interactions were subsequently realised for several 
species of alkali atoms, such as $^{85}$Rb \cite{CourteillePRL98,RobertsPRL98} 
and $^{133}$Cs \cite{VuleticPRL98,ChinPRL03}. Their applications involve the 
Bose-Einstein condensation of $^{85}$Rb \cite{CornishPRL00} and $^{133}$Cs 
\cite{WeberScience03} as well as studies of the collapse of condensates with 
negative scattering lengths \cite{RobertscollapsePRL00,DonleyNature01}. 

Quite generally, the zero energy resonance position exactly coincides with the 
field strength at which the energy of a diatomic vibrational bound state 
becomes degenerate with the threshold for dissociation into an atom pair at 
rest. On the side of positive scattering lengths, this Fesh\-bach molecular 
state describes a stable molecule in the absence of background gas collisions 
which ceases to exist at the position of the singularity. In the context of 
two-body systems involving the same species of atoms such a bound state is 
usually referred to as a dimer. The relation between collision resonances 
above threshold and bound states below it links the manipulation of 
interactions to the molecular conversion of separated atom pairs.

\begin{figure}[tbp]
  \includegraphics[width=\columnwidth,clip]{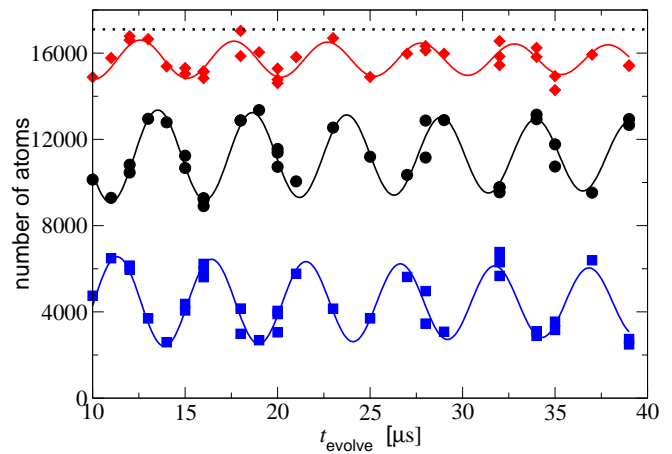}
  \caption{(Colour in online edition)
    Ramsey fringes between atomic and molecular components produced 
    from a Bose-Einstein condensate of 17100 $^{85}$Rb atoms 
    \cite{DonleyNature02} via a sequence of magnetic field pulses in the 
    vicinity of the 155\,G zero energy resonance. Circles and squares refer to 
    the number of particles in the remnant condensate and in an atomic burst 
    consisting of correlated pairs with a comparatively high relative 
    velocity, respectively. Diamonds indicate the total amount of particles in 
    both of these components. Its difference with respect to the total number 
    of atoms (dotted line) indicates the production of undetected Fesh\-bach 
    molecules. The fringes are shown as a function of the delay time of the 
    interferometer, $t_\mathrm{evolve}$, in which the atomic and molecular 
    states acquire their phase difference. The fringe frequency provides an 
    accurate measure of the binding energy \cite{ClaussenPRA03}. Adapted by 
    permission from Macmillan Publishers Ltd: Nature (London), 
    \cite{DonleyNature02}, copyright (2002).}
  \label{fig:DonleyFringe}
\end{figure}

Production of cold dimers was demonstrated first 
\cite{FiorettiPRL98,NikolovPRL99,TakekoshiPRA99,WynarScience00} via 
photo-association of atoms \cite{WeinerRMP99}. This achievement was followed 
by studies of condensed gases of $^{85}$Rb exposed to time dependent magnetic 
field variations consisting of pairs of pulses in the vicinity of the 155\,G 
zero energy resonance \cite{DonleyNature02}. These experiments temporarily 
probed the regime of strong interactions where the size of the scattering 
length was comparable to the average inter-atomic distance. Such perturbations 
led to the three distinct components of the gas illustrated in 
Fig.~\ref{fig:DonleyFringe}. The oscillatory behaviour of their occupations as 
a function of the time delay between the pulses implied an interpretation in 
terms of Ramsey interference fringes due to a superposition state of separated 
atoms and Fesh\-bach molecules.
 
\begin{figure}[htbp]
  \includegraphics[width=\columnwidth,clip]{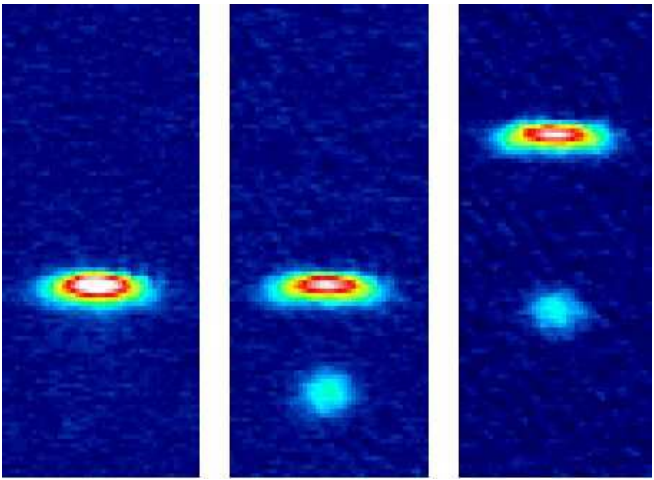}
  \caption{(Colour in online edition)
    A molecular component of about 3000 dimers is out-coupled from a 
    dilute cloud of 25000 ground state cesium atoms using Stern-Gerlach 
    separation in an inhomogeneous magnetic field \cite{HerbigScience03}. The 
    left and middle images refer to situations in which the magnetic field is 
    calibrated in such a way that it exactly compensates for the gravitational 
    force acting on the atoms in gases without and with a molecular component, 
    respectively. Due to their magnetic moment difference of 
    $-0.57\,\mu_\mathrm{Bohr}$
    ($\mu_\mathrm{Bohr}=9.27400949\times 10^{-24}\,$J/T is the Bohr magneton)
    with respect to separated atoms, the Fesh\-bach molecules in the middle 
    image drop down from the atomic cloud which is levitated and centred at 
    the same position as in the left reference image. Conversely, the right 
    image shows levitation of molecules and upward acceleration of the 
    separated atoms using a suitably adjusted inhomogeneous magnetic field.}
  \label{fig:Herbiglevitation}
\end{figure}

According to Fig.~\ref{fig:DonleyFringe}, the pulse sequence allowed a
conversion of up to about 16\,\% of the $^{85}$Rb atoms into Fesh\-bach 
molecules. Subsequent experiments improved the production efficiency by using 
magnetic field sweeps from negative to positive scattering lengths across a 
zero energy resonance. This technique was applied to cold gases consisting of 
incoherent two spin component mixtures of either $^{40}$K or $^{6}$Li atoms
\cite{RegalNature03,StreckerPRL03,CubizollesPRL03,JochimPRL03} as well as
Bose-Einstein condensates of $^{133}$Cs, $^{87}$Rb and $^{23}$Na
\cite{HerbigScience03,DuerrPRL04,XuPRL03}. In the context of these 
experiments, new schemes for the detection of Fesh\-bach molecules were 
developed. These techniques involve radio frequency (rf) photo-dissociation, 
atom loss and recovery, as well as the spatial separation of molecules from 
the remnant atomic cloud followed by their dissociation using magnetic field 
sweeps. Separation of Fesh\-bach molecules from an atomic gas, for instance, 
may be achieved via the Stern-Gerlach approach of 
Fig.~\ref{fig:Herbiglevitation} \cite{HerbigScience03,DuerrPRL04,ChinPRL05}, 
probing the magnetic moments of dimers at magnetic fields away from the zero 
energy resonance.

\begin{figure}[htbp]
  \includegraphics[width=\columnwidth,clip]{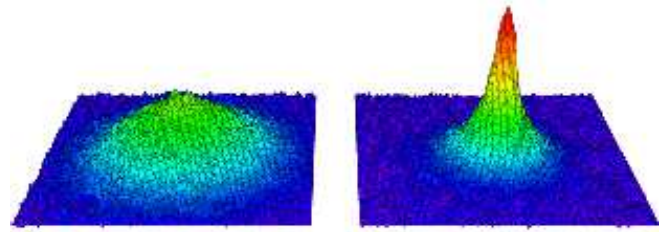}
  \caption{(Colour in online edition)
    Density distributions of dilute thermal (left image) and partially
    Bose-Einstein condensed (right image) gases of $^{40}$K$_2$ Fesh\-bach 
    molecules \cite{GreinerNature03}. Both molecular vapours were produced 
    using the technique of linear magnetic field sweeps across the 202\,G zero 
    energy resonance. The initial two spin component clouds of 470000 and 
    250000 Fermi atoms were prepared above ($250\,$nK) and below ($79\,$nK) 
    the associated critical temperatures for condensation, respectively. The
    condensate fraction in the right image is 12\,\%. Adapted by permission 
    from Macmillan Publishers Ltd: Nature (London), \cite{GreinerNature03}, 
    copyright (2003).}
  \label{fig:Greinercondensate}
\end{figure}
 
Near resonance, their large spatial extent associated with a loose bond
\cite{KoehlerPRA03,KoehlerPRL03} can lead to a remarkable stability of 
Fesh\-bach molecules with respect to inelastic collisions in the environment 
of a two spin component Fermi gas 
\cite{PetrovstabilityPRL04,PetrovstabilityPRA05}. The lifetimes of such 
dimers, ranging from about 100\,ms in the case of $^{40}$K$_2$ 
\cite{RegalstabilityPRL04} to several seconds in $^6$Li$_2$ gases 
\cite{JochimPRL03,CubizollesPRL03}, were sufficient for the observation of 
their Bose-Einstein condensation. This achievement was based on either the 
magnetic field sweep technique illustrated in Fig.~\ref{fig:Greinercondensate} 
\cite{GreinerNature03} or evaporative cooling of a molecular cloud 
\cite{ZwierleinPRL03,JochimScience03}. Such pioneering experiments gave rise 
to an ongoing series of studies 
\cite{LoftusPRL02,RegalBCStoBECPRL04,ZwierleinBCStoBECPRL04,BartensteinBCStoBECPRL04,BourdelBCStoBECPRL04,ChinScience04,KinastBCStoBECPRL04,OHaraScience02,Partridgeclosedchannel05,ZwierleinNature05}
probing the crossover from Bardeen-Cooper-Schrieffer (BCS) pairing at negative 
scattering lengths to molecular Bose-Einstein condensation
\cite{EaglesPR69,LeggettBCStoBEC80,NozieresJLowTempPhys85,Randeria95}.

This article reviews the theoretical background of the exotic species of 
highly excited Fesh\-bach molecules produced as well as their formation in the 
environment of a cold gas.  

Section~\ref{sec:weaklyboundmolecules} briefly introduces the concept of 
universality of weakly bound dimers, their relevant length and energy scales 
as well as the general form of their wave functions. This discussion 
integrates near resonant Fesh\-bach molecules into the general class of 
quantum halo systems whose classic examples are the deuteron of nuclear 
physics and the helium dimer $^4$He$_2$ molecule. More details about such 
exotic two-particle states as well as their extensions to few-body systems are 
given in a recent review \cite{JensenRMP04}.

Section~\ref{sec:Feshbachresonances} discusses those concepts of diatomic 
scattering and molecular physics that are particular to the applications of 
Fesh\-bach resonances in cold gases. Subsection~\ref{subsec:coupledchannels} 
introduces the microscopic origin of the resonance enhancement of interactions 
illustrated in Fig.~\ref{fig:Inouyeabyabg}, scattering channels and rigorous 
coupled channels theory \cite{StoofPRB88,MiesNIST96}, as well as the relation 
between zero energy resonances and molecular energy spectra. Several aspects 
of the rigorous method are well recovered in terms of two-channel approaches 
whose general concepts, such as the two-channel Hamiltonian and the 
meta-stable Fesh\-bach resonance state, are discussed in 
Subsection~\ref{subsec:twochannel}. On this basis, 
Subsection~\ref{subsec:dressedenergystates} introduces the bare and dressed 
bound and continuum energy levels. The universal properties of Fesh\-bach 
molecules near resonance are strictly derived in 
Subsection~\ref{subsec:universalFeshbach}. Their physical significance is 
illustrated in Subsection~\ref{subsec:universality} for several experimentally 
relevant examples, such as molecular binding energies and the lifetimes of 
dimers in cold Fermi and Bose gases. The size of the magnetic field range 
associated with universality implies a distinction between entrance- and 
closed-channel dominated zero energy resonances, whose physical origin is 
discussed in Subsection~\ref{subsec:classification}. Implementations of 
two-channel approaches are given in detail in 
Subsection~\ref{subsec:parameters}, describing properties of Fesh\-bach 
molecules close to as well as away from zero energy resonances. Their 
applications involve the Stern-Gerlach separation of dimers shown in 
Fig.~\ref{fig:Herbiglevitation}. Characteristic parameters relevant to 
two-channel approaches, such as, for instance, the magnetic moments associated 
with Fesh\-bach resonances of several atomic species, are summarised in 
Tables~\ref{tab:Vbgparameters} and \ref{tab:couplingparameters}. 

Section~\ref{sec:associationsweeps} reviews dynamical approaches describing 
the production and dissociation of cold Fesh\-bach molecules via linear 
magnetic field sweeps across zero energy resonances. Its introductory 
paragraphs outline, on the basis of the discrete energy spectrum of a trapped 
atom pair, the adiabatic transfer from quasi-continuum to dimer states. 
Molecular association of an atom pair via linear magnetic field sweeps falls 
into the category of rare dynamical two-body problems whose solutions can be 
represented in an analytic form \cite{DemkovJETP68,MacekPRA98}. A detailed 
derivation of the associated transition amplitudes in 
Subsection~\ref{subsec:linearsweeps} provides the foundation for all 
subsequent applications of the Landau-Zener approach 
\cite{LandauPZS32,ZenerPRS32}. These involve predictions of final populations 
of the quasi-continuum to dimer state transfer in tight atom traps, as well as 
fast sweep limits of molecule production in cold Bose and two spin component 
Fermi gases. Such conversion efficiencies can be sensitive to the quantum 
statistics associated with identical atoms whose effects are discussed in 
Subsection~\ref{subsec:BECsweeps}. The opposite saturation limits of molecule 
production via magnetic field sweeps in cold gases are a subject of ongoing 
research requiring dynamical descriptions of many-particle systems. Among such 
methods, the two-level mean field approach to Bose-Einstein condensates 
\cite{TimmermansPhysRep99} is outlined in detail, including its relation to 
the associated Landau-Zener theory \cite{MiesPRA00,GoralJPhysB04}. Intuitive 
as well as quantitative methods in the context of molecular Bose-Einstein 
condensation illustrated in Fig.~\ref{fig:Greinercondensate} are reviewed in 
Subsection~\ref{subsec:BoseFermi}, outlining the concept of an adiabatic 
production of dimers via magnetic field sweeps in cold gases with a 
significant momentum spread. Subsection~\ref{subsec:dissociation} discusses 
the theory of Fesh\-bach molecular dissociation, demonstrating the accuracy of 
analytic treatments of linear magnetic field sweeps across zero energy 
resonances.

Section~\ref{sec:AMcoherence} addresses the production of dimers via 
non-linear magnetic field variations, illustrated, for instance, in
Fig.~\ref{fig:DonleyFringe} in the context of Ramsey interferometry with atoms 
and Fesh\-bach molecules. The description of these experiments requires
general theoretical concepts, such as a precise treatment of molecular 
populations in gases, which are obscured in many applications involving linear 
magnetic field sweeps. Subsection~\ref{subsec:Ramsey} gives an intuitive 
explanation for the observations of the Ramsey fringes of
Fig.~\ref{fig:DonleyFringe} using a simplifying two-body approach. The general 
observable describing the number of molecules in the environment of a gas 
\cite{KoehlerPRA03} is introduced in Subsection~\ref{subsec:howtocount}. This
introduction includes simple applications to the fast sweep limits of dimer 
production in Bose and two spin component Fermi gases.
Subsection~\ref{subsec:BECdynamics} addresses the complete description of the
Ramsey interferometry experiments in terms of dynamical beyond mean field
approaches \cite{HollandPRL01,KoehlerPRA02}.

Section~\ref{sec:conclusions} concludes this review and provides an outlook
on related research, such as $p$-wave and optical Fesh\-bach resonances, cold 
dipolar and ground state molecules as well as extensions to few-body physics
and studies of Efimov's effect.

\section{Weakly bound diatomic molecules}
\label{sec:weaklyboundmolecules} 
\begin{figure}[htbp]
  \includegraphics[width=\columnwidth,clip]{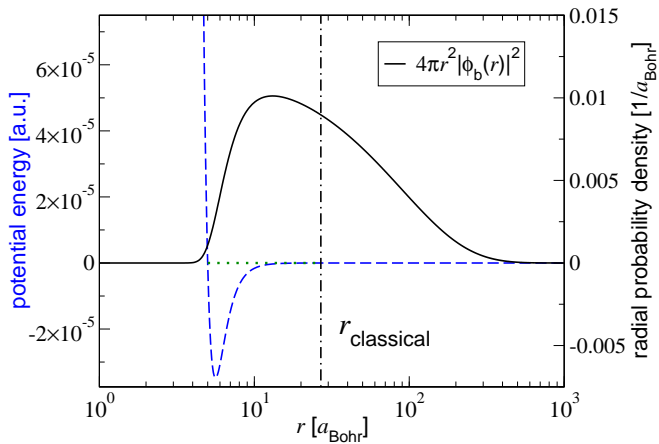}
  \caption{(Colour in online edition)
    Helium dimer interaction potential (dashed curve) \cite{TangPRL95}
    and radial probability density of the $^4$He$_2$ molecule (solid curve). 
    The dotted and dot-dashed lines indicate the bound state energy 
    $E_\mathrm{b}=-4.2\times 10^{-9}$\,a.u. 
    ($1\,$a.u.$=4.35974417\times 10^{-18}$\,J is the atomic unit of energy)
    and the classical turning point $r_\mathrm{classical}=27\,a_\mathrm{Bohr}$ 
    ($a_\mathrm{Bohr}=0.052917\,$nm is the Bohr radius), respectively. The 
    inter-atomic distance $r$ is given on a logarithmic scale.}
  \label{fig:He2}
\end{figure}

Weakly bound diatomic molecules are special cases of low energy halo systems 
\cite{JensenRMP04}. These remarkable quantum states are characterised by a 
large mean separation between the constituent particles which very much 
exceeds the outer classical turning point, $r_\mathrm{classical}$, determined 
by their binding energy and their attractive interaction. Halos are therefore 
quite distinct from spatially extended quasi classical states, such as the 
Rydberg levels associated with the long range Coulomb potential. Due to their 
non-classical but comparatively simple nature, two-body halos have played a 
significant role in the present understanding of composite few-particle 
systems. Classic examples are the deuteron in nuclear physics 
\cite{BlattWeisskopf52} as well as the weakly bound helium dimer $^4$He$_2$ 
molecule \cite{LuoJCP93,SchoellkopfScience94}, whose typical halo wave 
function is illustrated in Fig.~\ref{fig:He2}. We note that the modulus of the 
bound state energy of this halo molecule with respect to the dissociation 
threshold, i.e.~the zero of energy in Fig.~\ref{fig:He2}, is negligibly small 
compared to the well depth of the pair interaction. The $^4$He atoms are 
separated by distances larger than the classical turning point, 
$r_\mathrm{classical}$, with a probability of about 80\,\%. Such distances 
very much exceed the intuitive force range associated with the potential well. 

The characteristic long range of isotropic diatomic halo molecules, such as 
the helium dimer, is determined mainly by a single parameter of the 
interaction potential, the $s$-wave scattering length $a$. Whenever the 
inter-atomic interaction supports a weakly bound halo state the scattering 
length is positive. In accordance with effective range theory 
\cite{SchwingerERTnotes,SchwingerERTproceedings,BethePR49}, the length scale 
$a$ may be interpreted in terms of the radius of a hypothetical hard sphere 
which mimics the scattering properties of the real microscopic potential in 
the limit of zero collision energy. This radius is closely related to the 
bound state energy of a weakly bound diatomic halo molecule which is well 
approximated by the formula:
\begin{equation}
  E_\mathrm{b}=-\hbar^2/(ma^2).
  \label{Ebuniversal}
\end{equation}
Here $m$ is twice the reduced mass of the atom pair which in the case of 
identical particles coincides with the mass of a single atom. At inter-atomic 
distances, $r$, large compared to the classical turning point,
$r_\mathrm{classical}$, the associated isotropic bound state wave function 
assumes the following general form: 
\begin{equation}
  \phi_\mathrm{b}(r)=\frac{1}{\sqrt{2\pi a}}\frac{e^{-r/a}}{r}.
  \label{phibuniversal}
\end{equation}
This wave function gives the mean distance between the atoms, i.e.~the bond 
length of the molecule, to be:
\begin{equation}
  \langle r\rangle=4\pi\int_0^\infty r^2dr\,r|\phi_\mathrm{b}(r)|^2=a/2.
  \label{bondlength}
\end{equation}
For the typical example of the helium dimer in Fig.~\ref{fig:He2} the solution 
of the stationary two-body Schr\"odinger equation with a realistic pair 
interaction \cite{TangPRL95} predicts $\langle r\rangle=5.1\,$nm, while the 
estimate of Eq.~(\ref{bondlength}) yields $\langle r\rangle=5.2\,$nm. For 
comparison, measurements based on the diffraction of a helium molecular beam 
from a micro-fabricated material transmission grating \cite{GrisentiPRL00} 
determined the bond length of $^4$He$_2$ to be $\langle r\rangle=5.2(4)\,$nm.
 
A dependence of physical quantities, such as those of Eqs.~(\ref{Ebuniversal}) 
and (\ref{phibuniversal}), only on the scattering length rather than the 
details of the microscopic forces is usually referred to as universality. The 
relation between Eqs.~(\ref{Ebuniversal}) and (\ref{phibuniversal}) follows 
immediately from the stationary Schr\"odinger equation, 
$H_\mathrm{2B}\phi_\mathrm{b}(r)=E_\mathrm{b}\phi_\mathrm{b}(r)$, using the 
general two-body Hamiltonian associated with the relative motion of an atom 
pair,
\begin{equation}
  H_\mathrm{2B}=-\frac{\hbar^2}{m}\boldsymbol{\nabla}^2+V(r).
  \label{H2Bgeneral}
\end{equation} 
A typical realistic molecular potential $V(r)$ \cite{TangPRL95} is depicted in 
Fig.~\ref{fig:He2} for the example of the helium dimer. Its behaviour in the 
limit of large inter-atomic distances is dominated by the van der Waals 
interaction,
\begin{equation}
  V(r)\underset{r\to\infty}{\sim}-C_6/r^6.
  \label{potentialvdW}
\end{equation}
The constant $C_6$ is known as the van der Waals dispersion coefficient. 
In accordance with Eq.~(\ref{potentialvdW}), the outer classical turning point 
of the relative motion of an atom pair with the energy $E_\mathrm{b}$ of 
Eq.~(\ref{Ebuniversal}) is given by the following formula:
\begin{equation}
  r_\mathrm{classical}=[a (2 l_\mathrm{vdW})^2]^{1/3}. 
\end{equation}
Here $l_\mathrm{vdW}$ is a characteristic range associated with the van der 
Waals interaction between atoms, usually referred to as the van der Waals 
length,
\begin{equation}
  l_\mathrm{vdW}=\frac{1}{2}(mC_6/\hbar^2)^{1/4}.
  \label{lvdW}
\end{equation}
The characteristic property of diatomic halo molecules that their spatial
extent, determined by the scattering length $a$, very much exceeds 
$r_\mathrm{classical}$ therefore implies the condition $a\gg l_\mathrm{vdW}$
for their existence. In the range of such typically large inter-atomic 
distances, $r\gg l_\mathrm{vdW}$, the stationary Schr\"odinger equation for a 
halo molecule reduces to its interaction-free counterpart, i.e.
\begin{equation}
  E_\mathrm{b}\phi_\mathrm{b}(r)\underset{r\to\infty}{\sim} 
  -(\hbar^2/m)\boldsymbol{\nabla}^2\phi_\mathrm{b}(r),
\end{equation}
whose unit normalised solution associated with the bound state energy of 
Eq.~(\ref{Ebuniversal}) is given by Eq.~(\ref{phibuniversal}). 

The universal properties of Eqs.~(\ref{Ebuniversal}) and (\ref{phibuniversal})
and the length scales, $a$ and $l_\mathrm{vdW}$, associated with diatomic halo 
systems characterise, in the same manner, the helium dimer as well as highly 
excited long range Fesh\-bach molecules. Any other details of their binary 
interactions are obscured by the large spatial extent of the bound states. The 
size of the scattering length, however, which determines the long range of 
these states depends sensitively on the microscopic potential whose detailed 
structure varies considerably among different species.

\section{Fesh\-bach resonances}
\label{sec:Feshbachresonances}
Inter-atomic collisions in cold gases are characterised by de Broglie 
wavelengths much larger than the van der Waals length of the microscopic 
potential. Similarly to the universal properties of diatomic halo molecules, 
such a separation of length scales implies that the associated low energy 
interactions are determined mainly by the $s$-wave scattering length, $a$. 
Fesh\-bach resonances provide an opportunity of manipulating these 
inter-atomic forces by exposing a cold gas of alkali atoms to a spatially 
homogeneous magnetic field of strength $B$. Based on an introduction to the 
microscopic physics, this section describes the concept of magnetic tuning of 
the scattering length and the associated cross section (given, for instance, 
by $8\pi a^2$ in the case of identical Bose atoms) as well as its relation to 
the properties of molecular states.

\subsection{Molecular physics of resonances}
\label{subsec:coupledchannels}
Resonance enhancement of collision cross sections relies upon the existence 
of meta-stable states. The two-body molecular physics of such neutral atom 
Fesh\-bach resonance states is the subject of this subsection. This involves 
the microscopic origin of the inter-atomic interactions as well as the atomic 
and molecular symmetries that permit the classification of scattering channels 
and molecular bound and meta-stable energy levels. Such a basic understanding 
of low energy neutral atom scattering and bound states provides the grounding 
for practical two-channel approaches of the subsequent subsections.

\subsubsection{Inter-atomic interactions}
The helium atom of Section~\ref{sec:weaklyboundmolecules} has a $^1$S$_0$ 
configuration with no unpaired electron, and the interaction of two such atoms 
is represented by the single molecular Born-Oppenheimer potential illustrated 
in Fig.~\ref{fig:He2}. Such a system is too simple to have a magnetically 
tunable resonance state. It is therefore necessary to consider here the 
experimentally relevant case of a pair of $^2$S$_{1/2}$ atoms of the same 
species. The unpaired electron spins $\mathbf{s}_1$ and $\mathbf{s}_2$ from 
each atom can be coupled to a total spin 
$\mathbf{S}=\mathbf{s}_1+\mathbf{s}_2$ with the associated quantum numbers 
$S=$ 0 or 1. States with $S=0$ or $S=1$ are called singlet or triplet states, 
respectively. The electronic part of the inter-atomic interaction is 
represented, as for the simplest molecule H$_2$ \cite{Pauling39}, by singlet 
and triplet molecular Born-Oppenheimer potentials of $^1\Sigma_g^+$ and 
$^3\Sigma_u^+$ symmetry, corresponding respectively to $2S+1=$ 1 and 3. Here 
the notation ``$\Sigma$'' refers to zero projection of the electronic orbital 
angular momentum on the inter-atomic axis. The label ``+'' indicates that the 
electronic wave function is left unchanged upon reflection in a plane 
containing the nuclei. Finally, ``$g$'' and ``$u$'' are associated with the 
{\em gerade} (even) and {\em ungerade} (odd) symmetry upon inversion through 
the geometric centre of the molecule, respectively. Consequently, the latter 
symmetry is absent when the atoms are of different species. 

\begin{figure}[htbp]
  \includegraphics[angle=0,width=\columnwidth,clip]{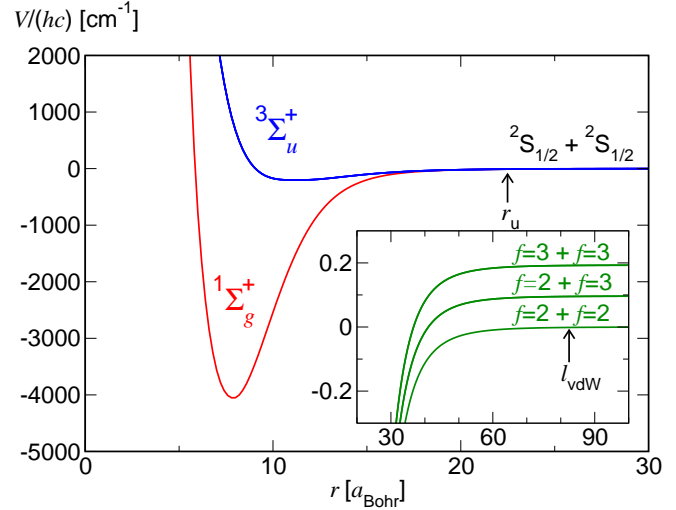}
  \caption{(Colour in online edition)
    The $^1\Sigma_g^+$ and $^3\Sigma_u^+$ Born-Oppenheimer potential 
    energy curves for the Rb$_2$ dimer molecule correlating with two separated 
    $^2$S$_{1/2}$ Rb atoms. An energy $E$ associated with the wave number unit
    $E/hc=1$\,cm$^{-1}$ corresponds to $E/h=29.979\,$GHz. The inset shows the 
    long range adiabatic potentials \cite{MiesNIST96,MiesPRA00} on a much 
    smaller energy scale with the zero field ($B=0$) hyperfine structure of 
    the $^{85}$Rb isotope with $E_\mathrm{hf}/h=3.035\,$GHz. The arrows show 
    the van der Waals length $l_\mathrm{vdW}$ of Eq.~(\ref{lvdW}) and the 
    uncoupling distance, $r_\mathrm{u}$, where the hyperfine energy 
    $E_\mathrm{hf}$ equals the difference between the $^3\Sigma_u^+$ and 
    $^1\Sigma_g^+$ Born-Oppenheimer potential curves. The $^{87}$Rb isotope 
    has the same Born-Oppenheimer potential energy curves, but the long range 
    curves would be different with atomic levels $f=$ 1 and 2 and 
    $E_\mathrm{hf}/h=6.835\,$GHz.}
  \label{fig:psj_Rb_V}
\end{figure}

Figure~\ref{fig:psj_Rb_V} shows potential energy curves $V_S(r)$ for $S=$ 0 
and 1 for two Rb atoms as an illustrative case. Similar potentials exist for 
any pair of like alkali metal atoms, and if the {\em gerade} and 
{\em ungerade} symmetry labels $g$ and $u$ are dropped, for pairs of unlike 
alkali metal atoms. For the case of two S-state atoms, the long range form of 
the potential is the van der Waals dispersion with the lead term $-C_6/r^6$ of 
Eq.~(\ref{potentialvdW}) that is identical for the $^1\Sigma_g^+$ and 
$^3\Sigma_u^+$ potentials. The splitting between these two potentials comes 
from the difference in chemical bonding interactions when the charge clouds 
of the two atoms overlap at small inter-atomic distances, $r<1$\,nm. The 
long-range form of this splitting is associated with the electron exchange 
interaction, which decreases exponentially as $r$ increases. This interaction 
is attractive for the $^1\Sigma_g^+$ state and repulsive for the 
$^3\Sigma_u^+$ state.   

\begin{figure}[htbp]
  \includegraphics[angle=0,width=\columnwidth,clip]{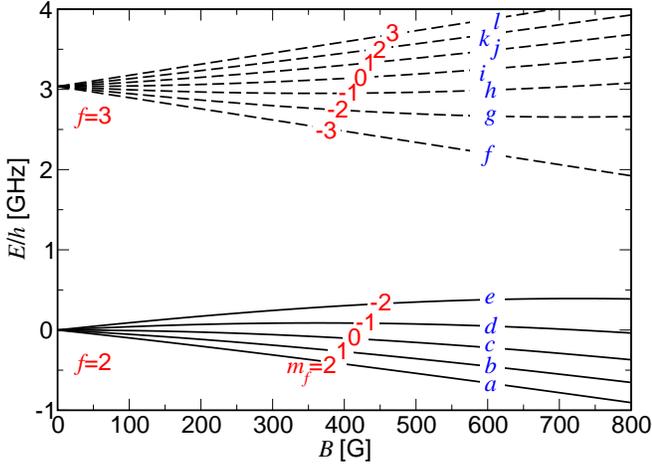}
  \caption{(Colour in online edition)
    Zeeman levels of the $^{85}$Rb $^2$S$_{1/2}$ atom versus the 
    magnetic field strength, $B$. Solid and dashed curves refer to the $f=2$ 
    and $f=3$ multiplets, respectively. The projection quantum numbers $m_f$ 
    and the alphabetic labels $a$, $b$, $c$, $\ldots$ are shown.}
  \label{fig:psj_Rb_Eat}
\end{figure}

More diatomic states need to be accounted for when the atomic nuclear spin 
quantum number $i$ is nonzero. At zero magnetic field, i.e.~$B=0$, the 
unpaired electrons in each atom with total electronic angular momentum 
$\mathbf{j}$ can interact with the nuclear spin $\mathbf{i}$ to produce the 
hyperfine levels of total angular momentum $\mathbf{f}=\mathbf{j}+\mathbf{i}$. 
For alkali metal atoms in their electronic ground states the identity 
$\mathbf{j}=\mathbf{s}$ gives the total angular momentum quantum number to be 
either $f=$ $i-1/2$ or $i+1/2$. The difference in energy between these two 
hyperfine levels is the hyperfine splitting $E_\mathrm{hf}$. If $f$ is an 
integer, the atom will be a composite boson, whereas if $f$ is half an odd 
integer, the atom will be a composite fermion. The interaction with a magnetic 
field will further split the energy of each $f$ state into its Zeeman 
components. Figure~\ref{fig:psj_Rb_Eat} illustrates the $B$-dependent energy 
levels of $^{85}$Rb, a bosonic isotope. In the limit of high magnetic field 
strengths the Zeeman energy of the unpaired electron becomes large compared 
to $E_\mathrm{hf}$. Consequently, $f$ will no longer be a good quantum number. 
Rotation about the magnetic field axis, however, is a symmetry transformation. 
This implies that $m_f$, the quantum number associated with the projection of 
the total angular momentum onto this axis, is conserved. 
Figure~\ref{fig:psj_Rb_Eat} also shows that the Zeeman levels of a $^{85}$Rb 
atom do not anti-cross in a range of magnetic fields up to a few hundred G. 
As for notation, it is therefore often convenient to label the Zeeman levels 
by $fm_f$ where $f$ is the total angular momentum quantum number at $B=0$ with 
which the level adiabatically correlates. We will use such an explicit 
reference to the different Zeeman states in the subsequent subsections. An 
alternative, brief notation \cite{StoofPRB88,MiesPRA00} indicated in 
Fig.~\ref{fig:psj_Rb_Eat} is to label the Zeeman components in order of 
increasing energy as $a, b, c,\ldots$ and shall be used throughout this 
subsection. 

\subsubsection{Coupled channels scattering}
\label{CC}
The hyperfine splitting $E_\mathrm{hf}$ and the magnetic Zeeman energy are 
typically large compared to the kinetic energy for cold collisions. 
Consequently, the scattering properties of two atoms depend strongly on the 
Zeeman levels in which they are prepared. A scattering channel is defined by 
specifying the quantum numbers that describe each of the two initially 
separated atoms. In the following it is convenient to perform a partial wave 
expansion of the stationary scattering wave function of the relative motion of 
an atom pair. Its components thus depend on the inter-atomic distance, $r$, in 
addition to the quantum numbers $\ell$ and $m_\ell$ associated with the 
angular momentum of the relative motion and its orientation with respect to 
the magnetic field axis. Given this convention, a scattering channel is 
specified by the channel index 
$\alpha=\{f_1 m_{f_1} f_2 m_{f_2} \ell m_\ell\}$. Here the braces signify 
that the wave function is (anti-)symmetric with respect to exchange of 
identical Bose (Fermi) particles \cite{StoofPRB88}. Consequently, the order 
of atomic indices 1 and 2 is unimportant. Only those channels with even (odd) 
$\ell$ exist for two identical bosons (fermions) in the same Zeeman level. 
All $\ell$ values are possible for identical bosons or fermions in different 
Zeeman levels or for nonidentical species. The partial waves associated with 
$\ell=0$, 1, or 2 are designated $s$, $p$, or $d$ waves. If the channel 
energy, $E_\alpha=E_{f_1m_{f_1}}+E_{f_2m_{f_2}}$, is less than the total 
energy of the system, $E$, the channel $\alpha$ is said to be open. 
Conversely, if $E_\alpha > E$, channel $\alpha$ is referred to as closed. When 
$E$ is lower than the lowest channel energy, all channels are closed and $E$
can refer only to one of the discrete molecular levels. If $E$ is higher 
than the lowest channel energy, then at least one channel is open, and $E$ is
associated with a stationary energy level in the scattering continuum.

In the case of the $^1$S$_0$ $^4$He atom of 
Section~\ref{sec:weaklyboundmolecules} the two-body interaction Hamiltonian is 
specified by a single-channel potential describing the helium dimer bound 
state as well as cold collisions. Such a treatment is not adequate for alkali 
atoms with nuclear spin because of intrinsic coupling among the various 
channels. It is therefore convenient to represent the wave functions for 
scattering or bound states as a coupled channels expansion into their 
components $\Psi_\alpha(r,E)$ in the separated atom spin basis labelled by 
$\alpha$ \cite{StoofPRB88,GaoPRA96,MiesNIST96}. The associated radial wave 
function $F_{\alpha}(r,E)$ of the relative motion of an atom pair with the 
energy $E$ is determined by the following formula:
\begin{equation}
  \Psi_\alpha(r,E)=F_{\alpha}(r,E)/r.
  \label{eq:psi}
\end{equation}
Substituting Eq.~(\ref{eq:psi}) into the stationary Schr{\"o}dinger equation 
gives the following coupled channels equations which, in combination with the 
boundary conditions imposed on $F_{\alpha}(r,E)$, determine both the continuum 
and bound states:
\begin{equation}
  \frac{\partial^2 F_{\alpha}(r,E)}{\partial r^2} + \frac{m}{\hbar^2} 
  \sum_\beta\left[E\delta_{\alpha\beta}-V_{\alpha\beta}(r)\right] 
  F_{\beta}(r,E)=0.
  \label{eq:SE}
\end{equation}
We note that $m$ should be identified in Eq.~(\ref{eq:SE}) with twice the 
reduced mass of an atom pair. The potential matrix, $V(r)$, takes on the 
following form in the separated atom spin basis:
\begin{equation}
  V_{\alpha\beta}(r) = \left[E_{f_1m_{f_1}} + E_{f_2m_{f_2}} + 
  \frac{\hbar^2 \ell(\ell+1)}{m r^2} \right] \delta_{\alpha\beta} 
  +V^\mathrm{int}_{\alpha \beta}(r).
  \label{eq:V}
\end{equation}
Here the atomic hyperfine and magnetic interaction terms are given by the 
experimentally known separated atom energies $E_{f_1m_{f_1}}$ and 
$E_{f_2m_{f_2}}$, and the kinetic energy of axis rotation is given by the 
centrifugal energy term proportional to $\ell(\ell+1)$. These terms are 
diagonal in the asymptotic basis. The complicated part of the scattering due 
to the electronic Born-Oppenheimer potentials and electron spin-dependent 
interactions is contained in the interaction matrix $V_\mathrm{int}(r)$
of Eq.~(\ref{eq:V}), which is comprised of two parts:
\begin{equation}
  V_\mathrm{int}(r)=V_\mathrm{el}(r)+V_\mathrm{ss}(r).
  \label{eq:Vint}
\end{equation}
The contribution $V_\mathrm{el}(r)$ represents the strong electronic 
interaction. It is diagonal in $\ell$ but non-diagonal in the atomic channel 
quantum numbers $f_1 m_{f_1} f_2 m_{f_2}$. Its elements diagonal in 
$f_1 m_{f_1} f_2 m_{f_2}$ depend on weighted sums of the two Born-Oppenheimer 
potentials $V_{S=0}(r)$ and $V_{S=1}(r)$, whereas the off-diagonal terms 
depend on the difference between $V_{S=0}(r)$ and $V_{S=1}(r)$. The strong 
electronic interaction, $V_\mathrm{el}(r)$, is responsible for elastic 
scattering and inelastic spin-exchange collisions 
\cite{BenderPR63,DalgarnoProcRoySoc65}, and gives rise to the broadest 
scattering resonances. 

The term $V_\mathrm{ss}(r)$ in Eq.~(\ref{eq:Vint}) represents the weak 
relativistic spin-spin potential energy \cite{StoofPRB88,MoerdijkPRA95}. It is 
due to the anisotropic dipolar interaction between the two electron spins, and 
is non-diagonal in both $f_1 m_{f_1} f_2 m_{f_2}$ and $\ell$, i.e., it couples 
different partial waves. In the limit of large inter-atomic distances 
$V_\mathrm{ss}(r)$ is proportional to $\alpha^2/r^3$, where 
$\alpha=1/137.0426$ is the fine structure constant. As $V_\mathrm{ss}(r)$ is a 
tensor operator of rank 2, only blocks that differ in $\ell$ by zero or two 
units have non-vanishing matrix elements, according to the Wigner-Eckart 
theorem \cite{WignerZPhys27,EckartRMP30}. In addition, there are no $s$-wave 
diagonal potentials varying as $1/r^3$.

At short range the spin-dipole interaction can be modified by second-order 
spin-orbit contributions, which are important for a heavy atom like Cs 
\cite{MiesNIST96,KotochigovaPRA01}. In general, the potential energy 
contribution $V_\mathrm{ss}(r)$ is responsible for weak inelastic spin-dipolar 
relaxation and gives rise to narrow scattering resonances.

The low energy collision physics of alkali atoms in specific hyperfine states 
is sensitive to the re-coupling of electron spins between the separated 
atoms and the short range zone of strong chemical interactions. At small
inter-atomic distances the potential energy scale is orders of magnitude 
larger than $E_\mathrm{hf}$. Whereas the electron spin is coupled to the 
nuclear spin on the same atom when the atoms are far separated, the electron 
spins become uncoupled from the nuclear spin and couple strongly to one 
another at small $r$ to make the $S=$ 0 and 1 states of the Born-Oppenheimer 
potentials. The distance where this re-coupling occurs is near $r_\mathrm{u}$ 
of Fig.~\ref{fig:psj_Rb_V} where the difference in Born-Oppenheimer potentials 
$V_{S=1}(r_\mathrm{u})-V_{S=0}(r_\mathrm{u})$ due to the exchange potential 
is equal to the atomic hyperfine energy $E_\mathrm{hf}$. This occurs typically 
in the distance range of 20 to 25\,$a_\mathrm{Bohr}$ for alkali atoms.
 
The coupled channels method \cite{StoofPRB88,MiesNIST96} of Eq.~(\ref{eq:SE}) 
properly accounts for the dynamical changes in the couplings among the five 
angular momenta $\mathbf{s}_1$, $\mathbf{s}_2$, $\mathbf{i}_1$, 
$\mathbf{i}_2$, and $\boldsymbol{\ell}$ as the atoms move through the region 
near $r_\mathrm{u}$. Basic symmetries of the coupling terms in 
$V_\mathrm{int}(r)$ of Eq.~(\ref{eq:Vint}) allow us to separate the 
interaction matrix into blocks, within which the coupling is strong and 
between which the coupling is intrinsically weak. Such a separation gives rise 
to classifications of the various stationary energy states in terms of their 
predominant symmetry properties. Projected levels that are bound just within a 
particular block are, in general, associated with scattering resonances when
their energy is above the scattering threshold. We note, however, that in 
contrast to the complete stationary states determined by Eq.~(\ref{eq:SE}), 
such Fesh\-bach resonance levels depend on the separation of 
$V_\mathrm{int}(r)$ into blocks or, equivalently, on the choice of basis set.

In this context, the separated atom spin basis referred to in Eq.~(\ref{eq:V}) 
is convenient at long range, but leads to off-diagonal elements at short 
range. One alternative, short range basis would first couple $\mathbf{s}_1$ 
and $\mathbf{s}_2$ to a resultant $\mathbf{S}$ and $\mathbf{i}_1$ and 
$\mathbf{i}_2$ to a resultant $\mathbf{I}$. Then $\mathbf{S}$ and $\mathbf{I}$ 
can be coupled to a resultant $\mathbf{F}$, which in turn couples to 
$\boldsymbol{\ell}$ to give the total angular momentum 
$\mathbf{F}_\mathrm{total}$ \cite{TiesingaPRA93,MoerdijkPRA95}. Then we could 
set up a molecular basis set with quantum numbers 
$(S I) F \ell F_\mathrm{total}M$. Here $M$ refers to the orientation quantum 
number associated with the projection of $\mathbf{F}_\mathrm{total}$ onto the 
magnetic field axis. This short range basis takes advantage of the fact that 
$V_\mathrm{el}(r)$ is diagonal in $S$. An alternative separated atom basis set 
could couple $\mathbf{f}_1$ and $\mathbf{f}_2$ to a resultant $\mathbf{F}$, 
and give the basis $(f_1 f_2)F \ell F_\mathrm{total}M$. This basis is useful 
at low $B$ fields at which the Zeeman levels do not anti-cross, and where $F$ 
may be viewed as a good quantum number. The unitary transformation between 
separated atom and molecular basis sets is called a frame transformation 
\cite{BenderPR63,DalgarnoProcRoySoc65,BurkePRL98,GaoPRA05}. Bound and 
meta-stable states of light elements like Li and Na are best classified by the 
molecular basis \cite{MoerdijkPRA95,SimonucciEurophysLett05} whereas the 
separated atom basis is better for heavy elements like Rb or Cs 
\cite{MartePRL02,ChinPRA04}. Similarly, the long-range part of a scattering 
wave function is best described in the separated atom basis set, whereas the 
molecular basis is more appropriate for the short range part. 

\subsubsection{Threshold collisions}
The quantum numbers $f_1m_{f_1}$ and $f_2m_{f_2}$ associated with the Zeeman 
states of the separated atoms in which a dilute gas is prepared determine the 
entrance channel of a two-body collision. In the context of cold collisions, 
it is usually sufficient to consider just the $s$-wave ($\ell=0$) component of 
an initial plane wave momentum state of the relative motion of an atom pair. 
To avoid inelastic processes known as spin relaxation, most experimental 
applications of Fesh\-bach resonances involve atom pairs in the lowest 
energetic Zeeman states for which $s$-wave scattering is allowed. In the 
following, we assume such a case and choose the zero of energy at the entrance 
channel energy, $E_\alpha$. This convention implies that the total energy $E$ 
of a colliding atom pair is identical to its positive kinetic energy, 
$\hbar^2k^2/m$. Here $k$ is referred to as the wave number and $p=\hbar k$ is 
the momentum of the relative motion. On the other hand, bound state energies, 
$E_\mathrm{b}$, are always negative.

The amplitudes associated with transitions between the initial and final 
states of a diatomic collision may be inferred from the asymptotic form of the 
scattering solutions to Eq.~(\ref{eq:SE}) in the limit of large distances, 
$r\to\infty$. As the only open channel is assumed to be the entrance channel, 
the long range boundary condition imposed on the associated component of the 
radial wave function reads 
\cite{Taylor72}:
\begin{equation}
   F_\alpha(r,\hbar^2k^2/m)\propto\sin\left[kr+\xi(\hbar k)\right].
   \label{boundaryradial}
\end{equation}
Here the absolute magnitude of $F_\alpha(r,\hbar^2k^2/m)$ is determined 
up to an overall energy dependent pre-factor whose value is a matter of 
convention. All information about a collision is contained in the scattering 
phase shift, $\xi(\hbar k)$. According to effective range theory 
\cite{SchwingerERTnotes,SchwingerERTproceedings,BethePR49}, 
$\xi(\hbar k)$ determines the scattering length, $a$, via the following low 
energy asymptotic expansion:
\begin{equation}
   k \cot\xi(\hbar k) = -\frac{1}{a} + \frac{1}{2}k^2 r_\mathrm{eff}.
\end{equation}
Here $r_\mathrm{eff}$ is known as the effective range. For most of the 
applications in this review, the scattering length alone is sufficient for the 
description of cold diatomic collisions.

In general, the scattering length is only weakly dependent on the magnetic 
field strength, $B$, unless $B$ can be tuned in such a way that a closed 
channel Fesh\-bach resonance level crosses the entrance channel scattering 
threshold. Such a scenario may occur due to a difference in magnetic moments, 
$\partial E_\alpha/\partial B$ and $\partial E_\beta/\partial B$, associated 
with the entrance and closed channels, respectively. The meta-stability of the 
resonance state leads to a time delay during a collision when the energies of 
the scattered atoms and of the resonance level are nearly matched. This 
results in an enhancement of the scattering cross section whose width in 
energy depends on the strength of the coupling between the entrance and closed 
channels via the lifetime of the resonance state. As the zero energy cross 
section is proportional to $a^2$, such a resonance enhancement of collisions 
may be used to widely tune the scattering length \cite{TiesingaPRA93} as 
illustrated in Fig.~\ref{fig:Inouyeabyabg}. We note, however, that because of 
the inter-channel coupling the magnetic field strength $B_0$ associated with 
the singularity of $a$ differs from the crossing point between the resonance 
energy and the scattering threshold. The magnetic field width and shift of 
such zero energy resonances will be the subject of 
Subsection~\ref{subsec:dressedenergystates}.

Singularities of the $s$-wave scattering length are always accompanied by the 
degeneracy of a bound vibrational level with the scattering threshold 
\cite{Taylor72}. In the context of magnetic Fesh\-bach resonances, the 
properties of such a coupled channels stationary energy state, termed the 
Fesh\-bach molecule, may be inferred from Eq.~(\ref{eq:SE}) in the zero 
bound state energy limit, $E_\mathrm{b}\to 0$. Similarly to the studies of 
Section~\ref{sec:weaklyboundmolecules}, the derivation just relies upon the 
fact that the potential matrix $V_\mathrm{int}(r)$ of Eq.~(\ref{eq:V}) 
vanishes at large separations, $r\to\infty$. In such an asymptotic distance 
range the atoms cease to interact, and the solution of Eq.~(\ref{eq:SE}) 
associated with any $s$-wave ($\ell=0$) channel is given by the following 
formula:
\begin{equation}
  F_\alpha(r,E_\mathrm{b})\propto
  \exp\left[-\sqrt{-m(E_\mathrm{b}-E_\alpha)}r/\hbar\right].
\end{equation}
Its pre-factor depends on the overall normalisation of the bound state and is 
thus determined by all components. In the limit $E_\mathrm{b}\to 0$, however, 
only the entrance channel radial wave function acquires a long range and, 
therefore, predominates all the others. The relation between the binding 
energy and the scattering length in the vicinity of a zero energy resonance 
will be the subject of Subsection~\ref{subsec:universalFeshbach}. This 
discussion will show, on the basis of general arguments, that $E_\mathrm{b}$ 
is determined by Eq.~(\ref{Ebuniversal}), while the Fesh\-bach molecular wave 
function reduces to its entrance channel component given by 
Eq.~(\ref{phibuniversal}).

If the entrance channel is not the lowest in energy the above scenario of zero 
energy resonances and Fesh\-bach molecules is only approximate. This implies
that the scattering length is always finite even when a closed channel 
Fesh\-bach resonance level is magnetically tuned to cross the entrance channel 
energy. The associated Fesh\-bach molecule can decay into the lower energetic 
open channels via spin relaxation \cite{ThompsonPRL05,KoehlerPRL05}. Such a 
rather general situation underlies, for instance, the experiments of 
Fig.~\ref{fig:DonleyFringe} involving gases of $^{85}$Rb, and serves as an 
example for the following explicit channel classification.

\subsubsection{Example of channel classification}
The most rigorous classification of scattering channels is by the total 
projection quantum number $M=m_{f_1}+m_{f_2}+m_\ell$. In the absence of 
external fields other than $B$, the symmetry with respect to rotation about 
the magnetic field axis implies that $M$ is strictly conserved during the 
course of a collision, i.e., states with different $M$ can not couple. The 
next useful classification is by the partial wave quantum number $\ell$. Only 
weak coupling is normally possible between states of different $\ell$, since 
it can only originate from the intrinsically small and anisotropic 
$V_\mathrm{ss}(r)$ matrix elements. 

\begin{table}[htdp]
  \caption{Separated atom quantum numbers for the $s$-wave ($\ell=0$) $M=-4$ 
    block of the coupling matrix $V_\mathrm{el}(r)$ for $^{85}$Rb, for which 
    $f$ assumes the values 2 and 3. The separated atom energies relative to 
    $E_{ee}=0$ are shown for $B=$ 0 and 160\,G (16.0\,mT).}
  \begin{ruledtabular}
    \begin{tabular}{c|c|c|c|cl}
      $(f_1\, f_2)$ & $m_{f_1}\, m_{f_2}$ & $\alpha$ & $E_{\alpha}/h$ [GHz] & 
      $E_{\alpha}/h$ [GHz] \\ & & & for $B=0$ & for $B=160$\,G \\
      \hline
      (2 2) & -2 -2 & $ee$ & 0 & 0 \\
      (2 3) & -1 -3 & $df$ & 3.035732 & 2.591623 \\ 
      (2 3) &-2 -2 & $eg$ & 3.035732 & 2.756754 \\ 
      (3 3) & -3 -1 & $fh$ & 6.071464 & 5.508558 \\ 
      (3 3) & -2 -2 & $gg$ & 6.071464 & 5.513508 \\
    \end{tabular}
  \end{ruledtabular}
  \label{tab:sVblock}
\end{table}

Table~\ref{tab:sVblock} shows an example of the quantum numbers needed to 
describe the $s$-wave channels associated with the interaction of a pair 
of $f=2$, $m_f=-2$ $^{85}$Rb atoms, which is the $e$ state in the alphabetic 
notation. This state is one for which there is a broad Fesh\-bach resonance 
close to the scattering threshold near 155\,G. Both the binding energies of 
the associated Fesh\-bach molecules and the scattering length may be inferred 
from solutions of Eq.~(\ref{eq:SE}), using just the matrix elements of the 
potential $V_\mathrm{el}(r)$ between the channel states of 
Table~\ref{tab:sVblock}. Since the nuclear spin of a $^{85}$Rb atom is 
$i=5/2$, the two ground state $f$ values are 2 and 3. The total projection 
quantum number for $s$ waves is $M=-4$ for any $B$. There are only five 
possible separated atom spin channels. As the hyperfine splitting is 
$E_\mathrm{hf}/h=3.035$\,Ghz for this species, cold collisions associated with 
temperatures $T$ on the order of 1\,$\mu$K (where $k_\mathrm{B}T/h = 21\,$kHz 
given the Boltzmann constant $k_\mathrm{B}=1.3806505 \times 10^{-23}\,$J/K) 
have only a single open $s$-wave channel, the $ee$ channel. All the other 
$\ell=0$ channels, $df$, $eg$, $fh$, and $gg$ are closed. 

\begin{table}[htdp]
  \caption{Separated atom quantum numbers for the open channel $d$-wave 
    ($\ell=2$) $M=-4$ block of the coupling matrix $V_\mathrm{ss}(r)$ for 
    $^{85}$Rb. The separated atom energies relative to $E_{ee}=0$ are shown 
    for $B=$ 0 and 160\,G (16.0\,mT). There are also 19 closed channels in the 
    $d$-wave $M=-4$ block.}
  \begin{ruledtabular}
    \begin{tabular}{c|c|c|c|cl}
      $(f_1\, f_2)$ & $m_{f_1}\, m_{f_2} \, m_\ell$ & $\alpha$ & 
      $E_{\alpha}/h$ [GHz] & $E_{\alpha}/h$ [GHz] \\ & & & for $B=0$ & 
      for $B=160$\,G \\
      \hline
      (2 2) & -1 -1  -2 & $dd$ & 0 & -0.161496 \\ 
      (2 2) &  \,\,0 -2  -2 & $ce$ & 0 & -0.157321 \\
      (2 2) & -1 -2  -1 & $de$ & 0 & -0.080748 \\ 
      (2 2) & -2 -2 \,\,0 & $ee$ & 0 & 0 \\ 
    \end{tabular}
  \end{ruledtabular}
  \label{tab:dVblock}
\end{table}

The anisotropic interaction $V_\mathrm{ss}(r)$ weakly couples the $s$-wave 
block to the $d$-wave block, and the $d$-wave block to the $g$-wave block, 
etc. This has two consequences: extra spin-relaxation channels are possible, 
and projected energy states of $d$-wave character (or even higher partial wave 
character) can give rise to scattering resonances for $s$-wave collisions. 
In the $^{85}$Rb case there are a total of 23 $M=-4$ $d$-wave channels that 
couple to the $s$-wave block illustrated in Table~\ref{tab:sVblock}. Only the 
four listed in Table~\ref{tab:dVblock} are open with respect to the $ee$ 
channel energy. These channels are all degenerate with the $ee$ $s$-wave 
channel at $B=0$. Such a degeneracy at zero magnetic field implies a 
suppression of inelastic collisions due to the Wigner threshold law 
\cite{WignerPR48} outlined in Subsection~\ref{subsec:universalFeshbach}. The 
$d$-wave channels of Table~\ref{tab:dVblock} become open, however, by a 
relatively large amount of energy on the cold $\mu$K temperature scale as $B$ 
increases. The associated energy gap gives rise to inelastic decay by which a 
pair of $e$-state atoms can relax in a collision when $B$ increases from zero 
\cite{RobertsPRL00}. These open $d$-wave channels are also responsible for the 
observed spontaneous dissociation of Fesh\-bach molecules with binding 
energies near the $ee$ scattering threshold \cite{ThompsonPRL05}. Associated 
lifetimes as a function of the magnetic field strength will be discussed in 
Subsection~\ref{subsec:universality}. 

\begin{table}[htdp]
  \caption{Separated atom quantum numbers $(f_1 f_2)F\ell$ for the $s$- and 
    $d$-wave part of $V_\mathrm{int}(r)$ for the bosonic isotope $^{85}$Rb, 
    for which $f$ assumes the values 2 and 3. Odd values of $F$ are missing 
    when $f_1$ equals $f_2$, because of Bose symmetry. Spin-exchange 
    interactions couple states in the same column, whereas states in different 
    columns can only be coupled by spin-dipolar interactions.}
  \begin{ruledtabular}
    \begin{tabular}{c|cccccccccccccc}
      $(f_1\, f_2)$ & & & & &  & &$F\ell$ & & & & &  & &\\
      \hline
      $(3\, 3)$ & $0s$ & & $2s$ & & $4s$ & & $6s$ & $0d$ & & $2d$ & & $4d$ 
      & & $6d$\\
      $(2\, 3)$ &  & $1s$ & $2s$ & $3s$ & $4s$ & $5s$ & & & $1d$ & $2d$ 
      & $3d$ & $4d$ & $5d$ &\\
      $(2\, 2)$ & $0s$ & & $2s$ & & $4s$ & & & $0d$ & & $2d$ & & $4d$ & &\\
    \end{tabular}
  \end{ruledtabular}
  \label{tab:fblocks}
\end{table} 

Table~\ref{tab:fblocks} illustrates the blocks of the matrix 
$V_\mathrm{int}(r)$ according to $\{f_1 f_2\}F \ell$ quantum numbers. The 
example refers to Bose atoms with a nuclear spin quantum number of $i=5/2$, 
such as $^{85}$Rb. Basis states in the same vertical column of 
Table~\ref{tab:fblocks}, that is, with the same $F\ell$ quantum numbers, are 
coupled by the strong exchange interactions in $V_\mathrm{el}(r)$. Basis 
states from different vertical columns can only be coupled by the weak 
interactions in $V_\mathrm{ss}(r)$. Classification using $F\ell$ blocks was 
used for Fesh\-bach resonance states with $\ell>0$ observed for the 
comparatively heavy alkali metal atoms $^{87}$Rb \cite{MartePRL02} and 
$^{133}$Cs \cite{ChinPRA04}. In particular, the $^{133}$Cs$_2$ Fesh\-bach 
molecules associated with the Stern-Gerlach separation experiments of 
Fig.~\ref{fig:Herbiglevitation} are mainly of $g$-wave character with $F=4$ 
and $\ell=4$. Such a separated atom classification shall be applied in the 
following to Fesh\-bach resonance and molecular states of $^{85}$Rb associated 
with the observations of Fig.~\ref{fig:DonleyFringe}.

\subsubsection{Near threshold bound states}
\begin{figure}[htbp]
\includegraphics[angle=0,width=\columnwidth,clip]{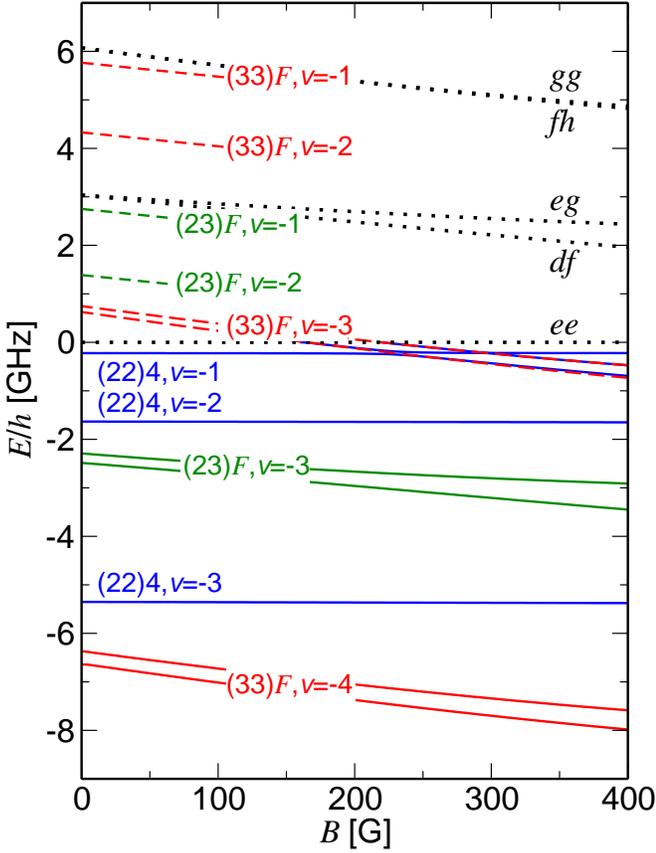}
  \caption{(Colour in online edition)
    Magnetic field dependent $M=-4$ $s$-wave energy levels of the  
    $^{85}$Rb$_2$ dimer. The dotted curves labelled by $ee$, $df$, $eg$, 
    $fh$ and $gg$ show the energies of the five separated atom channels of 
    Table~\ref{tab:sVblock}. Solid curves indicate the calculated $s$-wave 
    coupled channels bound state energies labelled by the vibrational quantum 
    number, $v$. Their symmetry labels refer to the set of quantum numbers 
    $(f_1f_2)F$. The $ee$ limit gives rise to a single vibrational progression 
    with $F=4$; the $df$ and $eg$ limits give rise to two $(23)F$ series with 
    $F=$ 4 and 5; and the $fh$ and $gg$ limits give rise to two $(33)F$ series 
    with $F=$ 4 and 6. In all cases, the $F=4$ level has the lower energy of 
    the $(f_1 f_2)$ pairs. At $B=0$ the $v=-1$ and $-2$, $(23)F$ levels and 
    the $v=-1$, $-2$, and $-3$, $(33)F$ levels represent meta-stable states 
    with $E>0$ embedded in the $(22)4$ $ee$ scattering continuum. These 
    meta-stable levels are represented by dashed curves, which give the 
    approximate positions of scattering resonances in the $ee$ channel. The 
    meta-stable levels with $F=4$ are coupled to the $ee$ $s$-wave scattering 
    continuum through the exchange interaction. The $F=$ 5 and 6 meta-stable 
    levels are not exchange coupled to the $ee$ entrance channel at $B=0$, 
    but become weakly coupled at higher fields where $F$ is no longer a good 
    quantum number.}
  \label{fig:psj_Rb85_Elev}
\end{figure}

The illustration of the relation between resonance and bound states on the one 
hand and singularities of the scattering length on the other hand relies upon
coupled channels calculations. Figure~\ref{fig:psj_Rb85_Elev} shows the five 
separated atom channel energies $E_\alpha(B)$ for the $s$-wave block described 
in Table~\ref{tab:sVblock}. In accordance with the convention used throughout 
this review, the scattering threshold associated with the $ee$ entrance 
channel defines the zero of energy. As the five channel energies just reflect 
the sums of their associated atomic Zeeman energies, they cluster into three 
groups. The lowest energy group consists just of the $ee$ entrance channel. 
The next group with energies on the order of $E_\mathrm{hf}$ includes the $df$ 
and $eg$ channels. Finally, the group with energies of about $2E_\mathrm{hf}$ 
refers to the $fh$ and $gg$ channels. Both channel energies associated with 
each one of the latter two groups are degenerate at $B=0$. At the low magnetic 
field strengths shown in Fig.~\ref{fig:psj_Rb85_Elev}, the choice of 
$(f_1 f_2)F$ quantum numbers well characterises the two-body physics. The 
$(22)$ entrance channel has $F=4$, the $(23)$ group has $F=$ 4 and 5 and the 
$(33)$ group has $F=$ 4 and 6. Only $F$ values of 4 or more are possible 
because the projection quantum number is $M=-4$; the odd value $F=5$ is ruled 
out for identical bosons with $(f_1 f_2)=(33)$.

The solid curves in Figure~\ref{fig:psj_Rb85_Elev} show the $s$-wave bound 
states of the $^{85}$Rb$_2$ molecular dimer with negative energies, 
$E_\mathrm{b}<0$. The curves virtually parallel to the $E=0$ threshold refer 
to the $v=$ $-1$, $-2$, and $-3$ states of the $ee$ entrance channel, labelled 
by their vibrational quantum numbers, $v$, starting with $v=-1$ for the 
highest excited level. In addition to $v$, their low field $(22)4$ quantum 
numbers are indicated in Fig.~\ref{fig:psj_Rb85_Elev}. These levels all have 
nearly the same magnetic moment as a pair of separated atoms in the entrance 
channel Zeeman state configuration. Each one of the other four, closed 
channels also has a vibrational series leading to each of the four 
closed-channel thresholds. The pair of $(23)F$, $v=-3$ levels with $F=$ 4 and 
5 and the pair of $(33)F$, $v=-4$ levels with $F=$ 4 and 6 are both bound with 
respect to the separated atoms. For both pairs the lowest level has $F=4$. 
These levels have different magnetic moments from a pair of $ee$ separated 
atoms. This implies that their bound state energies, $E_\mathrm{b}(B)$, 
relative to $E=0$ vary significantly with the magnetic field strength $B$.

\begin{figure}[htbp]
  \includegraphics[angle=0,width=\columnwidth,clip]{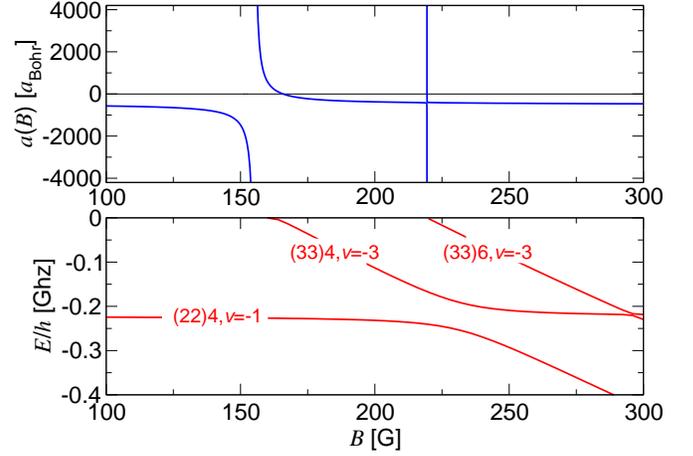}
  \caption{(Colour in online edition)
    Scattering length (upper panel) and bound state energy levels 
    (lower panel) versus $B$ for two $^{85}$Rb atoms in the $e$ state. The 
    separated atom energy of two $e$-state atoms is taken to be zero. A 
    singularity of the scattering length occurs where an energy level becomes 
    degenerate with the scattering threshold at $E=0$. The resolution of the 
    figure is not sufficient to show the variation of the bound state energies 
    with $B$ just below threshold; this variation is discussed in 
    Fig.~\ref{fig:85RbEbofB} of Subsection~\ref{subsec:universalFeshbach}. The 
    broad resonance is due to the threshold crossing of the strongly coupled 
    $(f_1f_2)F\ell=(33)4s$, $v=-3$ level, whereas the narrow resonance is due 
    to the crossing of the weakly coupled $(f_1f_2)F\ell=(33)6s$, $v=-3$ 
    level.}
  \label{fig:psj_Rb85_A.E}
\end{figure}

While the $^{85}$Rb$_2$ molecular dimer states involve the complete strong 
electronic interaction, $V_\mathrm{el}(r)$, the Fesh\-bach resonance levels 
of Fig.~\ref{fig:psj_Rb85_Elev} refer to blocks of the potential matrix 
associated with their symmetry labels. At zero magnetic field the $(23)$, $v=$ 
$-1$ and $-2$ levels and the $(33)$, $v=$ $-1$, $-2$, and $-3$ levels have 
positive energies so that they are meta-stable levels embedded in the $ee$ 
scattering continuum. The two $(33)F$ levels are especially interesting, in 
that they cross the $ee$ $E=0$ threshold and turn into bound states when $B$ 
is sufficiently large. Figure~\ref{fig:psj_Rb85_A.E} shows an expanded view of 
the bound state energy levels in the near-threshold crossing region as well as 
the $s$-wave scattering length $a$ of the $ee$ entrance channel as a 
function of $B$. The scattering length has a singularity at the magnetic field 
value $B_0$ where a new bound state appears with $E_\mathrm{b}(B_0)=0$. 
The $F=4$ $(33)$ level is coupled by the exchange interaction to the $F=4$ 
$ee$ entrance channel, and gives rise to a broad resonance with an associated 
singularity of $a$ near 155\,G. On the other hand, the $F=6$ $(33)$ level is 
only weakly coupled to the $F=4$ $ee$ entrance channel, and gives rise to a 
narrow zero energy resonance near 220\,G. In addition, the interaction between 
the $(33)F$ levels and the highest excited entrance channel bound state at 
$E_{-1}/h=-0.22$\,GHz is evident in Fig.~\ref{fig:psj_Rb85_A.E}. It results, 
for instance, in the avoided crossing of the $(22)4$, $v=-1$ and $(33)4$, 
$v=-3$ levels between 200\,G and 250\,G which is due to the same coupling that 
leads to the broad singularity of the scattering length near 155\,G. 

\subsection{Two-channel approach}
\label{subsec:twochannel}
Coupled channels calculations based on realistic molecular potentials and 
known atomic properties are capable of accounting for a variety of 
experimental collisional and bound state properties of alkali-metal species 
\cite{MartePRL02,ChinPRA04,BartensteinPRL05,MarcelisPRA04,LoftusPRL02,Abeelenscatteringlengths,LeoPRL00,HoubiersPRA98}. 
Their accuracy, including predictions of new resonances, is possible once the 
actual potentials have been calibrated in such a way that they recover the 
correct scattering lengths for the Born-Oppenheimer $^1\Sigma_g^+$ and 
$^3\Sigma_u^+$ potentials \cite{AbrahamPRA97,KempenPRL02} in addition to the 
van der Waals coefficient, $C_6$. Coupled channels approaches have the 
drawback, however, that they are not readily accessible. Consequently, it is 
desirable to find simpler approaches to Fesh\-bach resonance and molecular 
levels close to the dissociation threshold energy. To this end, the physics of 
binary collisions as well as the properties of the highly excited molecular 
bound states can usually be well described in terms of just two scattering 
channels \cite{Child74,MoerdijkPRA95,TimmermansPhysRep99,MiesPRA00}. In the 
following, we denote by the entrance channel the Zeeman state configuration of 
a pair of asymptotically separated atoms in which a dilute gas is initially 
prepared. Under the conditions of resonance enhancement, this spin 
configuration is strongly coupled, in general, to several energetically closed 
scattering channels. In idealised treatments, however, this coupling is 
usually due to the near degeneracy of the energies of a single meta-stable 
vibrational state, the Fesh\-bach resonance state $\phi_\mathrm{res}(r)$, and 
the colliding atoms. The spin configuration associated with the Fesh\-bach 
resonance state is referred to, in the following, simply as the closed 
channel. 

\subsubsection{Two-channel Hamiltonian}
\begin{figure}[htbp]
  \includegraphics[width=\columnwidth,clip]{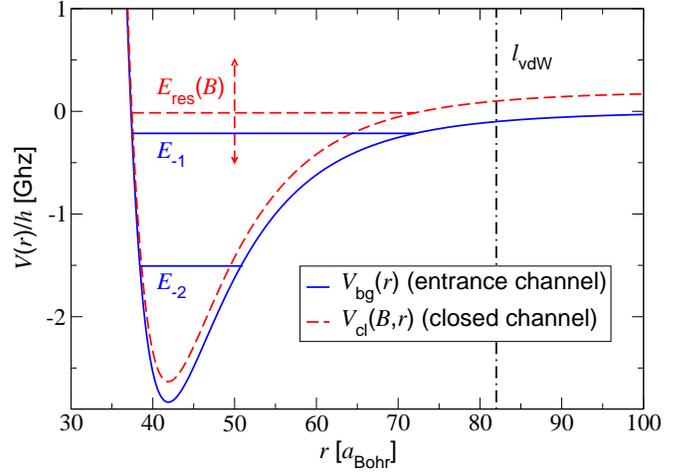}
  \caption{(Colour in online edition)
    Scheme of the entrance (solid curve) and closed channel 
    (dashed curve) potentials associated with a model representation of the 
    155 G zero energy resonance of $^{85}$Rb \cite{KoehlerPRL03}. The 
    horizontal dashed and vertical dot-dashed lines indicate the magnetic 
    field dependent energy $E_\mathrm{res}(B)$ of the Fesh\-bach resonance 
    level $\phi_\mathrm{res}(r)$ and the van der Waals length of 
    $l_\mathrm{vdW}=82\,a_\mathrm{Bohr}$, respectively. In this model the 
    outer classical turning point of the resonance level 
    $\phi_\mathrm{res}(r)$ associated with the energy $E_\mathrm{res}(B)$ in 
    the closed channel potential is located at about 
    $r_\mathrm{classical}=72\,a_\mathrm{Bohr}$. Its vibrational quantum number 
    $v=-1$ with respect to the dissociation threshold of $V_\mathrm{cl}(B,r)$ 
    is chosen arbitrarily. The horizontal solid lines indicate the energies 
    $E_{-1}$ and $E_{-2}$ of the bare vibrational bound states associated with 
    the entrance channel potential. This bare interaction is adjusted in such 
    way that it mimics the highest excited $(f_1f_2)F=(22)4$, $v=$ $-1$ and 
    $-2$ levels of Fig.~\ref{fig:psj_Rb85_Elev}. In this model the 
    off-diagonal spin exchange coupling has been chosen to be 
    $W(r)=\beta\exp(-r/\alpha)$, using $\alpha=5\,a_\mathrm{Bohr}$ and 
    $\beta/k_\mathrm{B}=38.5\,$mK to match the measured Fesh\-bach resonance 
    parameters \cite{ClaussenPRA03}.}
  \label{fig:85RbFeshbachmodel}
\end{figure}

The general Hamiltonian of the relative motion of an atom pair, i.e., the 
basis of all two-channel approaches
\cite{Child74,MoerdijkPRA95,TimmermansPhysRep99,MiesPRA00}, is given by the 
following matrix:
\begin{equation}
  H_\mathrm{2B}=
  \left(
  \begin{array}{cc}
    H_\mathrm{bg} & W(r)\\
    W(r) & H_\mathrm{cl}(B)
  \end{array}
  \right).
  \label{H2B2channel}
\end{equation}
Its off diagonal elements, $W(r)$, are the energies associated with the spin 
exchange (or dipole) interaction and provide the inter-channel coupling as a 
function of the distance $r$ between the atoms. The diagonal elements, 
$H_\mathrm{bg}$ and $H_\mathrm{cl}(B)$, can be interpreted in terms of 
entrance and closed channel Hamiltonians in the hypothetical absence of 
coupling, respectively. These single channel Hamiltonians consist of kinetic 
and potential energy contributions given by the following formulae:
\begin{align}
  H_\mathrm{bg}&=-\frac{\hbar^2}{m}\boldsymbol{\nabla}^2+V_\mathrm{bg}(r),
  \label{Hbg}\\
  H_\mathrm{cl}(B)&=-\frac{\hbar^2}{m}\boldsymbol{\nabla}^2+V_\mathrm{cl}(B,r).
  \label{Hcl}
\end{align}
Here and in the following we choose the zero of energy as the threshold for
dissociation of the entrance channel. Consequently, the background scattering 
potential $V_\mathrm{bg}(r)$ of Eq.~(\ref{Hbg}) vanishes in the limit of large 
inter-atomic separations, in accordance with the general asymptotic behaviour 
of the van der Waals interaction described by Eq.~(\ref{potentialvdW}). In
this approach the complete two-body Hamiltonian of Eq.~(\ref{H2B2channel}) 
depends on the magnetic field strength $B$ simply via an overall shift of the 
closed channel potential $V_\mathrm{cl}(B,r)$ with respect to the entrance
channel dissociation threshold. The amount of this shift is determined by the 
sum of the single particle Zeeman energies associated with the closed channel 
spin configuration of a pair of atoms. Typical diagonal potential energy 
contributions to the two-body Hamiltonian are illustrated schematically in 
Fig.~\ref{fig:85RbFeshbachmodel}, whereas special requirements for them will
be discussed in Subsection~\ref{subsec:parameters}. 

\subsubsection{The bare Fesh\-bach resonance state}
The closed channel Hamiltonian supports the bare Fesh\-bach resonance state in 
accordance with the following Schr\"odinger equation:
\begin{equation}
  H_\mathrm{cl}(B)\phi_\mathrm{res}(r)=E_\mathrm{res}(B)\phi_\mathrm{res}(r). 
  \label{SEphires}
\end{equation}
The associated resonance energy, $E_\mathrm{res}(B)$, can be tuned with 
respect to the entrance channel dissociation threshold, i.e.~the zero of 
energy in Fig.~\ref{fig:85RbFeshbachmodel}, by varying the magnetic field 
strength. In the hypothetical absence of inter-channel coupling the Fesh\-bach 
resonance level usually represents a comparatively tightly bound state of 
$H_\mathrm{cl}(B)$ with a classical radius $r_\mathrm{classical}$ smaller than 
the van der Waals length of Eq.~(\ref{lvdW}). Under realistic conditions, 
however, $\phi_\mathrm{res}(r)$ is only a meta-stable state with a decay width 
depending on the strength of the inter-channel coupling. Its precise form can 
not always be unambiguously identified from full coupled channels calculations
because the two-channel picture is an idealisation. The remarkable accuracy of 
such a simplifying approach is due to the fact that near resonance the 
physically relevant dressed stationary energy levels are largely insensitive 
to the detailed structure of the resonance state.

\subsection{Dressed energy states}
\label{subsec:dressedenergystates}
The dressed stationary energy states consist of two orthogonal components 
associated with the spin configurations $|\mathrm{bg}\rangle$ and 
$|\mathrm{cl}\rangle$ of the entrance and closed channels, respectively. Their 
wave functions are therefore of the following general form
\cite{DrummondPRL98}:
$|\mathrm{bg}\rangle\phi_\mathrm{bg}(\mathbf{r})
+|\mathrm{cl}\rangle\phi_\mathrm{cl}(\mathbf{r})$.
The components $\phi_\mathrm{bg}(\mathbf{r})$ and 
$\phi_\mathrm{cl}(\mathbf{r})$ depend on the relative position $\mathbf{r}$ of 
the atoms. In accordance with Eq.~(\ref{H2B2channel}), these components are 
determined by the following set of coupled stationary Schr\"odinger equations:
\begin{align}
  \label{SEphibg}
  H_\mathrm{bg}\phi_\mathrm{bg}(\mathbf{r})
  +W(r)\phi_\mathrm{cl}(\mathbf{r})&=E\phi_\mathrm{bg}(\mathbf{r}),\\
  \label{SEphicl}
  W(r)\phi_\mathrm{bg}(\mathbf{r})+H_\mathrm{cl}(B)
  \phi_\mathrm{cl}(\mathbf{r})&=E\phi_\mathrm{cl}(\mathbf{r}).
\end{align}
All applications in this review will involve either symmetric or 
anti-symmetric spin configurations of identical Bose and Fermi atom pairs, 
respectively. Their physical spatial wave functions will only consist of an
$s$-wave component. The solutions of Eqs.~(\ref{SEphibg}) and (\ref{SEphicl}) 
with a negative energy $E$ below the entrance channel dissociation threshold 
are associated with molecular bound states. Solutions with a positive energy 
belong to the continuum spectrum of the Hamiltonian and describe collisions of 
initially separated pairs of atoms.

\subsubsection{Dressed continuum states}
The dressed continuum states can be labelled by the relative momentum 
$\mathbf{p}$ of a pair of asymptotically separated atoms in the entrance 
channel spin configuration and are associated with the collision energies, 
$E=p^2/m$. Due to the continuous range of angles of incidence between the 
atoms, any given collision energy is infinitely degenerate. In this context it 
is convenient to postpone the introduction of any symmetry properties of the 
spatial wave functions associated with a possible identical nature of the 
atoms. Continuum states are usually chosen in such a way that their entrance 
channel component, $\phi_\mathbf{p}^\mathrm{bg}(\mathbf{r})$, behaves at 
asymptotically large inter-atomic distances like a superposition of an 
incident plane wave and an outgoing spherical wave \cite{Taylor72}. This 
choice of the wave function leads to the following boundary condition: 
\begin{equation}
  \phi_\mathbf{p}^\mathrm{bg}(\mathbf{r})
  \underset{r\to \infty}{\sim}\frac{1}{(2\pi\hbar)^{3/2}}
  \left[e^{i\mathbf{p}\cdot\mathbf{r}/\hbar}+f(\vartheta,p)
    \frac{e^{ipr/\hbar}}{r}\right].
  \label{BCphipbg}
\end{equation}
The quantity $f(\vartheta,p)$ is known as the scattering amplitude and depends 
on the modulus $p$ of the relative momentum of the atoms and on the scattering 
angle determined by the relation: 
$\cos\vartheta=\mathbf{p}\cdot\mathbf{r}/(pr)$. Due to the off diagonal 
coupling, $W(r)$, the stationary continuum states also have a closed channel 
component. The associated wave function, 
$\phi_\mathbf{p}^\mathrm{cl}(\mathbf{r})$, decays at asymptotically large 
inter-atomic distances because the cold collision energies of interest are 
below the closed channel dissociation threshold. This property reflects 
physical intuition as closed channel atom pairs are spatially confined by the 
potential $V_\mathrm{cl}(r)$ of Fig.~\ref{fig:85RbFeshbachmodel}.

Due to their long range of Eq.~(\ref{BCphipbg}), stationary continuum wave 
functions may be interpreted in terms of amplitudes for the density of 
particle flux rather than physical states \cite{Taylor72}. In the present 
context, the entrance channel component, 
$\phi_\mathbf{p}^\mathrm{bg}(\mathbf{r})$, determines observable low energy 
scattering properties of colliding atom pairs, such as, for instance, the 
differential cross section of distinguishable atoms, $|f(\vartheta,p)|^2$. 
These properties can be represented in terms of bare energy states associated 
with the Hamiltonians $H_\mathrm{bg}$ and $H_\mathrm{cl}(B)$. To this end, it 
is instructive to reformulate the coupled set of stationary Schr\"odinger 
equations~(\ref{SEphibg}) and (\ref{SEphicl}), including the boundary 
condition of Eq.~(\ref{BCphipbg}), in terms of the associated Green's 
functions in addition to the entrance channel continuum states. The bare 
Green's functions depend on a complex variable $z$ with the dimension of an 
energy in accordance with the following formulae:
\begin{align}
  \label{Gbg}
  G_\mathrm{bg}(z)&=
  \left(z-H_\mathrm{bg}\right)^{-1},\\
  G_\mathrm{cl}(B,z)&=
  \left[z-H_\mathrm{cl}(B)\right]^{-1}.
  \label{Gcl}
\end{align} 
The entrance channel continuum wave functions, 
$\phi_\mathbf{p}^{(+)}(\mathbf{r})$, also referred to as background scattering 
states, satisfy the stationary Schr\"odinger equation associated with the 
Hamiltonian $H_\mathrm{bg}$ of Eq.~(\ref{Hbg}), i.e.:
\begin{align}
  H_\mathrm{bg}\phi_\mathbf{p}^{(+)}(\mathbf{r})=
  \frac{p^2}{m}\phi_\mathbf{p}^{(+)}(\mathbf{r}).
  \label{barephip}
\end{align}
Their long range behaviour is determined by boundary conditions analogous to 
Eq.~(\ref{BCphipbg}) with $f(\vartheta,p)$ replaced by the bare amplitude, 
$f_\mathrm{bg}(\vartheta,p)$, associated with the background scattering.

Expressed in terms of the bare Green's functions and continuum states, the 
Schr\"odinger equations (\ref{SEphibg}) and (\ref{SEphicl}) read: 
\begin{align}
  \label{LSphipbg}
  |\phi_\mathbf{p}^\mathrm{bg}\rangle&=|\phi_\mathbf{p}^{(+)}\rangle+
  G_\mathrm{bg}(E+i0)W|\phi_\mathbf{p}^\mathrm{cl}\rangle,\\
  |\phi_\mathbf{p}^\mathrm{cl}\rangle&=
  G_\mathrm{cl}(B,E)W|\phi_\mathbf{p}^\mathrm{bg}\rangle.
  \label{LSphipcl}
\end{align}
The argument ``$z=E+i0$'' of the entrance channel Green's function in 
Eq.~(\ref{LSphipbg}) indicates that the physical collision energy $E=p^2/m$ is 
approached from the upper half of the complex plane. This choice of the energy 
argument ensures that the scattering wave function 
$\phi_\mathbf{p}^\mathrm{bg}(\mathbf{r})$ is compatible with 
Eq.~(\ref{BCphipbg}), in accordance with the following long range asymptotic 
behaviour of $G_\mathrm{bg}(z)$ in spatial coordinates:
\begin{equation}
  G_\mathrm{bg}(z,\mathbf{r},\mathbf{r}')\underset{r\to\infty}{\sim}-
  \frac{m(2\pi\hbar)^{3/2}}{4\pi\hbar^2}\frac{e^{ipr/\hbar}}{r}
  \left[\phi_\mathbf{p}^{(-)}(\mathbf{r}')\right]^*.
  \label{asympGbg}
\end{equation}
Here $\phi_\mathbf{p}^{(-)}(\mathbf{r}')=
\left[\phi_{-\mathbf{p}}^{(+)}(\mathbf{r}')\right]^*$ is the entrance channel 
continuum energy state with incoming spherical wave boundary conditions 
\cite{Taylor72}. Its label, $\mathbf{p}=(mE)^{1/2}\mathbf{r}/r$, may be 
interpreted as the asymptotic momentum associated with the relative motion of 
the scattered atoms. The closed channel continuum wave function 
$\phi_\mathbf{p}^\mathrm{cl}(\mathbf{r})$ of Eq.~(\ref{LSphipcl}) decays at 
asymptotically large inter-atomic distances because the bare Green's function 
on the right hand side is evaluated at the collision energy $E=p^2/m$ below 
the dissociation threshold of $V_\mathrm{cl}(r)$.

Direct application of the two-channel Hamiltonian (\ref{H2B2channel}) to 
Eqs.~(\ref{LSphipbg}) and (\ref{LSphipcl}) verifies their equivalence to the 
coupled Schr\"odinger equations (\ref{SEphibg}) and (\ref{SEphicl}). The 
associated set of integral equations for the spatial wave functions 
$\phi_\mathbf{p}^\mathrm{bg}(\mathbf{r})$ and
$\phi_\mathbf{p}^\mathrm{cl}(\mathbf{r})$ may be used, for instance, to 
numerically determine the exact dressed energy states in the two-channel 
approach \cite{KoehlerPRL03}. This formulation of the two-channel scattering 
problem is also particularly useful for an approximate treatment based on the 
singularities of the closed channel Green's function. The single resonance 
approach, underlying this treatment, provides analytic formulae for the 
dressed continuum wave functions as well as their associated collision 
parameters in terms of the bare energy states.

\subsubsection{Single resonance approach}
In accordance with Eqs.~(\ref{SEphires}) and (\ref{Gcl}), the closed channel 
Green's function has a singularity at the resonance energy 
$E_\mathrm{res}(B)$. Provided that the Fesh\-bach resonance state 
$\phi_\mathrm{res}(r)$ is unit normalised, 
i.e.~$\langle\phi_\mathrm{res}|\phi_\mathrm{res}\rangle=1$, the singular 
diagonal matrix element of $G_\mathrm{cl}(B,E)$ is given by the following 
formula:
\begin{equation} 
  \langle\phi_\mathrm{res}|G_\mathrm{cl}(B,E)|\phi_\mathrm{res}\rangle=
  \left[E-E_\mathrm{res}(B)\right]^{-1}. 
  \label{poleapproximation1}
\end{equation}
Based on Eq.~(\ref{poleapproximation1}), the single resonance approach takes 
advantage of the near degeneracy of the resonance energy with the entrance 
channel dissociation threshold illustrated in 
Fig.~\ref{fig:85RbFeshbachmodel}: At typical cold collision energies, 
$E=p^2/m$, the resonance detuning $\Delta E(B)=E-E_\mathrm{res}(B)$ in 
Eq.~(\ref{poleapproximation1}) is negligible compared to the spacings between 
the discrete energy levels of $H_\mathrm{cl}(B)$. Consequently, the closed 
channel Green's function in Eq.~(\ref{LSphipcl}) is dominated by its virtually 
singular diagonal contribution associated with the Fesh\-bach resonance level. 
These estimates thus lead to the following approximation 
\cite{Child74,GoralJPhysB04}: 
\begin{equation}
  G_\mathrm{cl}(B,E)\approx |\phi_\mathrm{res}\rangle
  \frac{1}{\Delta E(B)}
  \langle\phi_\mathrm{res}|.
  \label{poleapproximation}
\end{equation} 

The single resonance approach consists in replacing the bare Green's function 
in Eq.~(\ref{LSphipcl}) by the right hand side of 
Eq.~(\ref{poleapproximation}). This replacement implies that the functional 
form of the closed channel component of the dressed continuum wave function
$\phi_\mathbf{p}^\mathrm{cl}(\mathbf{r})$ is given by the Fesh\-bach resonance 
state $\phi_\mathrm{res}(r)$. The associated overlap factor 
$\langle\phi_\mathrm{res}|\phi_\mathbf{p}^\mathrm{cl}\rangle$ is determined by 
the energy and magnetic field dependent amplitude:
\begin{equation}
  A(B,E)=\langle\phi_\mathrm{res}|W|\phi_\mathbf{p}^\mathrm{bg}\rangle/
  \Delta E(B).
  \label{amplitude1}
\end{equation}
The dressed state component $|\phi_\mathbf{p}^\mathrm{cl}\rangle$ on the right 
hand side of Eq.~(\ref{LSphipbg}) can then also be eliminated in favour of the 
product of $|\phi_\mathrm{res}\rangle$ and $A(B,E)$. These replacements in 
Eqs.~(\ref{LSphipbg}) and (\ref{LSphipcl}) give the following explicit 
formulae for the dressed continuum states in terms of the bare Fesh\-bach 
resonance and background scattering states:
\begin{align}
  \label{phipbg}
  |\phi_\mathbf{p}^\mathrm{bg}\rangle&=|\phi_\mathbf{p}^{(+)}\rangle+
  G_\mathrm{bg}(E+i0)W|\phi_\mathrm{res}\rangle A(B,E),\\
  |\phi_\mathbf{p}^\mathrm{cl}\rangle&=|\phi_\mathrm{res}\rangle A(B,E).
  \label{phipcl}
\end{align}
The amplitude $A(B,E)$ of Eq.~(\ref{amplitude1}) may be expressed in terms of 
the same bare states by inserting Eq.~(\ref{phipbg}) into 
Eq.~(\ref{amplitude1}). This yields:
\begin{equation}
  A(B,E)=\frac{\langle\phi_\mathrm{res}|W|\phi_\mathbf{p}^{(+)}\rangle}
  {\Delta E(B)-
    \langle\phi_\mathrm{res}|WG_\mathrm{bg}(E+i0)W|\phi_\mathrm{res}\rangle}.
  \label{amplitude}
\end{equation} 
 
We note that the dressed continuum wave functions of Eqs.~(\ref{phipbg}) and 
(\ref{phipcl}) depend on the magnetic field strength only through the detuning 
$\Delta E(B)$ of the resonance energy $E_\mathrm{res}(B)$ in the denominator 
of Eq.~(\ref{amplitude}). Within the typical range of experimental magnetic 
field strengths $E_\mathrm{res}(B)$ is a linear function of $B$ to a good 
approximation. This linear dependence may be represented in terms of an 
expansion of $E_\mathrm{res}(B)$ about the resonant field, $B_\mathrm{res}$, 
at which the Fesh\-bach resonance level crosses the dissociation threshold of 
the entrance channel. In accordance with the choice of the zero of energy in 
Fig.~\ref{fig:85RbFeshbachmodel}, $B_\mathrm{res}$ is determined by the 
relation $E_\mathrm{res}(B_\mathrm{res})=0$. This implies the simple formula
\cite{MoerdijkPRA95}:
\begin{equation}
  E_\mathrm{res}(B)=\mu_\mathrm{res}
  (B-B_\mathrm{res}).
  \label{slope}
\end{equation}
Here the slope of the linear curve, $\mu_\mathrm{res}$, is the difference 
between the magnetic moments of the Fesh\-bach resonance state and a pair of 
asymptotically separated non-interacting atoms.
 
The single resonance approach of Eqs.~(\ref{phipbg}), (\ref{phipcl}) and 
(\ref{amplitude}) gives an exact representation of the dressed continuum 
states provided that the spatial configuration of a closed channel atom pair 
is restricted to the Fesh\-bach resonance state $\phi_\mathrm{res}(r)$. Such 
an assumption is associated with the following formal replacement of the 
closed channel Hamiltonian:
\begin{equation}
  H_\mathrm{cl}(B)\to 
  |\phi_\mathrm{res}\rangle E_\mathrm{res}(B)\langle\phi_\mathrm{res}|.
  \label{replacementHcl}
\end{equation}
This simplification of the complete two-channel Hamiltonian of 
Eq.~(\ref{H2B2channel}) yields low energy scattering amplitudes, 
$f(\vartheta,p)$, sufficiently accurate to determine the magnetic field 
dependence of the $s$-wave scattering length $a(B)$.

\subsubsection{Width and shift of a zero energy resonance}
The $s$-wave scattering length is determined by the long range asymptotic 
behaviour associated with the dressed continuum states in terms of the zero 
momentum limit of the scattering amplitude. This limit is well represented by 
the following partial wave analysis \cite{Taylor72}:
\begin{equation}
  f(\vartheta,p)=\sum_{\ell=0}^{\infty}(2\ell+1)f_\ell(p)P_\ell(\cos\vartheta)
  \underset{p\to 0}{\sim}-a.
  \label{definitiona}
\end{equation}
Here $\ell$ labels the quantum number associated with the orbital angular 
momentum and $P_\ell(\cos\vartheta)$ is a Legendre polynomial. The $s$-wave
scattering amplitude is related to the phase shift of 
Eq.~(\ref{boundaryradial}) through the formula 
$f_0(p)=\hbar \exp[i\xi(p)]\sin\xi(p)/p$. We note that the limit $p\to 0$ 
implies rotational symmetry of the entire wave function 
$\phi_\mathbf{p}^\mathrm{bg}(\mathbf{r})$, i.e.~independence of the scattering 
angle $\vartheta$, because the incident plane wave in Eq.~(\ref{BCphipbg}) is 
isotropic at zero momentum. The higher angular momentum components $f_\ell(p)$ 
of the scattering amplitude, besides the $s$ wave associated with $\ell=0$, 
are usually negligible in applications to binary collisions in cold gases, due 
to their proportionality to $p^{2\ell}$.

The behaviour of $\phi_\mathbf{p}^\mathrm{bg}(\mathbf{r})$ at asymptotically 
large inter-atomic distances $r$ can be inferred from Eq.~(\ref{asympGbg}) and 
from the bare amplitude, $f_\mathrm{bg}(\vartheta,p)$, associated with the 
background scattering wave function, $\phi_\mathbf{p}^{(+)}(\mathbf{r})$. In 
accordance with the explicit representation of the dressed continuum wave 
function of Eq.~(\ref{phipbg}), the scattering amplitude is thus given by the 
following expression:
\begin{equation}
  f(\vartheta,p)=f_\mathrm{bg}(\vartheta,p)-
  \frac{m(2\pi\hbar)^3
  \langle\phi_\mathbf{p}^{(-)}|W|\phi_\mathrm{res}\rangle}{4\pi\hbar^2} 
  A(B,E).
  \label{scatteringamplitude}
\end{equation}
In the zero momentum limit this expression recovers the resonance enhanced 
$s$-wave scattering length in terms of the formula \cite{MoerdijkPRA95}:
\begin{equation}
  a(B)=a_\mathrm{bg}\left(1-\frac{\Delta B}{B-B_0}\right).
  \label{aofB}
\end{equation}  
Its parameters are the background scattering length, $a_\mathrm{bg}$, the 
width, $\Delta B$, and the position of the zero energy resonance, $B_0$. The 
background scattering length is associated with the bare scattering amplitude, 
$f_\mathrm{bg}(\vartheta,p)$, via a relation analogous to 
Eq.~(\ref{definitiona}). The zero momentum limit of 
Eq.~(\ref{scatteringamplitude}) determines the width of the resonance to be:
\begin{equation}
  \Delta B=\frac{m(2\pi\hbar)^3}{4\pi\hbar^2 a_\mathrm{bg}\mu_\mathrm{res}}
    \left|\langle\phi_\mathrm{res}|W|\phi_0^{(+)}\rangle\right|^2.
       \label{resonancewidth}
\end{equation}

We note that Eq.~(\ref{aofB}) predicts the scattering length $a(B)$ to assume 
all values from $-\infty$ to $+\infty$ due to its singularity at $B_0$, the 
measurable magnetic field strength at which the zero energy resonance occurs. 
The width, $\Delta B$, characterises the distance in magnetic fields between 
the position of the singularity, $B_0$, and the zero of the scattering length. 
The denominator of the amplitude $A(B,E=0)$ of Eq.~(\ref{amplitude}) 
determines the physical resonance position, $B_0$, relative to the zero energy 
crossing point of the bare Fesh\-bach resonance level, $B_\mathrm{res}$, to 
be:
\begin{equation}
  B_0=B_\mathrm{res}-
  \langle\phi_\mathrm{res}|WG_\mathrm{bg}(0)W|\phi_\mathrm{res}\rangle/
  \mu_\mathrm{res}.
  \label{resonanceshift}
\end{equation}
The absolute magnitude of $B_0$ enters the two-channel approach as an 
adjustable parameter because the magnetic field strength $B_\mathrm{res}$ is 
not directly measurable. The resonance shift, $B_0-B_\mathrm{res}$, however, 
determines characteristic properties of the highest excited dressed 
vibrational bound state, such as, e.g., the admixture of the closed channel 
spin configuration to its wave function.

\subsubsection{Dressed molecular bound states}
In analogy to Eqs.~(\ref{LSphipbg}) and (\ref{LSphipcl}), the two-channel
dressed molecular bound states are determined in terms of the bare Green's 
functions by coupled matrix equations of the following form:
\begin{align}
  \label{LSphibbg}
  |\phi_\mathrm{b}^\mathrm{bg}\rangle&=
  G_\mathrm{bg}(E_\mathrm{b})W|\phi_\mathrm{b}^\mathrm{cl}\rangle,\\
  |\phi_\mathrm{b}^\mathrm{cl}\rangle&=
  G_\mathrm{cl}(B,E_\mathrm{b})W|\phi_\mathrm{b}^\mathrm{bg}\rangle.
  \label{LSphibcl}
\end{align}
The discrete bound state energies, $E_\mathrm{b}$, are negative, i.e.~below 
the dissociation threshold of the entrance channel. Both channels are 
therefore closed and, consequently, the atom pairs are confined by the 
potential wells of $V_\mathrm{bg}(r)$ and $V_\mathrm{cl}(r)$ in 
Fig.~\ref{fig:85RbFeshbachmodel}. Their associated wave functions, 
$\phi_\mathrm{b}^\mathrm{bg}(\mathbf{r})$ and 
$\phi_\mathrm{b}^\mathrm{cl}(\mathbf{r})$, vanish accordingly in the limit of 
large inter-atomic distances, similarly to the long range asymptotic behaviour 
of diatomic halo molecules of Eq.~(\ref{phibuniversal}). The components of the 
dressed two-channel bound states can therefore be interpreted in terms of 
probability amplitudes, subject to the normalisation condition: 
\begin{equation}
  \langle \phi_\mathrm{b}^\mathrm{bg}|
  \phi_\mathrm{b}^\mathrm{bg}\rangle+\langle \phi_\mathrm{b}^\mathrm{cl}|
  \phi_\mathrm{b}^\mathrm{cl}\rangle=1.
  \label{normalisation2ch}
\end{equation}

\begin{figure}[htbp]
  \includegraphics[width=\columnwidth,clip]{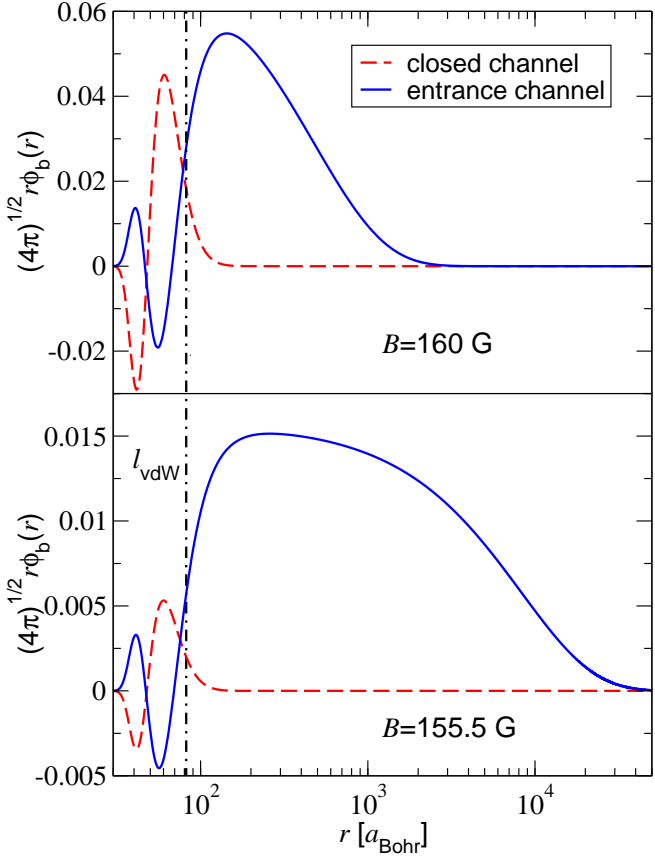}
  \caption{(Colour in online edition)
    Entrance and closed channel components of the highest excited 
    vibrational bound state associated with the 155\,G zero energy resonance 
    of $^{85}$Rb versus the inter-atomic distance $r$ \cite{KoehlerPRL03}. The 
    wave functions were determined using the model illustrated in 
    Fig.~\ref{fig:85RbFeshbachmodel} for magnetic field strengths of 160\,G 
    (upper panel) and 155.5\,G (lower panel), respectively. The nodes of the 
    wave functions at short distances are associated with the vibrational 
    states supported by the bare potentials. The long range behaviour of the 
    entrance channel wave functions, beyond the van der Waals length of 
    $82\,a_\mathrm{Bohr}$ (dot-dashed line), is largely determined just by 
    the binding energy of the Fesh\-bach molecule \cite{KoehlerPRA03}. We note 
    that the inter-atomic distance is given on a logarithmic scale.}
  \label{fig:85Rbboundstates}
\end{figure}

Figure~\ref{fig:85Rbboundstates} illustrates the magnetic field dependence of 
the highest excited dressed vibrational bound state, referred to in the 
following as the Fesh\-bach molecule, for the example of the 155\,G zero 
energy resonance of $^{85}$Rb. The wave functions 
$\phi_\mathrm{b}^\mathrm{bg}(\mathbf{r})$ and 
$\phi_\mathrm{b}^\mathrm{cl}(\mathbf{r})$ represent exact solutions of 
Eqs.~(\ref{LSphibbg}) and (\ref{LSphibcl}) associated with the model 
potentials of Fig.~\ref{fig:85RbFeshbachmodel} using 
$\mu_\mathrm{res}/h=-3.46\,$MHz/G. The entrance channel components 
$\phi_\mathrm{b}^\mathrm{bg}(\mathbf{r})$ of Fig.~\ref{fig:85Rbboundstates} 
are extended over a wide range of inter-atomic distances. This range increases 
beyond all limits as the magnetic field strength $B$ approaches the zero 
energy resonance position of $B_0\approx 155\,$G. The bond length of the 
Fesh\-bach molecules in Fig.~\ref{fig:85Rbboundstates}, i.e.~their mean 
inter-atomic distance, is given, for instance, by 
$\langle r\rangle=521\,a_\mathrm{Bohr}$ at 160\,G and $4255\,a_\mathrm{Bohr}$ 
at 155.5\,G. Figure~\ref{fig:85Rbboundstates} also suggests that the 
functional form of the closed channel component, 
$\phi_\mathrm{b}^\mathrm{cl}(\mathbf{r})$, is virtually independent of $B$. 
Numerical studies \cite{KoehlerPRL03} reveal that 
$\phi_\mathrm{b}^\mathrm{cl}(\mathbf{r})$ is proportional to the bare 
resonance wave function, $\phi_\mathrm{res}(r)$, which indicates the 
applicability of Eq.~(\ref{replacementHcl}). The probability of detecting an 
atom pair in the closed channel spin configuration,
i.e.~$\langle\phi_\mathrm{b}^\mathrm{cl}|\phi_\mathrm{b}^\mathrm{cl}\rangle
=\int d\mathbf{r}\,|\phi_\mathrm{b}^\mathrm{cl}(\mathbf{r})|^2$, decreases 
from about 4.7\,{\%} at $B=160\,$G to only 0.1\,{\%} at $B=155.5\,$G. Given 
that $\phi_\mathrm{b}^\mathrm{cl}(\mathbf{r})$ is spatially confined in the 
same manner as the resonance state, the suppression of the closed channel 
component is a consequence of the increasing range of 
$\phi_\mathrm{b}^\mathrm{bg}(\mathbf{r})$ and the normalisation condition of 
Eq.~(\ref{normalisation2ch}). All these trends reflect general properties of 
Fesh\-bach molecules which can be readily explained using the single resonance 
approach. 

In analogy to the treatment of the dressed continuum levels, the single 
resonance approach to the two-channel bound states is equivalent to the pole 
approximation to the closed channel Green's function of 
Eq.~(\ref{poleapproximation}). This approximation renders Eq.~(\ref{LSphibcl}) 
into a practical form. Its analytic solution may be inserted into 
Eq.~(\ref{LSphibbg}) which yields the following unit normalised dressed 
molecular two-component state:
\begin{equation}
  \left(
  \begin{array}{c}
    |\phi_\mathrm{b}^\mathrm{bg}\rangle\\
    |\phi_\mathrm{b}^\mathrm{cl}\rangle
  \end{array}
  \right)=\frac{1}{\mathcal{N}_\mathrm{b}}
  \left(
  \begin{array}{c}
    G_\mathrm{bg}(E_\mathrm{b})W|\phi_\mathrm{res}\rangle\\
    |\phi_\mathrm{res}\rangle
  \end{array}
  \right).
  \label{phib}
\end{equation}
Here $\mathcal{N}_\mathrm{b}$ is the associated normalisation constant whose 
explicit expression reads:
\begin{equation}
  \mathcal{N}_\mathrm{b}=\sqrt{1+
    \langle\phi_\mathrm{res}|W
    G_\mathrm{bg}^2(E_\mathrm{b})W|\phi_\mathrm{res}\rangle}.
  \label{twochannelnormalisation}
\end{equation} 
In the single resonance approach all bound state energies $E_\mathrm{b}$ are 
determined by a constraint on Eq.~(\ref{phib}) which can be derived by 
multiplying Eq.~(\ref{LSphibbg}) by $\langle\phi_\mathrm{res}|W$ from the 
left. This leads to the following formula:
\begin{equation}
  E_\mathrm{b}=\mu_\mathrm{res}(B-B_\mathrm{res})+
  \langle\phi_\mathrm{res}|WG_\mathrm{bg}(E_\mathrm{b})W
  |\phi_\mathrm{res}\rangle.
  \label{determinationEb}
\end{equation}
We note that Eq.~(\ref{determinationEb}) recovers Eq.~(\ref{resonanceshift}) 
in the limits $E_\mathrm{b}\to 0$ and $B\to B_0$, provided that the magnetic 
field strength approaches $B_0$ from the side of positive scattering lengths. 
This confirms that the binding energy of the Fesh\-bach molecule vanishes at 
the measurable resonance position $B_0$. Such a weak bond implies that the 
properties of the near resonant highest excited dressed vibrational state are 
determined solely by the scattering length, $a$, in analogy to the general 
findings with respect to halo dimers in 
Section~\ref{sec:weaklyboundmolecules}. The range of magnetic field strengths 
in which the Fesh\-bach molecular state as well as its energy depend just on 
$a$ is usually referred to as the universal regime.

\subsection{Universal properties of Fesh\-bach molecules}
\label{subsec:universalFeshbach}
In accordance with the results of Section~\ref{sec:weaklyboundmolecules}, the 
properties of halo dimers, such as, e.g., the large spatial extent of their 
wave functions in Fig.~\ref{fig:85Rbboundstates}, can all be inferred from 
Eq.~(\ref{Ebuniversal}), the universal formula for their binding energy. 

\subsubsection{Universal binding energy}
The formal derivation of Eq.~(\ref{Ebuniversal}) for Fesh\-bach molecules
\cite{GoralJPhysB04} relies upon an explicit determination of the matrix 
element involving $G_\mathrm{bg}(E_\mathrm{b})$ on the right hand side of 
Eq.~(\ref{determinationEb}). This may be performed on the basis of the 
following resolvent identity:
\begin{equation}
  G_\mathrm{bg}(E_\mathrm{b})=G_\mathrm{bg}(0)-
  E_\mathrm{b}G_\mathrm{bg}(0)G_\mathrm{bg}(E_\mathrm{b}).
  \label{resolventidentity}
\end{equation}
Multiplication of Eq.~(\ref{resolventidentity}) by 
$G_\mathrm{bg}^{-1}(0)=-H_\mathrm{bg}$ from the left and by 
$G_\mathrm{bg}^{-1}(E_\mathrm{b})=E_\mathrm{b}-H_\mathrm{bg}$ from the right 
readily verifies this identity. The contribution to 
Eq.~(\ref{determinationEb}) from the first term, $G_\mathrm{bg}(0)$, on the 
right hand side of Eq.~(\ref{resolventidentity}) yields the energy shift 
$\mu_\mathrm{res}(B_\mathrm{res}-B_0)$ due to Eq.~(\ref{resonanceshift}). To 
evaluate the contribution from the second term, 
$-E_\mathrm{b}G_\mathrm{bg}(0)G_\mathrm{bg}(E_\mathrm{b})$, it is instructive 
to employ the spectral decomposition of the bare entrance channel Green's 
function:
\begin{equation}
  G_\mathrm{bg}(z)=\int d\mathbf{p}\,
  \frac{|\phi_\mathbf{p}^{(+)}\rangle\langle\phi_\mathbf{p}^{(+)}|}{z-p^2/m}
  +G_\mathrm{bg}^\mathrm{b}(z).
  \label{Gbgspectral}
\end{equation}
Here the energy argument $z$ is either zero or $E_\mathrm{b}$. The quantity 
$G_\mathrm{bg}^\mathrm{b}(z)$ includes all contributions to $G_\mathrm{bg}(z)$ 
from the bare bound states whose energies, depicted in 
Fig.~\ref{fig:85RbFeshbachmodel}, are usually far detuned from typical 
energies of Fesh\-bach molecules in the universal regime. Thus neglecting 
$G_\mathrm{bg}^\mathrm{b}(z)$ in the product 
$-E_\mathrm{b}G_\mathrm{bg}(0)G_\mathrm{bg}(E_\mathrm{b})$ determines the zero 
bound state energy limit of Eq.~(\ref{determinationEb}) to be:
\begin{equation}
  \mu_\mathrm{res}(B-B_0)-m^2 E_\mathrm{b}
  \int d\mathbf{p}\,
  \frac{\left|\langle\phi_\mathrm{res}|W|\phi_\mathbf{p}^{(+)}
    \rangle\right|^2}
  {p^2(p^2+m|E_\mathrm{b}|)}\underset{E_\mathrm{b}\to 0}{\sim}0.
  \label{Ebuniversal1}
\end{equation}
The corrections neglected on the left hand side of Eq.~(\ref{Ebuniversal1}) 
are all linear in $E_\mathrm{b}$, while the leading contribution involving the 
momentum integral is proportional to $\sqrt{|E_\mathrm{b}|}$ in the limit 
$E_\mathrm{b}\to 0$. In particular, the asymptotic behaviour of the integral 
can be determined explicitly via a change of variable in spherical coordinates 
from the modulus of the momentum, $p$, to the dimensionless quantity 
$p/\sqrt{m|E_\mathrm{b}|}$. This yields: 
\begin{equation}
  \int d\mathbf{p}\,
  \frac{\left|\langle\phi_\mathrm{res}|W|\phi_\mathbf{p}^{(+)}
    \rangle\right|^2}
  {p^2(p^2+m|E_\mathrm{b}|)}
  \underset{E_\mathrm{b}\to 0}{\sim}
  \frac{2\pi^2\left|\langle\phi_\mathrm{res}|W|\phi_0^{(+)}
    \rangle\right|^2}{(m|E_\mathrm{b}|)^{1/2}}.
  \label{Ebuniversal2}
\end{equation}
Equation (\ref{resonancewidth}) may be used to eliminate the matrix element 
involving the coupling, $W(r)$, on the right hand side of 
Eq.~(\ref{Ebuniversal2}) in favour of the product 
$a_\mathrm{bg}\,\Delta B\,\mu_\mathrm{res}$. The solution of 
Eq.~(\ref{Ebuniversal1}) with respect to the bound state energy thus reads: 
\begin{equation}
  E_\mathrm{b}=-\frac{\hbar^2}{m[-a_\mathrm{bg}\,\Delta B/(B-B_0)]^2}.
  \label{Ebuniversal3}
\end{equation}
As anticipated \cite{DonleyNature02} for the general reasons outlined in 
Section~\ref{sec:weaklyboundmolecules}, this formula exactly recovers 
Eq.~(\ref{Ebuniversal}) in the limit $B\to B_0$, as the singular contribution 
to $a(B)$ on the right hand side of Eq.~(\ref{aofB}) exceeds 
$|a_\mathrm{bg}|$.

\subsubsection{The Wigner threshold law}
The preceding derivation of the low binding energy behaviour of the right hand 
side of Eq.~(\ref{determinationEb}) may be extended also to positive continuum 
energies, $E=\hbar^2k^2/m$, which determines the width of the Fesh\-bach 
resonance level \cite{MoerdijkPRA95,MiesPRA00}. This width in energy is 
related to the decay rate of the bare resonance state into a given continuum 
level associated with the wave number $k$ via Fermi's golden rule 
\cite{MukaiyamaPRL03,GoralJPhysB04,DuerrPRA04,HaquePRA05}. The approach 
underlying these derivations is based on the spectral decomposition of 
$G_\mathrm{bg}(z)$ in Eq.~(\ref{Gbgspectral}), in addition to the following 
formula:
\begin{equation}
  \frac{1}{z-p^2/m}=-i\pi\delta(E-p^2/m)+\mathcal{P}\frac{1}{E-p^2/m}.
  \label{principalvalue}
\end{equation}
Here ``$z=E+i0$'' is the complex argument of $G_\mathrm{bg}(z)$ introduced in 
Eq.~(\ref{LSphipbg}), and $\mathcal{P}$ indicates the principal value of the 
momentum integral in Eq.~(\ref{Gbgspectral}). In accordance with Fermi's 
golden rule, the width of the resonance level is given by the modulus of the 
imaginary part of the following matrix element of the bare entrance channel 
Green's function:
\begin{equation}
  \langle\phi_\mathrm{res}|WG_\mathrm{bg}(z)W
  |\phi_\mathrm{res}\rangle\underset{k\to 0}{\sim}
  \mu_\mathrm{res}(B_\mathrm{res}-B_0-i ka_\mathrm{bg}\Delta B).
  \label{Wignerlaw}
\end{equation}
We note, however, that Eq.~(\ref{principalvalue}) is not suitable for 
determining the left hand side of Eq.~(\ref{Wignerlaw}) in the case of 
negative bound state energies, $z=E_\mathrm{b}$, as the $\delta$ function 
contribution vanishes identically. The derivation of Eq.~(\ref{Ebuniversal3}) 
therefore relies upon Eq.~(\ref{resolventidentity}).

As the imaginary part of Eq.~(\ref{Wignerlaw}) is associated with the decay of 
the bare resonance state, the product
$a_\mathrm{bg}\mu_\mathrm{res}\,\Delta B$ is always positive and characterises 
the strength of the inter-channel coupling. Equation~(\ref{Wignerlaw}) also 
determines the low energy dressed continuum wave functions of 
Eqs.~(\ref{phipbg}) and (\ref{phipcl}) via Eq.~(\ref{amplitude}). In 
particular, the $s$-wave scattering amplitude of Eqs.~(\ref{definitiona}) and 
(\ref{scatteringamplitude}) is given by its well known general expansion about 
zero momentum \cite{Taylor72},
\begin{equation}
  f_0(\hbar k)\underset{k\to 0}{\sim} -a(1-ika),
  \label{Wignerlawf0}
\end{equation}
which is applicable provided that the condition $ka\ll 1$ is fulfilled. The 
proportionality to $k=\sqrt{mE}/\hbar$ of the imaginary parts on the right 
hand sides of Eqs.~(\ref{Wignerlaw}) and (\ref{Wignerlawf0}) reflects a 
general prediction by Wigner, known as the threshold law \cite{WignerPR48}. In 
the context of an analytic continuation to imaginary wave numbers,  
$k=i\sqrt{m|E_\mathrm{b}|}/\hbar$, Eq.~(\ref{Wignerlaw}) also yields the 
universal bound state energy of Eq.~(\ref{Ebuniversal3}) using 
Eq.~(\ref{determinationEb}) \cite{DuinePhysRep04}. The universal regime may 
therefore be considered as the range of magnetic field strengths in which the 
Wigner threshold law applies to the energy of the Fesh\-bach molecule in 
Eq.~(\ref{determinationEb}). Its extension about $B_0$ depends on the 
interplay between the spin exchange (or dipolar) and entrance channel 
interactions as well as on the properties of the Fesh\-bach resonance state
\cite{MiesPRA00,KoehlerPRL03,MarcelisPRA04,NygaardPRA06}.

\begin{figure}[htbp]
  \includegraphics[width=\columnwidth,clip]{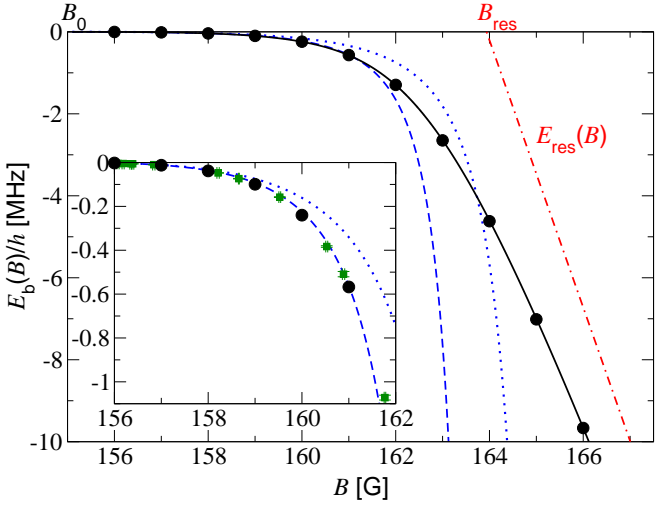}
  \caption{(Colour in online edition)
    The bound state energy of the $^{85}$Rb$_2$ Fesh\-bach molecule as a 
    function of the magnetic field strength in the vicinity of the 155\,G zero 
    energy resonance. The circles indicate a full coupled channels 
    calculation \cite{Kokkelmansprivate}. The solid curve results from a 
    two-channel approach \cite{GoralPRA05}. For comparison, the dotted curve 
    represents the universal estimate of $E_\mathrm{b}$, while the dashed 
    curve includes the leading correction to Eq.~(\ref{Ebuniversal}) due to 
    the van der Waals tail of the background scattering potential 
    \cite{GribakinPRA93,KoehlerPRA03} given by Eq.~(\ref{EbGF}) of 
    Subsection~\ref{subsec:classification}. We note the pronounced shift of 
    about 9\,G between the measurable resonance position, $B_0$, and the 
    magnetic field strength $B_\mathrm{res}$ at which the resonance energy 
    (dot-dashed line) crosses the dissociation threshold of the entrance 
    channel. The inset compares the theoretical approaches with measurements 
    \cite{ClaussenPRA03} indicated by squares.}
  \label{fig:85RbEbofB}
\end{figure}

Figure~\ref{fig:85RbEbofB}, for instance, illustrates several theoretical 
approaches to the bound state energy of the $^{85}$Rb$_2$ Fesh\-bach molecule 
in the vicinity of the 155\,G zero energy resonance. Although the universal 
formula of Eq.~(\ref{Ebuniversal}) determines the asymptotic behaviour of 
$E_\mathrm{b}$ in the limit $B\to B_0$, it provides a reasonable approximation 
only in a small region between $B_0\approx155$\,G and about 158\,G. Both the 
full coupled channels calculation (circles) and the two-channel 
predictions (solid curve) depend on specific properties of the particular 
$^{85}$Rb zero energy resonance besides the scattering length. These 
approaches fully recover the measured Fesh\-bach molecular energies (see the 
inset of Fig.~\ref{fig:85RbEbofB}) over the entire experimental range of 
magnetic field strengths from about 156\,G to 162\,G \cite{ClaussenPRA03}. The 
slope of the linear magnetic field dependence of $E_\mathrm{b}$ in the limit 
of high fields in Fig.~\ref{fig:85RbEbofB} is determined by the bare resonance 
energy, $E_\mathrm{res}(B)$ (dot-dashed line). This indicates an increasing 
admixture of the closed channel spin configuration to the dressed bound state. 

\subsubsection{Closed channel admixture}
In accordance with Eq.~(\ref{phib}) and the unit normalisation of the bare 
Fesh\-bach resonance state, the closed channel admixture to the Fesh\-bach 
molecule is determined by the wave function normalisation constant 
$\mathcal{N}_\mathrm{b}$ of Eq.~(\ref{twochannelnormalisation}) to be:
\begin{equation}
  \langle\phi_\mathrm{b}^\mathrm{cl}|\phi_\mathrm{b}^\mathrm{cl}\rangle
  =\mathcal{N}_\mathrm{b}^{-2}4\pi\int_0^\infty r^2 dr\,
  |\phi_\mathrm{res}(r)|^2
  =\mathcal{N}_\mathrm{b}^{-2}.
  \label{admixturecl}
\end{equation}
In the context of field theoretic approaches to the many-body physics of cold 
gases, the inverse square of the normalisation constant, 
$\mathcal{N}_\mathrm{b}^{-2}$, is sometimes referred to as the wave function 
renormalisation constant $Z(B)$ 
\cite{BraatenComment03,BruunPRL04,DuinePRA03,GoralJPhysB04}. The quantity 
$Z(B)$ can be expressed in terms of the difference in magnetic moments of the 
Fesh\-bach molecular state and a pair of separated atoms in the entrance 
channel spin configuration, i.e.~$\partial E_\mathrm{b}/\partial B$, via the 
following formula:
\begin{equation}
  Z(B)=\mathcal{N}_\mathrm{b}^{-2}=
  \mu_\mathrm{res}^{-1}\frac{\partial E_\mathrm{b}}{\partial B}.
  \label{ZofBgeneral}
\end{equation}
This exact result follows directly from Eqs.~(\ref{twochannelnormalisation})
and (\ref{determinationEb}) using the general relation
\begin{equation}
  G_\mathrm{bg}^2(E_\mathrm{b})=
  -\frac{\partial}{\partial E_\mathrm{b}}G_\mathrm{bg}(E_\mathrm{b}). 
  \label{dGbgdEb}
\end{equation}
Within the universal regime of magnetic field strengths the closed channel
admixture to the Fesh\-bach molecule can therefore be inferred from the 
derivative of Eq.~(\ref{Ebuniversal}) with respect to $B$. This yields:
\begin{equation}
  Z(B)\underset{B\to B_0}{\sim}
  \frac{2a\,\hbar^2/(ma^2)}{\mu_\mathrm{res}\,
  \Delta B\, a_\mathrm{bg}}.
  \label{wavefunctionrenormalisation}
\end{equation} 
Here the limit $B\to B_0$ is performed on the side of positive scattering 
lengths of the zero energy resonance.

In accordance with Eq.~(\ref{wavefunctionrenormalisation}), the wave function 
normalisation constant, $\mathcal{N}_\mathrm{b}$, diverges as the magnetic 
field strength approaches $B_0$, i.e.~in the limit $a\to\infty$, due to the 
proportionality of its leading contribution to $\sqrt{a}$. This implies that 
the closed channel admixture to the Fesh\-bach molecule of 
Eq.~(\ref{admixturecl}) is negligible in the universal regime and vanishes at 
the measurable zero energy resonance position. Within this limited range of 
magnetic field strengths about $B_0$ the Fesh\-bach molecule can therefore be 
described in terms of just its entrance channel component, in analogy to 
Section~\ref{sec:weaklyboundmolecules}. In particular, the wave function 
$\phi_\mathrm{b}^\mathrm{bg}(\mathbf{r})$ is determined by 
Eq.~(\ref{phibuniversal}) at inter-atomic distances large compared to the van 
der Waals length of Fig.~\ref{fig:85Rbboundstates}. In accordance with 
Eq.~(\ref{bondlength}), the mean distance between the atomic constituents of 
universal Fesh\-bach molecules is well estimated by its asymptotic value of 
one half of the scattering length in the limit $a\to\infty$ 
\cite{KoehlerPRL03}. In the universal regime of magnetic field strengths the 
Fesh\-bach molecules are therefore proper diatomic halo states with all the 
general properties described in Section~\ref{sec:weaklyboundmolecules}.

\subsection{Experimental signatures of universality}
\label{subsec:universality}
\begin{figure}[htbp]
  \includegraphics[width=\columnwidth,clip]{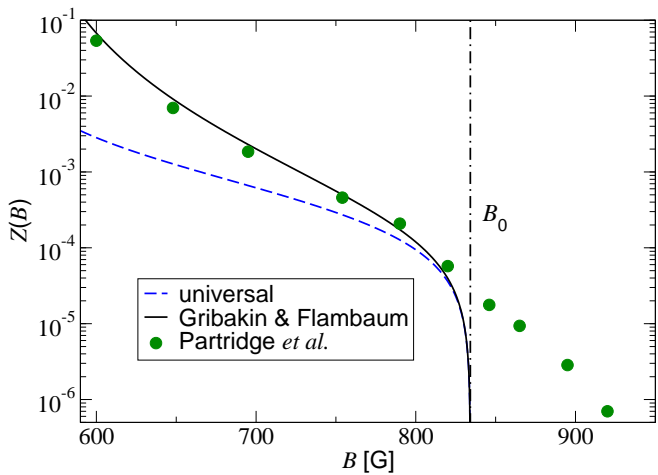}
  \caption{(Colour in online edition)
    Occupation of the closed channel resonance level relative to the 
    number of atom pairs in a balanced mixture of two spin components of 
    $^6$Li versus the magnetic field strength in the vicinity of the $834$\,G 
    zero energy resonance \cite{Partridgeclosedchannel05}. On the low field 
    side of the resonance position $B_0$ (dot-dashed line) the gas was 
    prepared as a Bose-Einstein condensate of $^6$Li$_2$ Fesh\-bach molecules. 
    The experimental data (circles) indicate the suppression of the closed 
    channel admixture to the dressed molecular state, 
    i.e.~$Z(B)=\mathcal{N}_\mathrm{b}^{-2}$, over three orders of magnitude in 
    agreement with Eq.~(\ref{ZofBGF}) of 
    Subsection~\ref{subsec:classification} (solid curve). The dashed curve 
    shows the universal estimate of Eq.~(\ref{wavefunctionrenormalisation}) 
    for comparison. We note that the bond length of the Fesh\-bach molecules 
    of Eq.~(\ref{bondlength}) reaches the order of magnitude of the average 
    inter-atomic spacing of the gas of $10^4\,a_\mathrm{Bohr}$ at a magnetic 
    field strength of about $800$\,G. In this strongly interacting regime the 
    purely two-body theories are bound to break down. On the high field side 
    of $B_0$ the remnant small resonance state population indicates pairing 
    phenomena in a cold, strongly correlated two spin component Fermi gas with 
    a negative scattering length \cite{Partridgeclosedchannel05}.} 
  \label{fig:6LiZofB}
\end{figure}

The suppression of the closed channel admixture to the Fesh\-bach molecular 
state has been directly observed for the example of near resonant $^6$Li$_2$ 
dimers \cite{Partridgeclosedchannel05}. Figure~\ref{fig:6LiZofB} shows a 
comparison of these measurements with the universal estimate of 
Eq.~(\ref{wavefunctionrenormalisation}) as well as with Eq.~(\ref{ZofBGF}) of 
Subsection~\ref{subsec:classification}. The estimate of Eq.~(\ref{ZofBGF}) is 
based on Eq.~(\ref{ZofBgeneral}) using the approximation for $E_\mathrm{b}$ of 
Eq.~(\ref{EbGF}), which is illustrated in the inset of 
Fig.~\ref{fig:85RbEbofB}. The underlying approach \cite{GribakinPRA93} 
accounts for the corrections to the universal bound state energy of 
Eq.~(\ref{Ebuniversal}) due to the van der Waals interaction of the background 
scattering potential in addition to the scattering length $a(B)$. Its 
predictions agree well with exact coupled channels calculations of the closed 
channel admixture to the Fesh\-bach molecular state 
\cite{Partridgeclosedchannel05}. Several experiments to date have also 
directly confirmed the universal limit of the binding energy, in analogy to 
the inset of Fig.~\ref{fig:85RbEbofB} \cite{DonleyNature02,ClaussenPRA03}, for 
a variety of alkali atomic species 
\cite{RegalNature03,BartensteinPRL05,MoritzPRL05}.

\subsubsection{Collisional relaxation}
Besides these demonstrations of universality, the long range nature of highly 
excited Fesh\-bach molecules is manifest in their lifetimes with respect to 
deeply inelastic collisions. Such molecular loss may occur due to relaxation 
into tightly bound diatomic states upon collisions with surrounding atoms or 
dimers \cite{MukaiyamaPRL03,CubizollesPRL03,RegalstabilityPRL04}. The 
associated loss mechanism was discussed first, in the context of cold gases, 
for collisions between H$_2$ dimers and hydrogen atoms 
\cite{BalakrishnanJChemPhys97}. In these threshold-less reactions the energy 
lost through de-excitation is transferred to the relative motion of the 
products in accordance with momentum conservation. The associated relative 
velocities are sufficiently high for the scattered particles to leave an atom 
trap. The density $n_\mathrm{d}$ of dimer molecules in a homogeneous gas is 
therefore depleted in accordance with the following rate equation: 
\begin{equation}
  \dot{n}_\mathrm{d}/n_\mathrm{d}=
  -K_\mathrm{ad} n_\mathrm{a}
  -K_\mathrm{dd} n_\mathrm{d}.
\end{equation}
Here $n_\mathrm{a}$ is the density of atoms and $K_\mathrm{ad}$ and 
$K_\mathrm{dd}$ denote the inelastic loss rate constants associated with 
atom-dimer and dimer-dimer collisions, respectively.

Several theoretical studies of collisional relaxation of alkali metal systems 
have been performed in the limit of low vibrational excitation of the initial 
dimer states \cite{CvitasPRL02,CvitasLiPRL05,CvitasheteroPRL05} as well as for 
Fesh\-bach molecules \cite{PetrovstabilityPRL04,PetrovstabilityPRA05}. In the 
case of collisions between alkali metal atoms and their dimers in excited 
states, whose bond lengths are smaller than $l_\mathrm{vdW}$, {\em ab initio} 
calculations suggest the inelastic rate constants, $K_\mathrm{ad}$, to be on 
the order of $10^{-10}\,$cm$^3$/s 
\cite{CvitasPRL02,CvitasLiPRL05,CvitasheteroPRL05}. For such species the 
collisional relaxation rates in cold gases would be too large for the 
observation of any phenomena relying upon the equilibration of the molecular 
component. The expected short lifetimes of alkali dimers have been confirmed, 
for instance, via photo-association of $^{87}$Rb atoms in a Bose-Einstein 
condensate \cite{WynarScience00}. 

\subsubsection{Lifetime of Fesh\-bach molecules in Fermi gases}
Therefore it came as a surprise that cold Fesh\-bach molecules produced from 
incoherent mixtures of two spin components of fermionic atoms could be 
stabilised for up to several seconds at densities of about 
$10^{13}$\,atoms/cm$^3$
\cite{JochimPRL03,CubizollesPRL03,RegalstabilityPRL04,StreckerPRL03}. Under 
such conditions the ratio of elastic to inelastic collisions is sufficiently 
large to allow for an efficient evaporative cooling of the dimers. As a 
consequence of their stability, even the Bose-Einstein condensation of 
Fesh\-bach molecules has been observed in the vicinity of the broad zero 
energy resonances of $^6$Li and $^{40}$K at about 830\,G and 202\,G, 
respectively \cite{ZwierleinPRL03,JochimScience03,GreinerNature03}. The 
associated resonance widths, $\Delta B$, are on the order of $300\,$G for 
$^6$Li \cite{BartensteinPRL05} and about 8\,G in the case of $^{40}$K 
\cite{GreinerNature03}. Such a broadness provides the opportunity of 
performing measurements in the universal regime of magnetic field strengths.

Both species of dimers in these experiments consist of pairs of unlike 
fermionic atoms in the lowest energetic Zeeman states. In the case of $^{40}$K 
the associated atomic spin components are determined by the quantum numbers 
$(f=9/2,m_f=-9/2)$ and $(f=9/2,m_f=-7/2)$. Here $f$ labels the angular 
momentum quantum number of the hyperfine level with which the Zeeman state 
correlates adiabatically at zero magnetic field, and $m_f$ indicates its spin 
orientation with respect to the field axis. Similarly, the pair of atomic 
Zeeman states associated with the 830\,G zero energy resonance of $^6$Li is 
described by the quantum numbers $(f=1/2,m_f=1/2)$ and $(f=1/2,m_f=-1/2)$. We 
note that the Pauli exclusion principle allows such atom pairs to 
interact via $s$ waves given that their spin wave functions are 
anti-symmetric. Consequently, all the results of 
Section~\ref{sec:weaklyboundmolecules} on isotropic halo dimers apply to these 
Fesh\-bach molecules consisting of unlike fermions. All experiments on 
molecular Bose-Einstein condensation have been performed at near resonant 
magnetic field strengths for which the associated dimer bond lengths can be as 
large as 100\,nm \cite{ZwierleinPRL03}.

The stability of such Fesh\-bach molecules with respect to collisional 
relaxation relies upon the Pauli exclusion principle in addition to the 
separation of length scales associated with the initial and final states 
\cite{PetrovstabilityPRL04,PetrovstabilityPRA05}. In accordance with 
Eq.~(\ref{bondlength}), the spatial extent of the initial halo wave function 
is determined by the scattering length, $a$. The bond length of the final, 
deeply bound target level may be estimated by the van der Waals length, 
$l_\mathrm{vdW}$, which is much smaller than $a$. Both dimer-dimer and 
atom-dimer relaxation therefore require at least three fermions to come 
together at short distances on the order of $l_\mathrm{vdW}$. One pair among 
these atoms necessarily shares the same spin state and can interact at most 
via $p$ waves. As the momentum scale associated with the initial Fesh\-bach 
molecular state is determined by the wave number $k\sim 1/a$, the inelastic 
loss rate constants are suppressed by powers of $kl_\mathrm{vdW}\ll 1$.

\begin{figure}[htbp]
  \includegraphics[width=\columnwidth,clip]{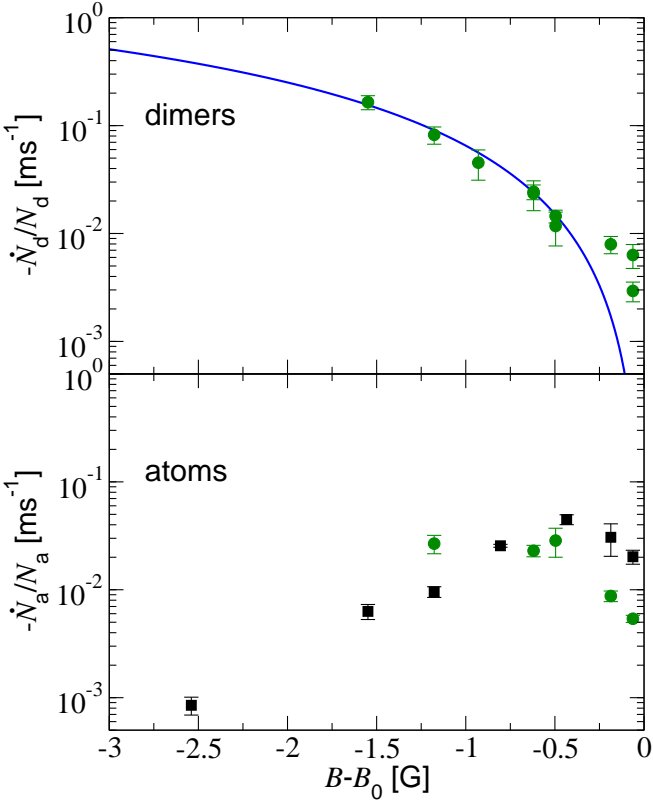}
  \caption{(Colour in online edition)
    Loss rates of diatomic Fesh\-bach molecules (upper panel) and 
    remnant atoms (lower panel) versus the magnetic field detuning, $B-B_0$, 
    in the vicinity of the 202\,G zero energy resonance in a cold gas of 
    $^{40}$K with a peak density of about $1.5\times 10^{13}$\,atoms/cm$^3$ 
    \cite{RegalstabilityPRL04}. The dimers were produced using the technique 
    of adiabatic sweeps of the magnetic field strength described in 
    Section~\ref{sec:associationsweeps}. Their number is denoted by 
    $N_\mathrm{d}$, while $N_\mathrm{a}$ refers to the number of remnant 
    unbound atoms. The circles in the upper panel indicate the measured dimer 
    loss, whose general trend is well fit by an $a^{-2.3\pm0.4}$ power law as 
    a function of the scattering length (solid curve). Significant deviations 
    from this scattering length dependence occur in a small region of magnetic 
    field strengths where the bond length of the dimers of 
    Eq.~(\ref{bondlength}) is comparable to the average distance of the atoms 
    in the gas. The lower panel shows the loss rate of remnant unbound atoms 
    in gases with (circles) or without (squares) deliberate production of 
    Fesh\-bach molecules. The associated trends suggest that the atomic loss 
    is largely unaffected by the presence of a dimer component in the gas 
    \cite{RegalstabilityPRL04}.}
  \label{fig:40Kdecay}
\end{figure}

Based on the halo wave function of Eq.~(\ref{phibuniversal}), the precise $a$ 
dependences of the suppression factors for atom-dimer and dimer-dimer 
relaxation have been predicted to be $a^{-3.33}$ and $a^{-2.55}$, respectively 
\cite{PetrovstabilityPRL04,PetrovstabilityPRA05}. While these predictions 
strictly apply just to the universal regime of magnetic field strengths, 
Fig.~\ref{fig:40Kdecay} illustrates that their general trends agree with 
measurements on cold gases of $^{40}$K with a component of Fesh\-bach 
molecules \cite{RegalstabilityPRL04}. These observations are consistent with 
the reduction of a predominant dimer-dimer relaxation in the limit of large 
scattering lengths, where the lifetimes reach about $100\,$ms. Conversely, as 
the magnetic field strength is tuned away from the zero energy resonance, the 
molecular lifetimes approach those small values of less than a millisecond, 
typical for short ranged alkali dimers in cold gases. 

Besides their scaling properties with respect to the scattering length, the 
inelastic loss rate constants of Fesh\-bach molecules depend sensitively on 
inter-atomic interactions at short distances below $l_\mathrm{vdW}$. The 
experimental trends regarding the collisional relaxation of alkali dimers 
confined to such length scales are inconclusive. Remarkably long lifetimes on 
the order of seconds were reported, for instance, even in the case of short 
ranged, closed-channel dominated Fesh\-bach molecules produced in a fermionic 
$^6$Li gas in the vicinity of the narrow 543\,G zero energy resonance 
\cite{StreckerPRL03}. Such alkali dimers, however, are not described by the 
halo wave function of Eq.~(\ref{phibuniversal}). Their observed stability, 
therefore, suggests a mechanism for the suppression of collisional relaxation 
beyond the scaling of the associated loss rate constants with powers of the 
inverse scattering length.

\subsubsection{Lifetime of Fesh\-bach molecules in Bose gases}
Fesh\-bach molecules consisting of identical Bose atoms generally tend to be 
less stable than their fermionic counterparts. Large collisional relaxation 
rate constants on the typical order of $10^{-10}\,$cm$^3/$s have been reported 
for those dimers associated in cold gases in the vicinity of comparatively 
narrow zero energy resonances of $^{87}$Rb, $^{133}$Cs and $^{23}$Na
\cite{MukaiyamaPRL03,DuerrPRL04,HerbigScience03,YurovskyPRA99,YurovskyPRA00}.
With a width of about 11\,G \cite{ClaussenPRA03} the 155\,G zero energy 
resonance of $^{85}$Rb is by far the broadest among the bosonic species from 
which Fesh\-bach molecules were produced. The atomic constituents of these 
dimers are prepared in the excited Zeeman state determined by the quantum 
numbers $(f=2,m_f=-2)$ which, in contrast to the electronic ground state, can 
be magnetically trapped. Pairs of such atoms are subject to inelastic spin 
relaxation collisions involving the threshold-less transition of the atomic 
energy level to a deeper Zeeman state \cite{RobertsPRL00}. The associated 
$d$-wave exit channels are illustrated in Table~\ref{tab:dVblock}. It is the 
spontaneous dissociation via spin relaxation, rather than collisional 
relaxation of the molecular vibrational state, that predominantly limits the 
observed lifetimes of Fesh\-bach molecules produced in dilute gases of 
$^{85}$Rb. This decay mechanism has been unequivocally identified through 
comparisons with quantitative predictions \cite{ThompsonPRL05,KoehlerPRL05}. 

\begin{figure}[htbp]
  \includegraphics[width=\columnwidth,clip]{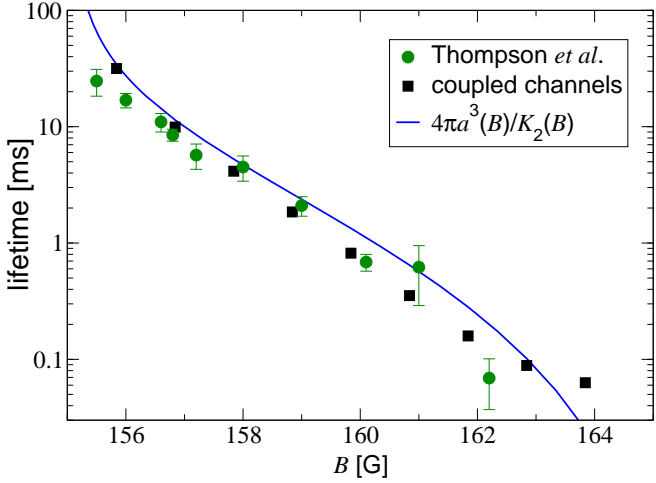}
  \caption{(Colour in online edition)
    The lifetime of $^{85}$Rb$_2$ dimers versus magnetic field strength 
    in the vicinity of the 155\,G zero energy resonance. The circles indicate 
    measurements in a cold thermal gas with a peak density of 
    $6.6\times10^{11}\,$atoms/cm$^3$ \cite{ThompsonPRL05}, while the squares 
    refer to exact coupled channels calculations accounting for the 
    spontaneous dissociation of Fesh\-bach molecules due to spin relaxation 
    \cite{KoehlerPRL05}. The solid curve represents an asymptotic estimate of 
    the lifetime based on the universal halo wave function of 
    Eq.~(\ref{phibuniversal}) and the spin relaxation loss rate constant, 
    $K_2(B)$, associated with a non-degenerate Bose gas in the limit of zero 
    collision energy. The two-body theory fails only close to resonance when 
    the bond length of the molecules becomes comparable to the average 
    inter-atomic distance of the gas.}
  \label{fig:85Rbdecay}
\end{figure}

Figure \ref{fig:85Rbdecay} shows the monotonic increase of the measured 
lifetimes over three orders of magnitude as the magnetic field strength 
approaches the resonance position, $B_0$. The lifetime of these Fesh\-bach 
molecules is density independent and reaches several tens of milliseconds. 
This behaviour of the molecular decay can be explained in terms of a 
probabilistic estimate based on the average volume, 
$\mathcal{V}=4\pi\langle r^3\rangle/3$, occupied by an atom pair in the 
universal halo state of Eq.~(\ref{phibuniversal}). In analogy to 
Eq.~(\ref{bondlength}) this volume is determined in terms of the scattering 
length $a$ by the formula $\mathcal{V}=\pi a^3$. The event rate, $\gamma$, 
associated with the spin relaxation of such a bound atom pair may be expressed 
in terms of the loss rate constant $K_2$ for spin relaxation in a 
non-degenerate Bose gas \cite{RobertsPRL00}. This yields 
$\gamma=K_2/(4\mathcal{V})$ \cite{StoofPRA89}. The lifetime, $\tau=1/\gamma$, 
is therefore proportional to the ratio of the magnetic field dependent 
quantities $a^3(B)$ and $K_2(B)$, i.e., $\tau=4\pi a^3(B)/K_2(B)$, where, near 
resonance, $K_2(B)$ can be evaluated at zero collision energy. The same 
formula for the molecular lifetime also follows rigorously from Fermi's golden 
rule \cite{KoehlerPRL05} and exactly recovers the results of the full coupled 
channels calculations of Fig.~\ref{fig:85Rbdecay} in the limit $B\to B_0$. The 
magnetic field dependence of $K_2(B)$ can be inferred from general properties 
of inelastic collisions \cite{BohnPRA97,BohnPRA99}. Consequently, the 
stability of Fesh\-bach molecules in both two spin component Fermi gases of 
$^{40}$K and dilute vapours of $^{85}$Rb directly probes the halo nature of 
these dimers near resonance, despite their different decay mechanisms.

\subsection{Classification of zero energy resonances}
\label{subsec:classification}
The singular behaviour of the scattering length as well as the associated 
existence of a universal regime of magnetic field strengths are common to all 
Fesh\-bach resonance phenomena in cold gases. Two-body universality implies, 
in particular, that the long range properties of the dressed diatomic bound 
states are well described solely in terms of an effective entrance channel 
interaction. The range of validity of such single channel approaches provides 
classification schemes for zero energy resonances.

\subsubsection{Size of the universal regime}
The halo wave function of Eq.~(\ref{phibuniversal}) may be interpreted as the 
only bound state supported by a contact pseudo potential, 
$g\delta(\mathbf{r})$, describing the low energy spectrum of an atom pair with 
a positive scattering length. Due to their simplicity, contact interactions 
are widely used in theoretical descriptions of cold gases 
\cite{DalfovoRMP99,Dalibard99,Randeria95}. In these approaches the properties 
of inter-atomic collisions enter the many-body Hamiltonian in terms of a 
coupling constant, $g=4\pi\hbar^2a/m$, depending on the microscopic potential 
just through the scattering length, $a$. The associated binary $s$-wave 
scattering amplitude is given by the following formula:  
\begin{equation}
   f_0(\hbar k)=-a/(1+ika).
   \label{f0contact}
\end{equation}
We note that Eq.~(\ref{f0contact}) also recovers the general low wave number 
expansion of Eq.~(\ref{Wignerlawf0}) in the Wigner threshold law regime. 
Consequently, the contact interaction approach provides a minimal 
implementation of the universal low energy two-body physics, applicable to 
both positive and negative scattering lengths. In the latter case the contact 
pseudo potential does not support any bound state.

The assumption of universality of the Fesh\-bach molecular state implies the 
closed channel admixture of Eq.~(\ref{admixturecl}) to be negligible. For 
magnetic field strengths, $B$, far inside the width of the zero energy 
resonance, i.e.~$|B-B_0|\ll \Delta B$, Eqs.~(\ref{aofB}) and 
(\ref{wavefunctionrenormalisation}) therefore yield the following condition 
necessary for the applicability of the contact interaction approach:
\begin{equation}
  \left|\frac{B-B_0}{\Delta B}\right|\ll 
  \frac{|\mu_\mathrm{res}\,\Delta B|}{2\hbar^2/(ma_\mathrm{bg}^2)}.
  \label{criterionuniversal}
\end{equation}
The energy ratio on the right hand side of Eq.~(\ref{criterionuniversal}) 
provides an upper estimate for the extent of the universal regime relative 
to the width of the zero energy resonance. A ratio small compared to unity 
indicates a weak coupling between the closed and entrance channels. 
Conversely, broad zero energy resonances with a large product 
$|\mu_\mathrm{res}\,\Delta B|$ tend to favour negligible admixtures of the 
bare resonance level to the dressed Fesh\-bach molecular state over a 
substantial fraction of their width. This trend may be enhanced by a small 
energy $\hbar^2/(ma_\mathrm{bg}^2)$ which indicates the presence of bound 
vibrational ($a_\mathrm{bg}>0$) or virtual ($a_\mathrm{bg}<0$) near resonant 
energy levels of the bare background scattering potential. 

A small closed channel admixture to the Fesh\-bach molecular state, however, 
is not necessarily identical to the universality of the binding energy. While 
for the example of the comparatively weakly coupled 1007\,G zero energy 
resonance of $^{87}$Rb \cite{DuerrPRA04} the right hand side of 
Eq.~(\ref{criterionuniversal}) gives $0.1$, it is as large as $81$ in the case 
of the 155\,G zero energy resonance of $^{85}$Rb. The estimate of 
Eq.~(\ref{criterionuniversal}) for $^{85}$Rb, however, is inaccurate with 
respect to the extension of the universal regime in units of the width, 
$\Delta B$, given the results of Fig.~\ref{fig:85RbEbofB}. The inset of 
Fig.~\ref{fig:85RbEbofB} reveals that this inaccuracy originates predominantly 
from corrections to the universal binding energy due to the van der Waals tail 
of the bare $^{85}$Rb background scattering potential \cite{GribakinPRA93}. In 
the case of such entrance-channel dominated zero energy resonances even the 
lowest order corrections to universality may be described just by a single 
channel approach \cite{KoehlerPRA03}. The minimal requirements on the 
associated effective potential, $V(B,r)$, besides its long range asymptotic 
behaviour of Eq.~(\ref{potentialvdW}) are closely related to general 
properties of alkali dimer energy wave functions.

\subsubsection{Entrance-channel dominated resonances}
In accordance with Fig.~\ref{fig:85Rbboundstates}, the vibrational bound state 
wave functions of $^{85}$Rb$_2$ Fesh\-bach molecules consist of short and long 
range contributions. The characteristic scale for such a spatial separation is 
the van der Waals length of Eq.~(\ref{lvdW}). Due to the deep wells of 
realistic background scattering potentials, the entrance channel wave 
functions of alkali dimers are well described by the semi-classical 
Wentzel-Kramers-Brillouin (WKB) approximation at short inter-atomic distances, 
$r\ll l_\mathrm{vdW}$. Their behaviour at large separations, 
$r\gg l_\mathrm{vdW}$, is determined mainly by the van der Waals interaction 
of Eq.~(\ref{potentialvdW}) in addition to the energy of the state. In both 
spatial regions the functional forms of the associated asymptotic solutions of 
the radial Schr\"odinger equation are known analytically 
\cite{GribakinPRA93,GaoC6PRA98,GaodefectPRA98}. As the interactions between 
alkali atom pairs are dominated by large van der Waals coefficients, $C_6$, on 
the order of thousands of atomic units (the atomic unit of $C_6$ is 
$9.5734\times10^{-26}\,$J\,nm$^6$), the asymptotic wave functions can be 
matched. Such a matching procedure provides the basis of accurate 
semi-classical treatments of bound as well as continuum entrance channel wave 
functions \cite{GribakinPRA93,FlambaumPRA99}. 

Provided that the interaction of a pair of alkali atoms is well described by 
an effective entrance channel potential, $V(B,r)$, the semi-classical approach 
to the zero energy wave function determines the scattering length to be 
\cite{GribakinPRA93}:
\begin{equation}
  a=\bar{a}\left[1-\tan(\varphi_\mathrm{WKB}-\pi/8)\right].
  \label{aGF}
\end{equation}
Here $\bar{a}$ is usually referred to as the mean scattering length and 
$\varphi_\mathrm{WKB}$ is the semi-classical phase shift. In accordance with 
the WKB approach, $\varphi_\mathrm{WKB}$ consists of the following integral 
between the zero energy classical turning point, $r_0$, associated with the 
effective potential, $V(B,r)$, and infinite distances: 
\begin{equation}
 \varphi_\mathrm{WKB}=\frac{1}{\hbar}\int_{r_0}^\infty
 dr\,\sqrt{-mV(B,r)}.
 \label{phiWKB}
\end{equation}
While $\varphi_\mathrm{WKB}$ is thus sensitive to the entire well of the 
interaction, from $V(B,r_0)=0$ to its long range tail of 
Eq.~(\ref{potentialvdW}), the coefficient $\bar{a}$ of Eq.~(\ref{aGF}) depends 
just on the van der Waals length, $l_\mathrm{vdW}$. Its explicit expression in 
terms of $l_\mathrm{vdW}$ and Euler's $\Gamma$ function reads:  
\begin{equation}
  \bar{a}=\frac{l_{\mathrm{vdW}}}{\sqrt{2}}\, 
  \frac{\Gamma(3/4)}{\Gamma(5/4)}\approx 0.95598\times l_{\mathrm{vdW}}.
  \label{meanscatteringlength}
\end{equation}
In accordance with Eq.~(\ref{aGF}), the parameter $\bar{a}$ determines the 
characteristic scale of the scattering length. This average potential range is 
modulated by the poles of the tangent function provided that its argument, 
$\varphi_\mathrm{WKB}-\pi/8$, is close to an odd integer multiple of $\pi/2$. 
As $\varphi_\mathrm{WKB}$ increases, each singularity of $a$ indicates the 
emergence of an additional vibrational bound state in the potential well.

The functional form of Eq.~(\ref{aGF}) using a realistic interaction is 
largely analogous to the formula for the scattering length of the simplified 
square well plus hard core model of an inter-atomic potential 
\cite{GribakinPRA93}. In the case of this exactly solvable model the mean 
scattering length, $\bar{a}$, recovers the finite outer radius of the well. 
The analogy between these realistic and simplified interactions therefore 
implies that in the limit $a\gg\bar{a}$ the energy of the highest excited 
vibrational state is well approximated by the following asymptotic formula:
\begin{equation}
  E_\mathrm{b}\approx -\hbar^2/\left[m(a-\bar{a})^2\right].
  \label{EbGF}
\end{equation}
An independent analysis based on effective range theory for realistic 
inter-atomic interactions \cite{GaodefectPRA98,FlambaumPRA99} confirms the 
magnitude of the range parameter $\bar{a}$ of Eq.~(\ref{EbGF}) up to a 
constant factor on the order of unity \cite{GaoJPhysB04}. The remnant 
uncertainties associated with Eq.~(\ref{EbGF}) may be related to the 
divergence of the effective range expansion for any potential with a long 
range van der Waals tail \cite{Taylor72}. Figure \ref{fig:85RbEbofB} 
illustrates the accuracy of Eq.~(\ref{EbGF}) within the experimentally 
relevant range of binding energies of near resonant $^{85}$Rb$_2$ Fesh\-bach 
molecules. 

According to the semi-classical approach, at magnetic field strengths in the 
vicinity of entrance-channel dominated zero energy resonances the diatomic 
bound and continuum spectra are determined mainly by $a(B)$ and $\bar{a}$.
In such a case, any potential, $V(B,r)$, which at each magnetic field 
strength, $B$, accounts for the scattering length as well as the van der Waals 
tail of Eq.~(\ref{potentialvdW}) provides a suitable description of the near 
resonant binary physics. This conclusion is rather intuitive because cold 
collisions are characterised by de Broglie wave lengths too large to resolve 
details of the effective entrance channel interaction besides its long range 
behaviour. The adjustment of $V(B,r)$ to recover the exact magnetic field 
dependence of the scattering length of Eq.~(\ref{aofB}) may be achieved, for 
instance, by varying the radius of its hard core. In the context of quantum 
defect theory such a description of alkali dimer spectra in terms of just the 
parameters $a$ and $C_6$ has been rigorously derived for a range of energies 
much wider than the cold regime \cite{GaodefectPRA98}. 

As $\bar{a}$ is positive, the semi-classical estimate of the bound state 
energy of Eq.~(\ref{EbGF}) is always below the universal prediction of 
Eq.~(\ref{Ebuniversal}) and becomes singular in the limit $a(B)\to\bar{a}$, 
outside the range of validity of Eq.~(\ref{EbGF}). This unphysical behaviour 
is counterbalanced in Eq.~(\ref{determinationEb}) by an increasing closed 
channel admixture to the Fesh\-bach molecule, in accordance with 
Eq.~(\ref{wavefunctionrenormalisation}), which tends to impose a linear slope 
on $E_\mathrm{b}(B)$. The principal question of the applicability of single 
channel approaches outside the universal regime of magnetic field strengths is 
decided by which one of these trends prevails near resonance
\cite{KoehlerPRA04}. Consequently, the bound state energy, $E_\mathrm{b}(B)$, 
in the vicinity of an entrance-channel dominated zero energy resonance is 
subject to the following inequality:
\begin{equation}
 \left|E_\mathrm{b}(B)+\frac{\hbar^2}{m[a(B)-\bar{a}]^2}\right|<
 \left|E_\mathrm{b}(B)+\frac{\hbar^2}{ma^2(B)}\right|.
 \label{conditionecd1}
\end{equation} 
Within the range of validity of Eq.~(\ref{EbGF}) the admixture of the closed 
channel resonance state to the Fesh\-bach molecule is small compared to unity 
and can be determined from the binding energy via Eq.~(\ref{ZofBgeneral}). 
This yields:
\begin{equation}
  Z(B)=\frac{2a\hbar^2/(ma^2)}{\mu_\mathrm{res}\Delta B a_\mathrm{bg}}
  \frac{(1-a_\mathrm{bg}/a)^2}{(1-\bar{a}/a)^3}.
  \label{ZofBGF}
\end{equation}
The accuracy of Eq.~(\ref{ZofBGF}) in applications to $^6$Li$_2$ Fesh\-bach 
molecules in the vicinity of the entrance-channel dominated 834\,G zero energy 
resonance is illustrated in Fig.~\ref{fig:6LiZofB}.

Based on Eq.~(\ref{conditionecd1}) a more practical criterion can be derived 
from a low energy expansion of the right hand side of 
Eq.~(\ref{determinationEb}) using a specific implementation of the general 
two-channel approach \cite{GoralJPhysB04}. This yields a dimensionless 
parameter, $\eta$, whose smallness indicates the validity of Eq.~(\ref{EbGF}) 
beyond the universal regime of magnetic field strengths \cite{StollPRA05}. 
Consequently, an entrance-channel dominated zero energy resonance fulfils the 
condition
\begin{equation}
  \eta=\frac{\bar{a}}{a_\mathrm{bg}}
  \frac{\hbar^2/(m\bar{a}^2)}{\mu_\mathrm{res}\, \Delta B}\ll 1.
  \label{conditionecd2}
\end{equation}
Such a criterion also results from an adiabatic description of Fesh\-bach 
resonances \cite{PetrovPRL04}. 

In the opposite limit, $\eta\gg 1$, a zero energy resonance shall be referred 
to, in the following, as closed-channel dominated. Closed-channel dominated 
zero energy resonances are typically narrow and their universal regime of 
magnetic field strengths is experimentally largely inaccessible. The 
description of their physical properties therefore crucially relies upon an 
explicit treatment of at least two scattering channels.

\subsection{Characteristic parameters of zero energy resonances}
\label{subsec:parameters}
For any two-channel approach to be sensible its implementation should describe 
the two-body energy spectrum beyond the Wigner threshold law domain. Otherwise 
the same physics could be captured simply by the contact pseudo interaction of 
Subsection~\ref{subsec:classification}. The approach should therefore recover 
both the scattering length of Eq.~(\ref{aofB}) and the binding energy of the 
Fesh\-bach molecule beyond the universal regime. Even within such a 
comparatively wide range of energies a variety of inter-atomic potentials are 
capable of describing the same two-body physics. It is the objective of this 
subsection to provide a minimal set of physical parameters that every 
two-channel approach should account for, and to illustrate a practical 
implementation for a typical experimental situation. The adjustment of the 
Hamiltonian of Eq.~(\ref{H2B2channel}) in the single resonance approach of 
Eq.~(\ref{replacementHcl}) will be performed on the basis of the general form 
of its energy spectrum derived in Subsection~\ref{subsec:dressedenergystates}. 
According to these derivations, the dressed two-channel energy states depend 
on the bare states associated with the background scattering via 
$G_\mathrm{bg}(z)$, on the resonance energy, $E_\mathrm{res}(B)$, and on the 
product $W|\phi_\mathrm{res}\rangle$ characterising the inter-channel 
coupling. Just these quantities need to be adjusted. The specific form of the 
resonance wave function, $\phi_\mathrm{res}(r)$, does not affect the two-body 
spectrum in the single resonance approach. 

\subsubsection{The background scattering potential}
In accordance with Subsection \ref{subsec:classification}, all implementations 
of $V_\mathrm{bg}(r)$ which recover $a_\mathrm{bg}$ in addition to 
Eq.~(\ref{potentialvdW}) yield equivalent energy spectra beyond the cold 
regime \cite{GaodefectPRA98}, provided that their number of levels is large 
compared to unity. A hard sphere in addition to the van der Waals tail of 
Eq.~(\ref{potentialvdW}), for instance, provides such a background scattering 
potential with a minimal number of parameters. Its explicit expression reads:
\begin{equation}
  V_\mathrm{bg}(r)=
  \left\{
  \begin{array}{ccc}
    +\infty & : & r<r_0\\
    -C_6/r^6 & : & r>r_0
  \end{array}
  \right..
  \label{VbgHCC6}
\end{equation}
The associated background scattering length is given by the following exact 
formula \cite{GribakinPRA93}:
\begin{equation}
  a_\mathrm{bg}=\bar{a}\left[1-\tan(\varphi_\mathrm{WKB}-3\pi/8)\right].
  \label{aHCC6}
\end{equation}
Here $\bar{a}$ is the mean scattering length of 
Eq.~(\ref{meanscatteringlength}) and 
$\varphi_\mathrm{WKB}=2 l_\mathrm{vdW}^2/r_0^2$ is the semi-classical phase 
shift of Eq.~(\ref{phiWKB}). The difference of $\pi/4$ between the arguments 
of the tangent functions in Eqs.~(\ref{aGF}) and (\ref{aHCC6}) is due to the 
discontinuity of Eq.~(\ref{VbgHCC6}) at the core radius $r_0$. 
Equation~(\ref{aHCC6}) may be used to determine $r_0$ in such a way that 
$V_\mathrm{bg}(r)$ of Eq.~(\ref{VbgHCC6}) exactly recovers the background 
scattering length. Similar procedures have been performed in two-channel 
approaches using continuous implementations of $V_\mathrm{bg}(r)$ 
\cite{MiesPRA00,KoehlerPRL03,MarcelisPRA04,NygaardPRA06}. An example of 
such an effective interaction is illustrated in 
Fig.~\ref{fig:85RbFeshbachmodel} in addition to its energy levels. 
Table~\ref{tab:Vbgparameters} provides the parameters $a_\mathrm{bg}$ and 
$C_6$ characterising the background scattering potential for several 
experimentally relevant zero energy resonances.

\begin{table*}
  \caption{Parameters characterising the background scattering potential 
    associated with several experimentally relevant zero energy resonances. 
    Those values of $a_\mathrm{bg}$ and $E_{-1}$ that are unreferenced refer 
    to calculations performed for this review. The energy, $E_\mathrm{-1}$, 
    associated with the highest excited vibrational state of the background
    scattering potential is given only in the cases of isolated resonances.
    The atomic unit of the van der Waals dispersion coefficient, $C_6$, is 
    $9.5734\times 10^{-26}\,$J\,nm$^6$, $a_\mathrm{Bohr}=0.052917\,$nm is the 
    Bohr radius and 1\,G=$10^{-4}$\,T.}
  \label{tab:Vbgparameters}
  \begin{ruledtabular}
    \begin{tabular}{ccccc}
      Species 
      & $B_0$ [G]
      & $a_\mathrm{bg}$ [$a_\mathrm{Bohr}$] 
      & $C_6$ [a.u.]
      & $|E_{-1}|/h$ [MHz]\\
      \hline
      $^6$Li 
      & 543.25(5) \cite{StreckerPRL03} 
      & 59 
      & 1393.39 \cite{YanPRA96} 
      &\\
      & 834.149 \cite{BartensteinPRL05} 
      & $-1405$ \cite{BartensteinPRL05}
      & 1393.39 \cite{YanPRA96} 
      &\\
      $^{23}$Na 
      & 853 \cite{StengerPRL99} 
      & 63.9 \cite{MiesPRA00}
      & 1561 \cite{KharchenkoPRA97} 
      & 208\\
      & 907 \cite{StengerPRL99} 
      & 62.8 \cite{MiesPRA00}
      & 1561 \cite{KharchenkoPRA97} 
      & 218\\
      $^{40}$K  
      & 202.10(7) \cite{RegalpairingPRL04} 
      & 174(7) \cite{LoftusPRL02} 
      & 3897 \cite{DereviankoPRL99} 
      & 8.6\\
      & 224.21(5) \cite{RegalPRL03} 
      & 174(7) \cite{LoftusPRL02} 
      & 3897 \cite{DereviankoPRL99} 
      & 8.6\\
      $^{85}$Rb 
      & 155.0 \cite{ThompsonRFPRL05} 
      & $-443(3)$ \cite{ClaussenPRA03}
      & 4703 \cite{KempenPRL02} 
      & 218\\
      $^{87}$Rb 
      & 1007.40(4) \cite{VolzPRA03} 
      & 100.5 \cite{VolzPRA03} 
      & 4703 \cite{KempenPRL02} 
      & 24.0\\
      $^{133}$Cs 
      & 19.90(3) \cite{ChinPRA04} 
      & 163 \cite{JulienneJModOpt04} 
      & 6890(35) \cite{LeoPRL00} 
      &\\
      & 47.97(3) \cite{ChinPRA04} 
      & 905 \cite{JulienneJModOpt04} 
      & 6890(35) \cite{LeoPRL00} 
      & 0.045
    \end{tabular}
  \end{ruledtabular}
\end{table*}

Equation~(\ref{VbgHCC6}) as well as those equivalent implementations of 
$V_\mathrm{bg}(r)$ that explicitly include the long range asymptotic van der 
Waals interaction of Eq.~(\ref{potentialvdW}) are, in principle, suited to 
describe several bare energy levels. The number of bound states supported by 
Eq.~(\ref{VbgHCC6}), for instance, can be arbitrarily increased by decreasing 
the core radius $r_0$ under the constraint of a fixed background scattering 
length. The description of such wide energy ranges, however, is usually beyond 
the scope of two-channel approaches. In addition, the explicit treatment of 
the van der Waals tail of the background scattering potential is largely 
impractical in applications to the many-body physics of dilute gases. Several 
implementations of two-channel approaches, therefore, use effective low energy 
interactions to recover different aspects of the cold collision physics under 
the conditions of resonance enhancement \cite{KokkelmansPRA02,BruunPRL04,
DuinePhysRep04,GoralJPhysB04,DrummondPRA04,Chincondmat05}.  

\begin{figure}[htbp]
  \includegraphics[width=\columnwidth,clip]{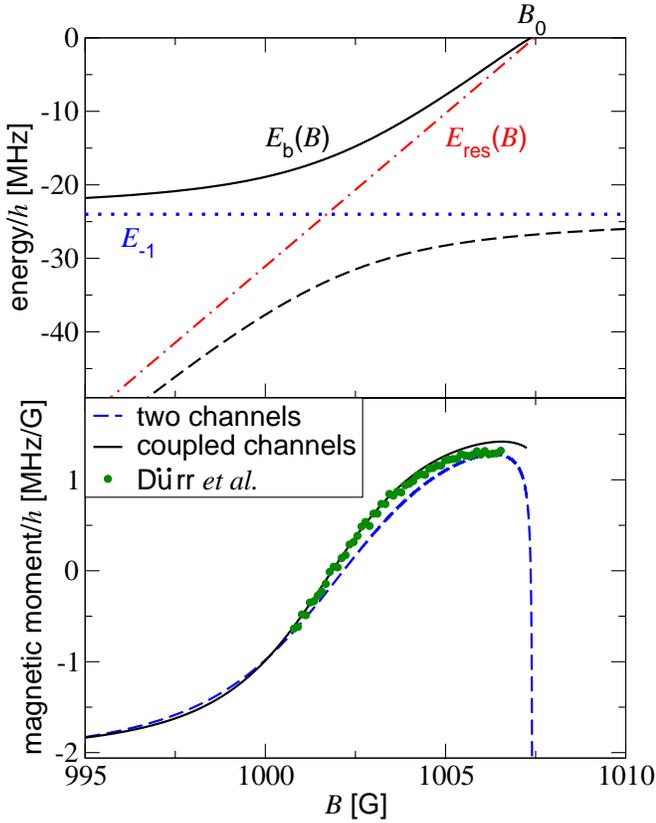}
  \caption{(Colour in online edition)
    Avoided crossing of the highest excited vibrational levels of 
    $^{87}$Rb$_2$ (upper panel) and the magnetic moment of the Fesh\-bach 
    molecule (lower panel) versus the magnetic field strength in the vicinity 
    of the 1007\,G zero energy resonance. The solid curve of the upper panel 
    indicates the bound state energy $E_\mathrm{b}(B)$ of the Fesh\-bach 
    molecule, while the dashed curve refers to the next more tightly bound 
    dressed vibrational level. The dotted and dot-dashed lines are associated 
    with the energies $E_{-1}/h=-24\,$MHz of the bare highest excited 
    vibrational level of $V_\mathrm{bg}(r)$ and $E_\mathrm{res}(B)$ of the 
    closed channel resonance state, respectively. The crossing between the 
    bare levels at 1001.7\,G leads to the measured variation of the magnetic 
    moment of the Fesh\-bach molecule indicated by the circles in the lower 
    panel \cite{DuerrPRL04}. For comparison, the solid and dashed curves refer 
    to exact coupled channels calculations \cite{vanKempenslope04} and a 
    two-channel approach \cite{GoralJPhysB04}, respectively.}
  \label{fig:87Rbcrossing}
\end{figure}

Figure~\ref{fig:87Rbcrossing} shows a typical range of energies relevant to 
the Stern-Gerlach separation of Fesh\-bach molecules from a $^{87}$Rb 
Bose-Einstein condensate \cite{DuerrPRL04}. This experimental technique based 
on exposing a mixture of atoms and dimers to an inhomogeneous magnetic field 
\cite{HerbigScience03,DuerrPRL04,ChinPRL05} is illustrated in 
Fig.~\ref{fig:Herbiglevitation} for the example of $^{133}$Cs$_2$. The 
relative force between the two components of the gas is proportional to the 
field gradient as well as to the difference in magnetic moments of a 
Fesh\-bach molecule and a pair of separated atoms, i.e., 
$\partial E_\mathrm{b}/\partial B$. In the rubidium case the magnetic field 
dependence of $E_\mathrm{b}(B)$ in Fig.~\ref{fig:87Rbcrossing} is sensitive to 
the avoided crossing of dressed energy levels of $^{87}$Rb$_2$ due to the 
highest excited bare vibrational level of $V_\mathrm{bg}(r)$. In accordance 
with the size of the positive background scattering length of about 
$100\,a_\mathrm{Bohr}$ \cite{VolzPRA03}, this level is sufficiently close to 
the dissociation threshold that its energy, $E_{-1}$, is directly probed by 
the experiment \cite{DuerrPRL04}.

An effective entrance channel interaction suitable for a description of this 
experiment needs to recover both the precise value of $a_\mathrm{bg}$ and the 
bare vibrational level with the energy $E_{-1}$, beyond the Wigner threshold 
law regime. The associated minimal bare Green's function, $G_\mathrm{bg}(z)$, 
is therefore required to reproduce the bound state pole at the energy argument 
$z=E_{-1}$. Similarly to Eq.~(\ref{poleapproximation}), this requirement leads 
to the following separable representation \cite{GoralJPhysB04}:
\begin{equation}
  G_\mathrm{bg}(z)=G_0(z)+
  G_0(z)|\chi_\mathrm{bg}\rangle\tau_\mathrm{bg}(z)
  \langle\chi_\mathrm{bg}|G_0(z).
  \label{Gbgseparable}
\end{equation}
Here the form factor, $|\chi_\mathrm{bg}\rangle$, needs to recover, via the 
relation $G_0(E_{-1})|\chi_\mathrm{bg}\rangle\propto|\phi_{-1}\rangle$, the 
bare vibrational state satisfying the Schr\"odinger equation 
$H_\mathrm{bg}|\phi_{-1}\rangle=E_{-1}|\phi_{-1}\rangle$. The resonance term, 
$\tau_\mathrm{bg}(z)$, may be represented, by the following ratio:
\begin{equation}
  \tau_\mathrm{bg}(z)=\frac{\xi_\mathrm{bg}}{1-\xi_\mathrm{bg}
    \langle\chi_\mathrm{bg}|G_0(z)|\chi_\mathrm{bg}\rangle}. 
  \label{taubg}
\end{equation}
As $\tau_\mathrm{bg}(z)$ is required to reproduce the singularity of the bare 
Green's function in the limit $z\to E_{-1}$, the amplitude, $\xi_\mathrm{bg}$, 
is given by $1/\langle\chi_\mathrm{bg}|G_0(E_{-1})|\chi_\mathrm{bg}\rangle$.

The separable representation of $G_\mathrm{bg}(z)$ of Eq.~(\ref{Gbgseparable}) 
provides the exact Green's function associated with an effective potential, 
\begin{equation}
  V_\mathrm{bg}^\mathrm{eff}=|\chi_\mathrm{bg}\rangle\xi_\mathrm{bg}
  \langle\chi_\mathrm{bg}|,
  \label{Vseparable}
\end{equation}
determined by the amplitude, $\xi_\mathrm{bg}$, and the form factor,
$|\chi_\mathrm{bg}\rangle$. This follows directly from the resolvent identity 
\cite{Taylor72}, i.e.,
\begin{equation}
  G_\mathrm{bg}(z)=G_0(z)+G_0(z)V_\mathrm{bg}^\mathrm{eff}G_\mathrm{bg}(z),
  \label{resolventbg}
\end{equation}
which is readily verified upon multiplication by 
$G_\mathrm{bg}^{-1}(z)=z-H_\mathrm{bg}$ from the right and by 
$G_0^{-1}(z)=z+\hbar^2\boldsymbol{\nabla}^2/m$ from the left. Iterating 
Eq.~(\ref{resolventbg}) yields the Born series which reduces to a geometric 
series in the special case of the separable potential of
Eq.~(\ref{Vseparable}). Its exact sum is given by Eq.~(\ref{Gbgseparable})
with the resonance term of Eq.~(\ref{taubg}). 

Such effective interactions are commonly employed in the theory of 
few-particle systems \cite{YamaguchiPR54,MitraPR62,LovelacePR64} as well as in 
condensed matter physics \cite{Schrieffer64}. In applications to cold gases 
the form factor is not resolved because it is sensitive only to the physics on 
distance scales on the order of the van der Waals length. Its associated 
functional form is therefore arbitrary and may be chosen, for instance, to be 
Gaussian \cite{GoralJPhysB04}. In the momentum space representation this 
yields:
\begin{equation}
  \langle\mathbf{p}|\chi_\mathrm{bg}\rangle=\chi_\mathrm{bg}(p)=
  \frac{\exp\left(-\frac{p^2\sigma_\mathrm{bg}^2}{2\hbar^2}\right)}
       {(2\pi\hbar)^{3/2}}.
       \label{formfactorGaussian}
\end{equation}
Here $\sigma_\mathrm{bg}$ accounts for the range of the interaction, and 
$\langle\mathbf{r}|\mathbf{p}\rangle=\exp(i\mathbf{p}\cdot\mathbf{r}/\hbar)
/(2\pi\hbar)^{3/2}$ denotes the plane wave associated with the relative 
momentum $\mathbf{p}$. Given this choice of form factor, the parameters 
$\sigma_\mathrm{bg}$ and $\xi_\mathrm{bg}$ are determined by the requirement 
that the bare Green's function of Eq.~(\ref{Gbgseparable}) exactly recovers 
$a_\mathrm{bg}$ and $E_{-1}$. The zero energy limit of the scattering 
amplitude associated with Eq.~(\ref{Gbgseparable}), i.e., 
$f_\mathrm{bg}(0)=-m\tau_\mathrm{bg}(0)/(4\pi\hbar^2)$, yields the condition
\begin{equation}
  a_\mathrm{bg}=\sigma_\mathrm{bg}\frac{x}
  {1+x/\sqrt{\pi}},
  \label{condabg}
\end{equation}
where $x=m\xi_\mathrm{bg}/(4\pi\hbar^2\sigma_\mathrm{bg})$ is a dimensionless
variable. In addition, the bare energy level determined by the pole of the 
resonance term of Eq.~(\ref{taubg}) gives the following relation:
\begin{equation}
  1-\frac{x}{\sqrt{\pi}}
  \left[\sqrt{\pi}ye^{y^2}\mathrm{erfc}(y)-1\right]=0.
  \label{condEminusone}
\end{equation}
Here $\mathrm{erfc}(y)=\frac{2}{\sqrt{\pi}}\int_y^\infty e^{-u^2} du$ is the 
complementary error function with the argument
$y=\sqrt{m|E_{-1}|}\sigma_\mathrm{bg}/\hbar$.

For the example of the 1007\,G zero energy resonance of $^{87}$Rb the energy 
$E_{-1}$ can be determined using the potential of Eq.~(\ref{VbgHCC6}). To this 
end, its parameter $r_0$ should be chosen in such a way that the number of 
vibrational levels is large compared to unity and that the known quantities 
$a_\mathrm{bg}$ and $C_6$ of Table~\ref{tab:Vbgparameters} are recovered. This 
yields $E_{-1}/h=-24\,$MHz in agreement with coupled channels calculations 
\cite{vanKempenslope04}. Based on the precise values of $E_{-1}$ and 
$a_\mathrm{bg}$, Eqs.~(\ref{condabg}) and (\ref{condEminusone}), in turn, give 
the range parameter and the amplitude of the separable potential to be 
$\sigma_\mathrm{bg}=44\,a_\mathrm{Bohr}$ and 
$m\xi_\mathrm{bg}/(4\pi\hbar^2)=-339\,a_\mathrm{Bohr}$, respectively.

We note that the above adjustment of the effective entrance channel 
interaction is restricted to zero energy resonances with a positive background 
scattering length, i.e., $a_\mathrm{bg}>0$. In the opposite case, 
$a_\mathrm{bg}<0$, separable interactions do not support any bound state, 
similarly to the universal contact pseudo potential of 
Subsection~\ref{subsec:classification}. Among the experimentally relevant 
examples of Table~\ref{tab:Vbgparameters} only the broad, entrance-channel 
dominated zero energy resonances of $^6$Li and $^{85}$Rb have a negative 
background scattering length. Associated two-channel approaches describing 
the dressed energy levels beyond the Wigner threshold law regime exist, at 
least, in applications to the broad resonance of $^{85}$Rb 
\cite{KokkelmansPRL02,KoehlerPRA04,GoralPRA05}. In the case of $^6$Li a double 
resonance approach suitable for the description of many-body systems has been 
suggested \cite{KokkelmansPRA02}. 

Figures~\ref{fig:85RbEbofB} and \ref{fig:6LiZofB} reveal, however, that in 
both cases, $^{85}$Rb and $^6$Li, the Fesh\-bach molecule is well described by 
an effective single channel interaction, over a wide range of magnetic field 
strengths, in accordance with Subsection~\ref{subsec:classification}. An 
appropriate magnetic field dependent single channel separable potential, 
suitable for applications to few-body bound states \cite{StollPRA05} as well 
as the dynamics of cold gases \cite{KoehlerPRA03}, may be constructed on the 
basis of Eqs.~(\ref{Vseparable}) and (\ref{formfactorGaussian}). The 
associated constant range parameter of the form factor is given by 
$\sigma_\mathrm{bg}\approx\sqrt{\pi}\bar{a}/2$. The adjustment of  
$\xi_\mathrm{bg}$ via Eq.~(\ref{condabg}) to the magnetic field dependent 
scattering length $a(B)$ of Eq.~(\ref{aofB}) instead of just $a_\mathrm{bg}$ 
ensures, in turn, that the near resonant energy $E_\mathrm{b}(B)$ of the 
Fesh\-bach molecule recovers Eq.~(\ref{EbGF}).

\subsubsection{The resonance energy}
In accordance with the single resonance approach of 
Eq.~(\ref{replacementHcl}), the difference in energies of the resonance level 
and the entrance channel dissociation threshold, $E_\mathrm{res}(B)$, 
characterises the closed channel part of the Hamiltonian. The associated 
relative magnetic moment, 
$\mu_\mathrm{res}=\partial E_\mathrm{res}/\partial B$, may be inferred from a 
Stern-Gerlach separation experiment, relying upon the force experienced by an 
atom due to the inhomogeneous magnetic field, $\mathbf{B}$. Given any definite 
Zeeman state, this force is of the general form 
$\mathbf{F}_\mathrm{a}=-\boldsymbol{\nabla} E_\mathrm{a}$ where $E_\mathrm{a}$ 
is the Zeeman energy. The magnitude of 
$\mathbf{F}_\mathrm{a}=-(\partial E_\mathrm{a}/\partial B)
\boldsymbol{\nabla}|\mathbf{B}|$ depends on the orientation quantum number 
$m_f$ of the total angular momentum with respect to the field direction. This 
dependence is well described by the Breit-Rabi formula \cite{BreitPR31}: 
\begin{equation}
  \mathbf{F}_\mathrm{a}=\pm\frac{\frac{2m_f}{2i+1}+x}
	{2\left(1+\frac{4m_f}{2i+1}x+x^2\right)^{1/2}}
	 g_j\mu_\mathrm{Bohr}\boldsymbol{\nabla}|\mathbf{B}|.
	 \label{BreitRabi}
\end{equation}
Here $i$ is the quantum number associated with the nuclear spin,
$\mu_\mathrm{Bohr}=9.27400949\times 10^{-24}\,$J/T denotes the Bohr magneton, 
and the Land{\'e} factor, $g_j\approx 2$, refers to the electronic magnetic 
moment. The dimensionless variable, 
$x=g_j\mu_\mathrm{Bohr} |\mathbf{B}|/E_\mathrm{hf}$, depends on the field 
strength, $|\mathbf{B}|$, in addition to the hyperfine energy splitting in the 
absence of magnetic fields, $E_\mathrm{hf}$ \cite{ArimondoRMP77}. For any 
given orientation quantum number, $m_f$, and atomic species, the sign of the 
force of Eq.~(\ref{BreitRabi}) is determined by the Zeeman multiplet.

\begin{figure}[htbp]
  \includegraphics[width=\columnwidth,clip]{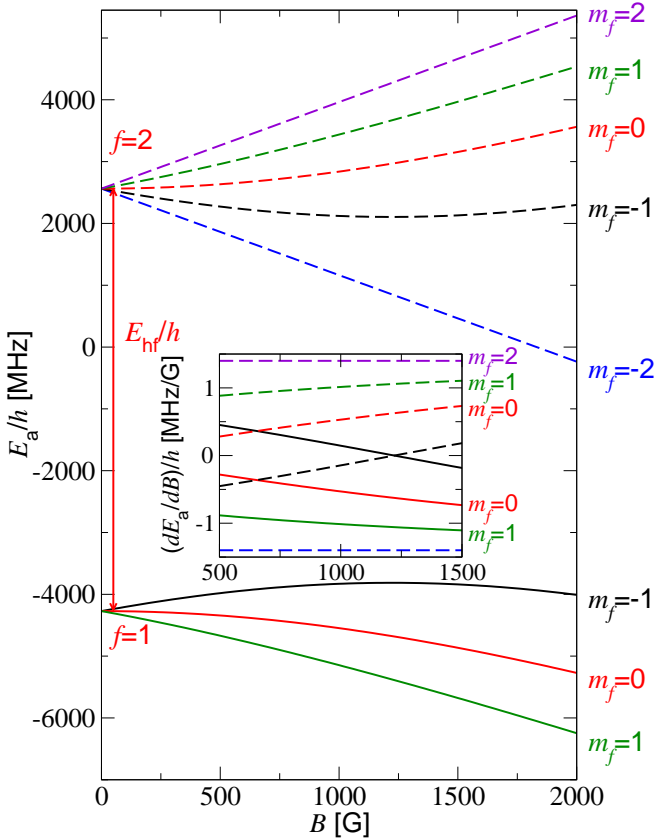}
  \caption{(Colour in online edition)
    The Zeeman multiplets of $^{87}$Rb associated with the total 
    angular momentum quantum numbers $f=1$ (solid curves) and $f=2$ 
    (dashed curves). The inset shows the magnetic moments, 
    $\partial E_\mathrm{a}/\partial B$, determined by Eq.~(\ref{BreitRabi}) 
    and indicates the orientation quantum numbers $m_f$ of those Zeeman states 
    that are relevant to Fig.~\ref{fig:87Rbcrossing}.}
  \label{fig:87RbEZeeman}
\end{figure}

For the example illustrated in Fig.~\ref{fig:87RbEZeeman}, the nuclear spin 
quantum number $i=3/2$ of $^{87}$Rb \cite{ArimondoRMP77} in addition to the 
electronic spin give rise to two Zeeman multiplets. These states adiabatically 
correlate with the hyperfine levels of total angular momentum quantum numbers 
$f=1$ and $f=2$ in the limit of zero magnetic field. The $s$-wave entrance 
channel spin configuration of any pair of the identical Bose atoms in the 
experiments of Fig.~\ref{fig:87Rbcrossing} is characterised by the quantum 
numbers $(f_1=1,m_{f_1}=+1;f_2=1,m_{f_2}=+1)$, referring to their electronic 
ground states. In accordance with Eq.~(\ref{BreitRabi}) and the zero field 
hyperfine structure splitting of $E_\mathrm{hf}/h=6834\,$MHz 
\cite{ArimondoRMP77}, the total magnetic moment associated with these ground 
state atoms at $B_0=1007.4\,$G is determined by 
$2\partial E_\mathrm{a}/\partial B=-h\times 2\,$MHz/G. This derivative varies 
by less than 1\,\% over the range of magnetic field strengths displayed in 
Fig.~\ref{fig:87Rbcrossing}.

The predominant spin exchange interaction couples the pairs of $^{87}$Rb 
ground state atoms to four $s$-wave scattering channels characterised by pairs 
of Zeeman levels, $(f_1,m_{f_1};f_2,m_{f_2})$, whose total angular momentum 
orientation quantum number is conserved, i.e., $m_{f_1}+m_{f_2}=+2$. These 
closed channels are therefore described by $(1,1;2,1)$, $(1,0;2,2)$, 
$(2,0;2,2)$ and $(2,1;2,1)$. In accordance with the inset of 
Fig.~\ref{fig:87RbEZeeman}, their associated total magnetic moments at 
resonance amount to $0$, $h\times0.9\,$MHz/G, $h\times1.9\,$MHz/G, and 
$h\times2\,$MHz/G, respectively. The magnetic moment of the $^{87}$Rb$_2$ 
Fesh\-bach molecule in the lower panel of Fig.~\ref{fig:87Rbcrossing} 
consists, in principle, of contributions from all five channels weighted by 
their admixtures. Both the experimental data \cite{DuerrPRL04} and the coupled 
channels predictions \cite{vanKempenslope04}, however, are consistent with a 
Fesh\-bach resonance state predominantly consisting of the pairs of Zeeman 
states just from the $f=2$ multiplet. The Breit-Rabi formula therefore yields 
an estimated difference in the magnetic moments of the resonance state and the 
entrance channel spin configuration of about 
$\mu_\mathrm{res}=h\times4\,$MHz/G in the vicinity of the zero energy 
resonance position. In an analogous manner the size of the parameter 
$\mu_\mathrm{res}$ may be inferred from Eq.~(\ref{BreitRabi}) for a variety of 
species. This procedure is particularly useful in the case of genuinely 
two-channel problems, such as, for instance, the spin configurations relevant 
to the 202\,G zero energy resonance of $^{40}$K \cite{BruunPRL04}. The 
recommended values of $\mu_\mathrm{res}$ of Table~\ref{tab:couplingparameters} 
are all based on coupled channels predictions. 

\begin{table*}
  \caption{Parameters characterising the inter-channel coupling associated 
    with several experimentally relevant zero energy resonances. Those values 
    of $\Delta B$ and $\mu_\mathrm{res}$ that are unreferenced refer to 
    calculations performed for this review. In accordance with 
    Eq.~(\ref{conditionecd2}), the size of the parameter $\eta$ indicates 
    whether a zero energy resonance is closed- or entrance-channel dominated.}
  \label{tab:couplingparameters}
  \begin{ruledtabular}
    \begin{tabular}{ccccc}
      Species 
      & $B_0$ [G] 
      & $\Delta B$ [G] 
      & $\mu_\mathrm{res}/h$ [MHz/G]
      & $\eta$\\
      \hline
      $^6$Li 
      & 543.25(5) \cite{StreckerPRL03} 
      & 0.1 
      & 2.8 
      & 1215\\
      & 834.149 \cite{BartensteinPRL05} 
      & $-300$ \cite{BartensteinPRL05}
      & 2.8 
      & 0.02\\
      $^{23}$Na 
      & 853 \cite{StengerPRL99} 
      & 0.01 \cite{MiesPRA00}
      & 5.24 \cite{MiesPRA00} 
      & 1090\\
      & 907 \cite{StengerPRL99} 
      & 1.0 \cite{MiesPRA00}
      & 5.24 \cite{MiesPRA00} 
      & 11\\
      $^{40}$K  
      & 202.10(7) \cite{RegalpairingPRL04} 
      & 7.8(6) \cite{GreinerNature03}
      & 2.35 \cite{NygaardPRA06}
      & 0.46\\
      & 224.21(5) \cite{RegalPRL03} 
      & 9.7(6) \cite{RegalPRL03} 
      & 2.35 \cite{NygaardPRA06}
      & 0.37\\
      $^{85}$Rb 
      & 155.0 \cite{ThompsonRFPRL05} 
      & 10.71(2) \cite{ClaussenPRA03}
      & $-3.26$ \cite{Kokkelmansprivate} 
      & 0.04\\
      $^{87}$Rb 
      & 1007.40(4) \cite{VolzPRA03} 
      & 0.21 \cite{DuerrPRA04}
      & 4.2 \cite{DuerrPRL04} 
      & 5.9\\
      $^{133}$Cs 
      & 19.90(3) \cite{ChinPRA04} 
      & 0.005 \cite{JulienneJModOpt04} 
      & 0.798 \cite{JulienneJModOpt04} 
      & 437\\
      & 47.97(3) \cite{ChinPRA04} 
      & 0.15 \cite{JulienneJModOpt04} 
      & 2.09 \cite{JulienneJModOpt04} 
      & 0.99
    \end{tabular}
  \end{ruledtabular}
\end{table*}

\subsubsection{Inter-channel coupling} 
The inter-channel coupling leads to the decay width and energy shift of the 
bare resonance level due to its interaction with the background scattering 
continuum and entrance channel bare vibrational states. In accordance with 
Eq.~(\ref{Wignerlaw}), both quantities are determined by the parameters 
$\Delta B$ of Eq.~(\ref{resonancewidth}) and $B_0-B_\mathrm{res}$ of 
Eq.~(\ref{resonanceshift}) in the Wigner threshold law domain. While the width 
in the magnetic field strength, $\Delta B$, is measured routinely, the 
associated shift is not directly observable. Its magnitude, however, may be 
inferred using general ideas developed in the context of multichannel quantum 
defect theory \cite{MiesRaoultPRA00,MiesRaoultPRA04,JulienneJOptSocAm89}. This 
yields the following approximate formula:
\begin{equation}
   B_0-B_\mathrm{res}=\Delta B\frac{a_\mathrm{bg}}{\bar{a}}
   \left[
     \frac{1-a_\mathrm{bg}/\bar{a}}{1+\left(1-a_\mathrm{bg}/\bar{a}\right)^2}
   \right].
   \label{magicformula}
\end{equation}
Here $\bar{a}$ is the mean scattering length of 
Eq.~(\ref{meanscatteringlength}). Consequently, the size of the resonance 
shift depends just on the quantities $a_\mathrm{bg}$, $\Delta B$, and $C_6$ 
which are all accessible to experimental studies. Entrance-channel dominated 
zero energy resonances, such as the example of $^{85}$Rb illustrated in 
Fig.~\ref{fig:85RbEbofB}, tend to have large shifts comparable to the size of 
$\Delta B$. The predicted magnitude of $B_0-B_\mathrm{res}=9\,$G in 
Fig.~\ref{fig:85RbEbofB} is consistent with the coupled channels binding 
energies \cite{Kokkelmansprivate}. In the case of the closed-channel dominated 
1007\,G zero energy resonance of $^{87}$Rb in the upper panel of 
Fig.~\ref{fig:87Rbcrossing} the value of $B_0-B_\mathrm{res}=0.07\,$G of 
Eq.~(\ref{meanscatteringlength}) is significantly smaller than the width. For 
all the examples given in Tables~\ref{tab:Vbgparameters} and 
\ref{tab:couplingparameters} the corrections to Eq.~(\ref{magicformula}) are 
negligible. 

The quantity $W|\phi_\mathrm{res}\rangle$, characterising the inter-channel 
coupling in the single resonance approach, should be adjusted in such a way 
that the matrix element of the bare Green's function in Eq.~(\ref{Wignerlaw}) 
recovers both $\Delta B$ and $B_0-B_\mathrm{res}$. Such an adjustment ensures, 
in particular, that the bound state energy of Eq.~(\ref{determinationEb}) 
properly interpolates between the universal and asymptotic regimes of magnetic 
field strengths. An explicit minimal implementation of the two-channel 
Hamiltonian of Eq.~(\ref{H2B2channel}) may therefore be based, for instance, 
on the separable background scattering potential of Eq.~(\ref{Vseparable}) in 
addition to the following general expression:
\begin{equation}
  W|\phi_\mathrm{res}\rangle =|\chi\rangle \zeta.
\end{equation}
Here the amplitude $\zeta$ determines the overall magnitude of the 
inter-channel coupling and $|\chi\rangle$ its functional form. As the off 
diagonal potential, $W(r)$, is not resolved by the typically large de Broglie 
wave lengths associated with cold collisions, the form factor $|\chi\rangle$ 
may be chosen as a Gaussian function in momentum space, similarly to 
Eq.~(\ref{formfactorGaussian}). This yields \cite{GoralJPhysB04}:
\begin{equation}
  \langle\mathbf{p}|\chi\rangle=\chi(p)=
  \frac{\exp\left(-\frac{p^2\sigma^2}{2\hbar^2}\right)}
       {(2\pi\hbar)^{3/2}}.
       \label{couplingGaussian}
\end{equation}
The associated range parameter, $\sigma$, and the amplitude, $\zeta$, are 
completely determined by the requirement that the imaginary and real parts of 
Eq.~(\ref{Wignerlaw}) recover the physical quantities $\Delta B$ and 
$B_0-B_\mathrm{res}$, respectively. For the purpose of this adjustment, it is 
convenient to introduce the average range parameter
$\overline{\sigma}=\frac{1}{\sqrt{2}}(\sigma^2+\sigma_\mathrm{bg}^2)^{1/2}$.
The imaginary part of the matrix element of the bare Green's function on the
left hand side of Eq.~(\ref{Wignerlaw}) yields the first condition
\begin{equation}
  \Delta B=\frac{m|\zeta|^2}{4\pi\hbar^2a_\mathrm{bg}\mu_\mathrm{res}}
  \left(
  1-\frac{a_\mathrm{bg}}{\sqrt{\pi}\overline{\sigma}}
  \right)^2.
  \label{deltaBofzetasigma}
\end{equation}
The associated widths of several experimentally relevant zero energy 
resonances are summarised in Table~\ref{tab:couplingparameters}. Given the 
bare Green's function of Eq.~(\ref{Gbgseparable}) and Gaussian form factors,
also the real part of Eq.~(\ref{Wignerlaw}) can be evaluated analytically.
This leads to the second condition,
\begin{equation}
  B_0-B_\mathrm{res}=\Delta B\,\frac{a_\mathrm{bg}}{\sqrt{\pi}\sigma}
  \frac{1-\frac{a_\mathrm{bg}}{\sqrt{\pi}\sigma}
    \left(\frac{\sigma}{\overline{\sigma}}\right)^2}
       {\left(
	 1-\frac{a_\mathrm{bg}}{\sqrt{\pi}\sigma}
	 \frac{\sigma}{\overline{\sigma}}
	 \right)^2},
       \label{shiftozetasigma}
\end{equation}
whose value for the resonance shift on the left hand side is provided by 
Eq.~(\ref{magicformula}). Equations~(\ref{deltaBofzetasigma}) and 
(\ref{shiftozetasigma}), in turn, simultaneously determine the parameters
$\sigma$ and $\zeta$ characterising the inter-channel coupling. We note that 
the overall phase of the amplitude $\zeta$ is irrelevant to the physics 
described by the associated Hamiltonian.

For the example of the 1007\,G zero energy resonance of $^{87}$Rb such an 
adjustment yields $\sigma=22\,a_\mathrm{Bohr}$ and 
$m|\zeta|^2/(4\pi\hbar^2\sigma)=h\times 10\,$MHz. Given the effective 
background scattering potential of Eq.~(\ref{Vseparable}) and 
$\mu_\mathrm{res}$ of Table~\ref{tab:couplingparameters}, this procedure 
provides a complete implementation of the two-channel single resonance 
approach. Its predictions with respect to dressed binding energies are 
illustrated in the upper panel of Fig.~\ref{fig:87Rbcrossing} as well as in 
Fig.~\ref{fig:85RbEbofB}. The lower panel of Fig.~\ref{fig:87Rbcrossing} shows 
the magnetic moment of $^{87}$Rb Fesh\-bach molecules determined from the 
product $\mu_\mathrm{res}Z(B)$ via Eqs.~(\ref{twochannelnormalisation}) and 
(\ref{ZofBgeneral}) using the two-channel single resonance approach. The 
overall offset of the data of about $-2\,$MHz refers to the magnetic moment 
associated with the entrance channel spin configuration. The comparisons with 
the measurements \cite{DuerrPRL04} and with the results of coupled channels 
calculations \cite{vanKempenslope04} indicate that the effective two-channel 
approach can fully recover the microscopic physics within the experimental 
energy range. 

\section{Association of Fesh\-bach molecules} 
\label{sec:associationsweeps}
The different experimental techniques for molecular association in cold gases 
all depend in one way or another on the properties of the diatomic energy 
spectra. Several approaches to the production of cold Fesh\-bach molecules are 
based on the relaxation of an atomic gas into dimers near resonance 
\cite{JochimPRL03,ZwierleinPRL03} or on dynamical sweeps of the magnetic field 
strength across $B_0$
\cite{CubizollesPRL03,XuPRL03,RegalNature03,StreckerPRL03,HerbigScience03,DuerrPRL04}. 
Both techniques take advantage of the degeneracy of the Fesh\-bach molecular 
energy, $E_\mathrm{b}$, and the threshold for dissociation into free atoms in 
the limit $B\to B_0$. Relaxation of an atomic gas into dimer molecules 
requires collisions of at least three atoms to balance the energies and is 
commonly employed, to date, just in two spin component mixtures of $^6$Li 
Fermi gases. The conceptually simpler molecular association via magnetic field 
sweeps seems to be more generally applicable to both Bose and Fermi gases and 
will therefore be the main subject of this section. 
Figure~\ref{fig:87Rbcrossing} illustrates its principle which relies, for the 
example of the 1007\,G zero energy resonance of $^{87}$Rb, upon the adiabatic 
transition from the diatomic zero energy continuum level to the bound state 
energy $E_\mathrm{b}(B)$ with decreasing $B$. Conversely, 
Fig.~\ref{fig:85RbEbofB} suggests that in the case of the 155\,G zero energy 
resonance of $^{85}$Rb free atom pairs may be associated to Fesh\-bach 
molecules by increasing $B$ across $B_0$. The difference in energy of the 
colliding atoms and diatomic molecules is absorbed by the time dependent 
magnetic field. As, in general, the field is spatially homogeneous, the 
association process does not affect the centre of mass momentum of the atom 
pairs. From this viewpoint, the Fesh\-bach molecules produced by magnetic 
field sweeps or other related dynamical techniques 
\cite{DonleyNature02,ThompsonRFPRL05} are as cold as the atomic gas they 
originate from. 

\subsection{Linear sweeps of the magnetic field strength}
\label{subsec:linearsweeps}
In an idealised treatment of the molecular association the magnetic field 
strength may be assumed to vary linearly in time. This implies the relation
\begin{equation}
  B(t)=B_\mathrm{res}+\dot{B}\,(t-t_\mathrm{res})
\end{equation} 
where $\dot{B}$ is usually referred to as the ramp speed and $t_\mathrm{res}$ 
is chosen to be the time at which the bare resonance energy, $E_\mathrm{res}$, 
crosses the dissociation threshold of the entrance channel. The field 
strength, $B_\mathrm{res}$, associated with $t_\mathrm{res}$ is indicated in 
Fig.~\ref{fig:85RbEbofB}. In accordance with Eq.~(\ref{slope}), the resonance 
energy is then also a linear function of time,
\begin{equation} 
  E_\mathrm{res}(t)=\dot{E}_\mathrm{res}\,(t-t_\mathrm{res}),
  \label{Eressweep}
\end{equation}
with a constant derivative, $\dot{E}_\mathrm{res}=\mu_\mathrm{res}\dot{B}$. 
This formula presupposes that the magnetic field sweep is sufficiently slow
for the electronic degrees of freedom of the atoms to adiabatically adjust to 
the changes in the magnetic field strength. To achieve molecular association 
of an atom pair, $\dot{E}_\mathrm{res}$ needs to be negative \cite{MiesPRA00}. 
Consequently, the Fesh\-bach resonance level is swept downward in time. This 
requirement determines $\dot{B}$ through the sign of $\mu_\mathrm{res}$. 
Conversely, upward sweeps of the Fesh\-bach resonance level across $B_0$ lead 
to heating of the atomic cloud. This general principle may be readily verified 
on the basis of the magnetic field dependence of the discrete spectrum of 
dressed energy levels of a trapped atom pair.

\subsubsection{Adiabatic association of Fesh\-bach molecules}
Tight traps containing just a single pair of atoms may be realised, for 
instance, by the individual sites of an optical lattice 
\cite{TiesingaPRA00,JakschPRL02} or by micro-fabricated materials 
\cite{LongPhilTrans03,FolmanAdvAtomMolOptPhys02,HindsPRL01,MuellerPRL99,ThywissenPRL99,WeinsteinPRA95}. 
The two-body energy spectra associated with periodic potentials of optical 
lattices have been studied theoretically in the Wigner threshold law domain 
\cite{OrsoPRL05,WoutersPRA06} as well as observed experimentally 
\cite{MoritzPRL05}. While in the limit of high excitations the tunnelling of 
atoms is significant, the deepest localised diatomic levels of tight lattice 
sites are well described by the harmonic oscillator approximation. As a 
consequence, the centre of mass and relative motions of an atom pair confined 
to a single site can be treated separately, similarly to the two-body problem 
in free space. 

A spherically symmetric harmonic confinement modifies the bare entrance 
channel Hamiltonian associated with the resonance enhanced interaction as 
follows:
\begin{equation}
  H_\mathrm{bg}=-\frac{\hbar^2}{m}\boldsymbol{\nabla}^2+V_\mathrm{bg}(r)
  +V_\mathrm{trap}(r).
\end{equation}
Here $V_\mathrm{trap}(r)$ denotes the potential energy of the isotropic 
harmonic oscillator in the barycentric frame which is given in terms of the 
reduced mass, $m/2$, and the angular frequency, $\omega_\mathrm{ho}$, to be:
\begin{equation}
  V_\mathrm{trap}(r)=\frac{1}{2}\frac{m}{2}\omega_\mathrm{ho}^2 r^2.
\end{equation}
Similarly to Eq.~(\ref{barephip}), the associated bare vibrational energy 
levels are determined by the stationary Schr\"odinger equation,
\begin{equation}
  H_\mathrm{bg}\phi_v(r)=E_v\phi_v(r).
\end{equation}
Here the index $v=\ldots,-2,-1,0,1,2,\ldots$ labels the vibrational excitation 
in such a way that $v=0$ correlates adiabatically, in the limit 
$\omega_\mathrm{ho}\to 0$, with the dissociation threshold of the entrance 
channel. According to this counting scheme, negative indices, $v<0$, are 
associated with the bare vibrational levels $E_{-1}$, $E_{-2}$, etc., 
indicated in Fig.~\ref{fig:85RbFeshbachmodel}. In the limit of low vibrational 
excitation, $v\gtrsim 0$, the spatial extents of the bare states, including 
$\phi_0(r)$, are characterised by the trap length,
$a_\mathrm{ho}=\sqrt{\hbar/(m\omega_\mathrm{ho})}$. This length scale usually
very much exceeds the modulus of the background scattering length, 
$|a_\mathrm{bg}|$. To first order in $a_\mathrm{bg}/a_\mathrm{ho}$ the 
energies of the excited levels are well approximated by the following formula 
\cite{BuschFoundPhys98}:
\begin{equation}
  E_v\approx
  \left[
    \frac{3}{2}+2v+\sqrt{\frac{2}{\pi}}\binom{v+1/2}{v}
    \frac{a_\mathrm{bg}}{a_\mathrm{ho}}
    \right]\hbar\omega_\mathrm{ho}.
  \label{Evcontact}
\end{equation}
Here $\binom{v+1/2}{v}$ is a combinatorial. The dressed energy levels may be 
determined via the two-channel Hamiltonian of Eq.~(\ref{H2B2channel}) using 
the single resonance approach of Eq.~(\ref{replacementHcl}) \cite{MiesPRA00}, 
or via a single channel energy dependent contact interaction 
\cite{BoldaPRA02,BlumePRA02}. Both methods yield spectra consistent with full 
coupled channels calculations \cite{TiesingaPRA00}.

\begin{figure}[htbp]
  \includegraphics[width=\columnwidth,clip]{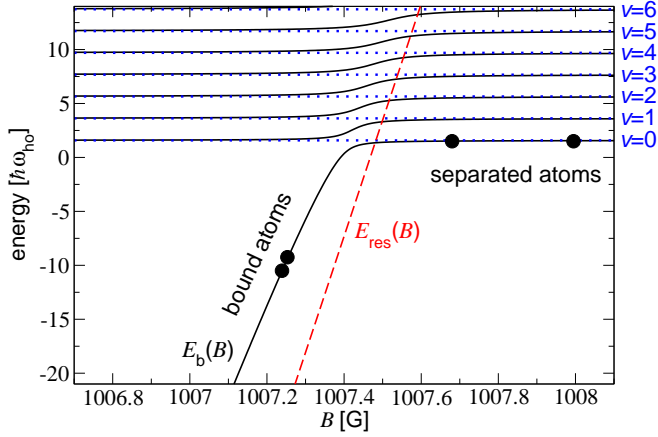}
  \caption{(Colour in online edition)
    Schematic illustration of the molecular association of a pair of 
    ground state $^{87}$Rb atoms via a downward magnetic field sweep in a 
    spherical harmonic atom trap with an oscillator frequency of 
    $\nu_\mathrm{ho}=39\,$kHz \cite{ThalhammerPRL06}. The bare vibrational 
    levels ($v=0,\ldots, 6$) associated with the background scattering quasi 
    continuum and the Fesh\-bach resonance energy, $E_\mathrm{res}(B)$, are 
    indicated by dotted and dashed lines, respectively. Solid curves refer to 
    the magnetic field dependence of the dressed energy levels in the vicinity 
    of the zero energy resonance position of $B_0=1007.4\,$G.}
  \label{fig:87Rb39kHz}
\end{figure}

Figure~\ref{fig:87Rb39kHz} shows the magnetic field dependence of the dressed 
two-body energy levels of ground state $^{87}$Rb atoms predicted by the 
two-channel approach for a tight trap with a frequency of 
$\nu_\mathrm{ho}=\omega_\mathrm{ho}/(2\pi)=39\,$kHz \cite{ThalhammerPRL06}. An 
atom pair from a rubidium Bose-Einstein condensate prepared on the high field 
side of the 1007\,G zero energy resonance and loaded adiabatically into an 
optical lattice site is well described by the $v=0$ state. According to 
Fig.~\ref{fig:87Rb39kHz}, this level adiabatically correlates with the 
Fesh\-bach molecule on the low field side of $B_0$. Consequently, a magnetic 
downward sweep across $B_0$ associates the separated rubidium atoms to 
molecules with certainty in the limit of zero ramp speed, i.e.~$\dot{B}\to 0$. 
An excited trap level with the vibrational quantum number $v$ is transferred 
to $v-1$. Conversely, an adiabatic upward magnetic field sweep across $B_0$ 
dissociates $^{87}$Rb$_2$ Fesh\-bach molecules and leads to heating 
transitions from $v$ to $v+1$ in the excited trap levels. In general, the ramp 
direction for such cooling or heating transitions is determined just by the 
sign of $\mu_\mathrm{res}$ or, equivalently, by the time variation of the 
resonance energy, $\dot{E}_\mathrm{res}$. 

\subsubsection{Exact time evolution of a single atom pair}
While the dressed energy levels of Fig.~\ref{fig:87Rb39kHz} reveal the 
mechanism of Fesh\-bach molecular association in the limit of zero ramp speed, 
the dynamics of the diatomic wave function is described by the Schr\"odinger 
equation,
\begin{equation}
  i\hbar\frac{\partial}{\partial t}|\Psi(t)\rangle=
  H_\mathrm{2B}(t)|\Psi(t)\rangle.
  \label{SEPsidynamical}
\end{equation}
Here $H_\mathrm{2B}$ is the two-channel Hamiltonian of Eq.~(\ref{H2B2channel}) 
in the single resonance approach of Eq.~(\ref{replacementHcl}). Its time 
dependence is determined by the linear variation of the resonance energy of 
Eq.~(\ref{Eressweep}). The diatomic state, $|\Psi(t)\rangle$, has components 
in the entrance and closed channels, whose wave functions may be expanded into 
bare states, in accordance with the formulae
\begin{align}
  \Psi_\mathrm{bg}(r,t)&=\sum_v\phi_v(r) C_v(t),\\
  \Psi_\mathrm{cl}(r,t)&=\phi_\mathrm{res}(r)C_\mathrm{res}(t).
\end{align}
Such a basis set expansion of Eq.~(\ref{SEPsidynamical}) leads to the 
following dynamical equations for the associated time dependent coefficients
\begin{align}
  \label{CIdynamicstrap}
  i\hbar\dot{C}_v(t)&=E_vC_v(t)+\langle\phi_v|W|\phi_\mathrm{res}\rangle
  C_\mathrm{res}(t),\\
  i\hbar\dot{C}_\mathrm{res}(t)&=E_\mathrm{res}(t)C_\mathrm{res}(t)
  +\sum_v\langle\phi_\mathrm{res}|W|\phi_v\rangle C_v(t).
  \label{CIdynamicsres}
\end{align}
This configuration interaction approach is particularly useful for numerical 
treatments of Fesh\-bach molecular association using any form of time 
dependent magnetic field variation \cite{MiesPRA00}. The special case of 
linear sweeps belongs to those quantum mechanical problems whose exact 
dynamics can be treated analytically \cite{DemkovJETP68,MacekPRA98}. 

To this end, it is useful to split $H_\mathrm{2B}(t)$ into stationary and time 
dependent contributions:
\begin{equation}
  H_\mathrm{2B}(t)=H_\mathrm{stat}+H_\mathrm{cl}(t).
  \label{separationH2B}
\end{equation}
Here the stationary Hamiltonian, 
$H_\mathrm{stat}=H_\mathrm{2B}(t_\mathrm{res})$, is associated with the 
magnetic field strength $B_\mathrm{res}$. In the single resonance approach the 
dynamical contribution, $H_\mathrm{cl}(t)$, is given by the following formula:
\begin{equation}
  H_\mathrm{cl}(t)=|\phi_\mathrm{res},\mathrm{cl}\rangle
  E_\mathrm{res}(t)\langle\phi_\mathrm{res},\mathrm{cl}|.
  \label{Hintseparable}
\end{equation}
Similarly to the effective potential of Eq.~(\ref{Vseparable}), the time 
dependent interaction, $H_\mathrm{cl}(t)$, involves only a single state 
associated with the closed channel, the product 
$|\phi_\mathrm{res},\mathrm{cl}\rangle=|\phi_\mathrm{res}\rangle
|\mathrm{cl}\rangle$. The following derivations will show that it is the 
separable form of Eq.~(\ref{Hintseparable}) in addition to the linear 
dependence of the resonance energy $E_\mathrm{res}$ on $t$ which allow for the 
analytic treatment of the time evolution.

The complete dynamics of an atom pair exposed to a linear magnetic field sweep 
may be inferred from the time evolution operator determined by the 
Schr\"odinger equation, 
\begin{equation}
  i\hbar \frac{\partial}{\partial t}U_\mathrm{2B}(t,t')
  =H_\mathrm{2B}(t)U_\mathrm{2B}(t,t'),
  \label{SEU2B}
\end{equation}
in addition to the boundary condition $U_\mathrm{2B}(t,t)=1$. The diatomic 
state at time $t$ is thus given in terms of $U_\mathrm{2B}(t,t')$ and the 
state at time $t'$ by  
\begin{equation}
  |\Psi(t)\rangle=U_\mathrm{2B}(t,t')|\Psi(t')\rangle.
\end{equation} 
Similarly, the dynamics in the absence of the interaction of 
Eq.~(\ref{Hintseparable}) is described by 
$\exp[-iH_\mathrm{stat}(t-t')/\hbar]$. Associated with these free and complete 
time evolution operators are the retarded Green's functions,
\begin{align}
  \label{Grefretarded}
  G_\mathrm{stat}^{(+)}(t-t')=&\frac{1}{i\hbar}\theta(t-t')
  \exp[-iH_\mathrm{stat}(t-t')/\hbar],\\
  G_\mathrm{2B}^{(+)}(t,t')=&\frac{1}{i\hbar}\theta(t-t')U_\mathrm{2B}(t,t'),
  \label{G2Bretarded}
\end{align}
where $\theta(t-t')$ is the step function which yields unity if $t>t'$ and 
zero elsewhere. The Schr\"odinger equation (\ref{SEU2B}) may be represented in 
terms of Eqs.~(\ref{Grefretarded}) and (\ref{G2Bretarded}) via the following 
formula:
\begin{align}
  \nonumber
  G_\mathrm{2B}^{(+)}(t,t')=&G_\mathrm{stat}^{(+)}(t-t')\\
  &+\int d\tau \,G_\mathrm{stat}^{(+)}(t-\tau)H_\mathrm{cl}(\tau)
  G_\mathrm{2B}^{(+)}(\tau,t').
  \label{LSG2Bretarded}
\end{align}
Differentiation with respect to the variable $t$ readily verifies this 
relation by recovering the time derivative of the complete retarded Green's 
function of Eq.~(\ref{G2Bretarded}) which is directly determined by 
Eq.~(\ref{SEU2B}). 
The integral representation of Eq.~(\ref{SEU2B}) chosen in 
Eq.~(\ref{LSG2Bretarded}) is usually referred to as the post-form of the 
dynamical equation. The associated pre-form reads
\begin{align}
  \nonumber
  G_\mathrm{2B}^{(+)}(t,t')=&G_\mathrm{stat}^{(+)}(t-t')\\
  &+\int d\tau \,G_\mathrm{2B}^{(+)}(t,\tau)H_\mathrm{cl}(\tau)
  G_\mathrm{stat}^{(+)}(\tau-t')
  \label{LSG2Bretardedpre}
\end{align}
and can be verified similarly to the derivation of Eq.~(\ref{LSG2Bretarded}).

While, in general, the operator equation (\ref{LSG2Bretarded}) requires a 
complete basis set expansion for its numerical solution, the separable form of 
Eq.~(\ref{Hintseparable}) reduces this problem to the calculation of just the 
pair of matrix elements
\begin{align}
  \label{gref}
  g_\mathrm{stat}^{(+)}(t-t')&=\langle\phi_\mathrm{res},\mathrm{cl}|
  G_\mathrm{stat}^{(+)}(t-t')|\phi_\mathrm{res},\mathrm{cl}\rangle,\\
  g_\mathrm{2B}^{(+)}(t,t')&=\langle\phi_\mathrm{res},\mathrm{cl}|
  G_\mathrm{2B}^{(+)}(t,t')|\phi_\mathrm{res},\mathrm{cl}\rangle.
  \label{g2B}
\end{align}
This follows from Eq.~(\ref{Hintseparable}) via multiplying 
Eq.~(\ref{LSG2Bretarded}) by $\langle\phi_\mathrm{res},\mathrm{cl}|$ from the 
left and by $|\phi_\mathrm{res},\mathrm{cl}\rangle$ from the right which, in 
turn, determines Eq.~(\ref{g2B}) through the following integral equation: 
\begin{align}
  \nonumber
  g_\mathrm{2B}^{(+)}(t,t')=&g_\mathrm{stat}^{(+)}(t-t')\\
  &+\int d\tau \,g_\mathrm{stat}^{(+)}(t-\tau)E_\mathrm{res}(\tau)
  g_\mathrm{2B}^{(+)}(\tau,t').
  \label{inteqg2B}
\end{align}
Given the free retarded Green's function of Eq.~(\ref{Grefretarded}), the 
complete time evolution operator can be inferred from the solution of 
Eq.~(\ref{inteqg2B}) by inserting Eq.~(\ref{LSG2Bretardedpre}) into 
Eq.~(\ref{LSG2Bretarded}) and performing the time integrations. 

Using the convolution theorem, a Fourier transform turns the integral on the 
right hand side of Eq.~(\ref{inteqg2B}) into a product of functions. In the 
case of a linear magnetic field sweep such a procedure allows  
Eq.~(\ref{inteqg2B}) to be solved analytically. To this end, it is instructive 
to introduce the energy dependent matrix elements,
\begin{align}
  g_\mathrm{stat}(z)=&\int dt\,e^{iz(t-t')/\hbar} 
  g_\mathrm{stat}^{(+)}(t-t'),\\
  g_\mathrm{2B}(z,t')=&\int dt\,e^{iz(t-t')/\hbar}
  g_\mathrm{2B}^{(+)}(t,t'),
\end{align}
associated with the free and complete retarded Green's functions. Here the 
regularised argument ``$z=E+i0$'' ensures the convergence of the time 
integrals in the limit $t\to +\infty$ by approaching the real energy $E$ from 
the upper half of the complex plane. Given the resonance energy of 
Eq.~(\ref{Eressweep}), a Fourier transform renders Eq.~(\ref{inteqg2B}) into 
the following inhomogeneous first order linear differential equation:
\begin{equation}
  i\hbar\frac{\partial g_\mathrm{2B}(z,t')}{\partial E}=
  \hbar\varphi'(z,t') g_\mathrm{2B}(z,t')+
  1/\dot{E}_\mathrm{res}.
  \label{deg2Bofz}
\end{equation}
Its dependence on the inter-atomic interaction is incorporated in the energy 
derivative $\varphi'(z,t')=\partial\varphi(z,t')/\partial E$ of the complex 
phase
\begin{equation}
  \varphi(z,t')=
  -\frac{1}{\hbar\dot{E}_\mathrm{res}}\int_0^E\frac{dE'}{g_\mathrm{stat}(z')}
  +E(t'-t_\mathrm{res})/\hbar,
  \label{definintionphase}	
\end{equation}
where ``$z'=E'+i0$'' denotes a regularised integration variable with the same 
imaginary part as $z$. 

The imaginary part of $\varphi(z,t')$ can be inferred from the Hamiltonian 
$H_\mathrm{stat}$ using a spectral decomposition of $g_\mathrm{stat}(z)$ 
analogous to Eq.~(\ref{Gbgspectral}). In particular, 
Eq.~(\ref{principalvalue}) determines the sign of $\mathrm{Im}\varphi(z,t')$ 
to be: 
\begin{equation}
  \mathrm{sign}[\mathrm{Im}\varphi(z,t')]=-\mathrm{sign}(\dot{E}_\mathrm{res})
  \mathrm{sign}(\mathrm{Im}z).
\end{equation}
Consequently, in the case of a downward ramp of the Fesh\-bach resonance 
level, i.e.~$\dot{E}_\mathrm{res}<0$, the damped, retarded solution to 
Eq.~(\ref{deg2Bofz}) is given by
\begin{equation}
  g_\mathrm{2B}(z,t')=-\int_E^\infty dE'\,
  \frac{e^{-i[\varphi(z,t')-\varphi(z',t')]}}{i\hbar\dot{E}_\mathrm{res}}.
  \label{solutiong2Bofz}
\end{equation}
This formula may be verified from Eq.~(\ref{deg2Bofz}) by differentiation with
respect to $E$. The matrix element of the complete retarded Green's function 
of Eq.~(\ref{g2B}) is just the inverse Fourier transform of 
Eq.~(\ref{solutiong2Bofz}) which yields:
\begin{equation}
  g_\mathrm{2B}^{(+)}(t,t')=-\int \frac{dE}{2\pi\hbar}\int_E^\infty dE'\,
  \frac{e^{-i[\varphi(z,t)-\varphi(z',t')]}}{i\hbar\dot{E}_\mathrm{res}}. 
  \label{solutiong2Bdownward}
\end{equation}
The exact time evolution operator determined by 
Eq.~(\ref{solutiong2Bdownward}) is applicable to the association of Fesh\-bach 
molecules in free space as well as to the case of a trapped atom pair 
illustrated in Fig.~\ref{fig:87Rb39kHz}.

\subsubsection{Landau-Zener approach}
Already in 1932 Landau and Zener have independently derived the foundation for 
a simple estimate of the probability for molecule production in linear 
magnetic field sweeps \cite{LandauPZS32,ZenerPRS32}. Their generic approaches 
may be interpreted in terms of a coupled system of two channels, each of which 
containing just a single state. In applications to Fesh\-bach molecular 
association such a treatment is equivalent to the single resonance approach of 
Eq.~(\ref{replacementHcl}) in addition to the following replacement of the 
entrance channel Hamiltonian:
\begin{equation}
  H_\mathrm{bg}\to|\phi_0\rangle E_0\langle\phi_0|.
\end{equation}
Here $|\phi_0\rangle$ may be interpreted, for instance, in terms of the zeroth 
vibrational state of the relative motion of a trapped atom pair with the 
energy $E_0$. This reduction of the two-channel continuum to a two-level 
system gives rise to analytic solutions of the coupled set of stationary 
Schr\"odinger equations~(\ref{SEphibg}) and (\ref{SEphicl}). Given the simple
form of the entrance channel Green's function in the Landau-Zener approach, 
i.e.
\begin{equation}
  G_\mathrm{bg}(z)=|\phi_0\rangle\frac{1}{z-E_0}\langle\phi_0|,
\end{equation}
Eq.~(\ref{determinationEb}) determines the two-level dressed energies to be:
\begin{equation}
  E_{\pm}=\frac{E_0+E_\mathrm{res}}{2}
  \pm\frac{|E_0-E_\mathrm{res}|}{2}
  \sqrt{1+
  4\frac{|\langle\phi_\mathrm{res}|W|\phi_0\rangle|^2}
  {(E_0-E_\mathrm{res})^2}}.
  \label{Eplusminus}
\end{equation}
The magnetic field dependent slopes of the levels $E_+$ and $E_-$ indicate, 
quite generally, a crossing of the discrete bare entrance- and closed-channel 
energies, $E_0$ and $E_\mathrm{res}$, respectively. Analytic representations 
of the associated dressed two-component stationary states, $|\phi_+\rangle$ 
and $|\phi_-\rangle$, can be inferred from Eq.~(\ref{phib}). 

In accordance with Eq.~(\ref{Eplusminus}), a downward sweep of 
$E_\mathrm{res}$ across $E_0$ transfers a pair of atoms in the initial state 
$|\phi_0,\mathrm{bg}\rangle$ into the final state 
$|\phi_\mathrm{res},\mathrm{cl}\rangle$ in the limit of zero ramp speed. Their 
energies adiabatically follow the $E_-$ curve. Given the linear variation of 
$E_\mathrm{res}$ of Eq.~(\ref{Eressweep}), such a simplified scenario 
presupposes the sweep to start and end asymptotically far from the crossing 
point of the bare levels. This, in turn, implies the formal limits 
$t_\mathrm{i}\to-\infty$ and $t_\mathrm{f}\to\infty$ of the initial and final 
times, respectively. The adiabatic energy variation of the Landau-Zener 
two-level approach is similar to the harmonic trap case illustrated in 
Fig.~\ref{fig:87Rb39kHz}, except that the quasi continuum of excited levels 
with the indices $v>0$ is neglected.

Finite ramp speeds allow an atom pair to end up in a superposition of the 
entrance and closed channel bare states $|\phi_0,\mathrm{bg}\rangle$ and 
$|\phi_\mathrm{res},\mathrm{cl}\rangle$, respectively. In accordance with 
Eq.~(\ref{g2B}) and the assumption of a two-level system, the probability for 
an asymptotic transition between the bare levels is given by
\begin{align}
  \nonumber
  p_{0,\mathrm{res}}&=\left|\langle\phi_\mathrm{res},\mathrm{cl}|
  U_\mathrm{2B}(t_\mathrm{f},t_\mathrm{i})
  |\phi_0,\mathrm{bg}\rangle\right|^2\\
  &=1-|i\hbar g_\mathrm{2B}(t_\mathrm{f},t_\mathrm{i})|^2
  \label{p0res}
\end{align}
in the limits $t_\mathrm{i}\to-\infty$ and $t_\mathrm{f}\to\infty$. These time 
limits can be determined analytically using the stationary phase condition for 
the energy integrals over the oscillatory functions on the right hand side of 
Eq.~(\ref{solutiong2Bdownward}). This exact approach relies upon the 
observation that asymptotically only those regions close to zeros of the 
derivatives of the phases $\varphi'(z,t_\mathrm{i})$ and 
$\varphi'(z,t_\mathrm{f})$ contribute to the integrals. All the remaining 
energy ranges in which the complex exponentials are rapidly oscillating yield 
negligible averages.  

The positions of the stationary phases can be readily found in the case of a 
two-level system. An explicit determination of the stationary Green's 
function, $G_\mathrm{stat}(z)=(z-H_\mathrm{stat})^{-1}$, associated with the 
Hamiltonian $H_\mathrm{stat}$ of Eq.~(\ref{separationH2B}) yields
\begin{equation}
  1/g_\mathrm{stat}(z)=z-\langle\phi_\mathrm{res}|WG_\mathrm{bg}(z)
  W|\phi_\mathrm{res}\rangle.
  \label{grefinverse}
\end{equation}
This formula and Eq.~(\ref{definintionphase}) give the derivative of the phase 
to be
\begin{equation}
  \varphi'(z,t)=
  -\frac{E-E_\mathrm{res}(t)-\langle\phi_\mathrm{res}|WG_\mathrm{bg}(z)
  W|\phi_\mathrm{res}\rangle}{\hbar\dot{E}_\mathrm{res}}.
  \label{phasederivative} 
\end{equation} 
Consequently, $\varphi'(z,t)$ vanishes at the dressed energies $E_\pm$ 
associated with the magnetic field strength at time $t$, in accordance with 
Eq.~(\ref{determinationEb}). 

As only those regions of parameters $E$ and $E'$ in the close proximity of 
$E_\pm$ significantly contribute to Eq.~(\ref{solutiong2Bdownward}), the 
phases may be expanded to second order about their stationary points. This 
yields:
\begin{equation}
  \varphi(z,t)\approx\varphi(z_\pm,t)
  +\frac{1}{2}\varphi''(E_\pm,t)(E-E_\pm)^2.
\end{equation}
Here $z_\pm=E_\pm+i0$ denotes the regularised energy parameter associated with
$E_\pm$ at time $t$. The second derivative,
\begin{equation}
  \varphi''(E_\pm,t)=-\frac{1}{\hbar\dot{E}_\mathrm{res}}
  \left[1+\langle\phi_\mathrm{res}|WG_\mathrm{bg}^2(E_\pm)
    W|\phi_\mathrm{res}\rangle\right],
\end{equation}
is always positive in the case of a downward ramp of $E_\mathrm{res}(t)$ and
its inverse plays the role of a variance in the complex Gaussian integrals 
over $E$ and $E'$. Performing the Gaussian integration associated with 
$t_\mathrm{i}$ on the right hand side of Eq.~(\ref{solutiong2Bdownward}) gives
\begin{align}
  \int_E^\infty dE'\, e^{i\varphi(z',t_\mathrm{i})}
  \approx\sqrt{2\pi}\sum_{n=\pm}
  \frac{e^{i[\varphi(z_n^\mathrm{i},t_\mathrm{i})+\pi/4]}}
       {\sqrt{\varphi''(E_n^\mathrm{i},t_\mathrm{i})}}
  \theta(E_n^\mathrm{i}-E).
  \label{phaseintegral}
\end{align}
Here $z_\pm^\mathrm{i}$ and $E_\pm^\mathrm{i}$ refer to the dressed energies
at the initial magnetic field strength. Their asymptotic values in the case of 
a downward sweep of $E_\mathrm{res}$ are given by $E_+^\mathrm{i}\to\infty$ 
and $E_-^\mathrm{i}\to E_0$ in the limit $t_\mathrm{i}\to-\infty$. This 
implies 
$\varphi''(E_+^\mathrm{i},t_\mathrm{i})\to -1/(\hbar \dot{E}_\mathrm{res})$ 
and $\varphi''(E_-^\mathrm{i},t_\mathrm{i})\to \infty$. Consequently, only the 
term associated with $E_+^\mathrm{i}$ significantly contributes to the sum of
Eq.~(\ref{phaseintegral}). The integral over the parameter $E$ on the right 
hand side of Eq.~(\ref{solutiong2Bdownward}) can be evaluated similarly to 
Eq.~(\ref{phaseintegral}). This yields
\begin{equation}
  i\hbar g_\mathrm{2B}^{(+)}(t_\mathrm{f},t_\mathrm{i})=
  e^{-i[\varphi(E_-^\mathrm{f},t_\mathrm{f})-
      \varphi(E_+^\mathrm{i},t_\mathrm{i})]}
  \label{g2BLZasymp}    
\end{equation}
in the limits $t_\mathrm{i}\to-\infty$ and $t_\mathrm{f}\to\infty$. Here 
$E_-^\mathrm{f}$ refers to the dressed energy with the asymptotic behaviour 
$E_-\to-\infty$ at the final time of the magnetic field sweep. The imaginary 
parts of the phases $\varphi(E_-^\mathrm{f},t_\mathrm{f})$ and 
$\varphi(E_+^\mathrm{i},t_\mathrm{i})$ relevant to the transition probability 
of Eq.~(\ref{p0res}) may be obtained from Eqs.~(\ref{definintionphase}) and 
(\ref{principalvalue}). This leads to the Landau-Zener formulae,
\begin{align}
  \label{resultp0res}
  p_{0,\mathrm{res}}&=1-e^{-2\pi\delta_\mathrm{LZ}},\\
  p_{0,0}&=e^{-2\pi\delta_\mathrm{LZ}}.
  \label{resultp00}
\end{align}
Here $p_{0,0}$ denotes the probability for detecting the atom pair in the bare 
state $|\phi_0,\mathrm{bg}\rangle$ at the end of the magnetic field sweep, and 
\begin{equation}
  \delta_\mathrm{LZ}=\frac{|\langle\phi_\mathrm{res}|W|\phi_0\rangle|^2}
	{\hbar|\dot{E}_\mathrm{res}|}
  \label{deltaLZ}
\end{equation}
is the Landau-Zener parameter.

\begin{figure}[htbp]
  \includegraphics[width=\columnwidth,clip]{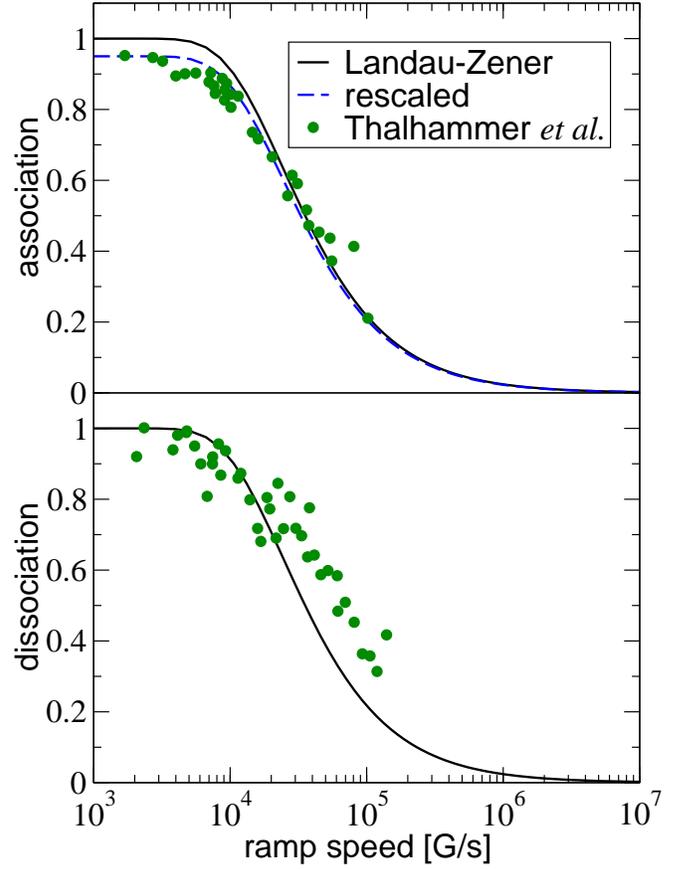}
  \caption{(Colour in online edition)
    Efficiency of Fesh\-bach molecular association (upper panel) and 
    dissociation into the lowest energy band of an optical lattice 
    (lower panel) via linear magnetic field sweeps across the 1007\,G zero 
    energy resonance of $^{87}$Rb versus the ramp speed 
    \cite{ThalhammerPRL06}. The solid curves refer to the Landau-Zener formula 
    of Eq.~(\ref{resultp0res}) and the coefficient for a harmonic oscillator 
    of Eq.~(\ref{deltaLZtrap}) using the measured frequency of 
    $\nu_\mathrm{ho}=39\,$kHz. The dashed curve indicates the same 
    Landau-Zener prediction scaled by a factor of 0.95 which accounts for 
    possible deficits in the ideal diatomic filling, for instance, due to 
    tunnelling between lattice sites \cite{ThalhammerPRL06}.}
  \label{fig:LZ87Rb39kHz}
\end{figure}

For a pair of atoms in an isotropic harmonic trap the matrix element involving 
the inter-channel coupling $W$ can be inferred from Eq.~(\ref{resonancewidth}) 
via the general approximate relation
\begin{equation}
  |\langle\phi_\mathrm{res}|W|\phi_v\rangle|^2\approx 2\pi m\sqrt{mE_v}
  \left|\langle\phi_\mathrm{res}|W|\phi_0^{(+)}\rangle\right|^2
  \frac{\partial E_v}{\partial v}.
  \label{couplingelement}
\end{equation}
Here $E_v$ is the energy of the bare oscillator level of Eq.~(\ref{Evcontact}) 
in the limit of low excitations in the Wigner threshold law regime. Under the 
assumption that the single particle trap length very much exceeds the 
background scattering length, i.e.~$|a_\mathrm{bg}|/a_\mathrm{ho}\ll 1$, the 
density of states is given by 
$\partial E_v/\partial v=2\hbar\omega_\mathrm{ho}$. Consequently, 
Eqs.~(\ref{resonancewidth}) and (\ref{couplingelement}) determine the 
Landau-Zener parameter associated with a trapped atom pair to be 
\cite{JulienneJModOpt04}:
\begin{equation}
  \delta_\mathrm{LZ}^\mathrm{ho}=\frac{\sqrt{6}\hbar}{\pi m a_\mathrm{ho}^3}
  \left|
  \frac{a_\mathrm{bg}\Delta B}{\dot{B}}
  \right|.
  \label{deltaLZtrap}
\end{equation}
Figure~\ref{fig:LZ87Rb39kHz} illustrates the validity of the Landau-Zener 
approach in applications to Fesh\-bach molecular association and dissociation 
in an optical lattice where the sites are filled with two atoms
\cite{ThalhammerPRL06}.

On the basis of Eq.~(\ref{solutiong2Bdownward}), the preceding determination 
of the asymptotic behaviour of transition amplitudes may be extended to an 
arbitrary number of energy states of the entrance channel interaction 
\cite{DemkovJETP68,MacekPRA98,YurovskyJPhysB98,YurovskyJPhysB99}. This shows 
that despite the fact that the Landau-Zener approach neglects the bare excited 
and molecular levels, it gives the exact probability for the loss of atom 
pairs from the $v=0$ mode of a harmonic trap. At the end of an asymptotic 
linear downward sweep of $E_\mathrm{res}$ all population either remains in the 
initial $v=0$ state or is transferred either into energetically lower levels, 
$v<0$, or into the resonance state. A similar statement applies in the case of 
asymptotic upward sweeps which are also exactly described by 
Eqs.~(\ref{resultp0res}), (\ref{resultp00}) and (\ref{deltaLZtrap}). 
Consequently, in the limits $t_\mathrm{i}\to-\infty$ and 
$t_\mathrm{f}\to\infty$ transitions between states occur only in the intuitive 
direction of the sweep. The intermediate dynamics, however, involves all 
levels, i.e., the amplitudes associated with unintuitive transitions interfere 
away only at asymptotically large times. 

\subsection{Magnetic field sweeps in Bose-Einstein condensates}
\label{subsec:BECsweeps}
Several pioneering studies on the properties of zero energy resonances of 
alkali atom pairs \cite{InouyeNature98,StengerPRL99,CornishPRL00} have been 
performed in dilute Bose-Einstein condensates 
\cite{AndersonScience95,DavisPRL95,BradleyPRL95,BradleyPRL97}. Condensation of 
Bose atoms \cite{BoseZPhys24,EinsteinSBPreuss24,EinsteinSBPreuss25} occurs 
when the occupation number $N_\mathrm{c}$ of a particular single particle 
state becomes comparable to the total number of atoms, $N$, in such a way that 
$N_\mathrm{c}/N$ remains finite in the thermodynamic limit. The signatures of 
this phenomenon in dilute alkali gases have been identified, for instance, 
through a specific, narrow momentum distribution of the atoms or a 
characteristic spectrum of collective excitations \cite{DalfovoRMP99}. While 
low collision momenta facilitate the theoretical description of the two-body 
physics of molecular association, the complex many-particle nature of 
Bose-Einstein condensates subject to dynamically resonance enhanced 
interactions becomes particularly significant on long time scales. In 
addition, the Fesh\-bach molecules of some species of alkalis associated with
closed-channel dominated resonances proved to be unstable in the environment 
of a gas \cite{HerbigScience03,DuerrPRL04,XuPRL03}. For this reason, such 
experiments were performed, in part, under conditions of free expansion 
\cite{DuerrPRL04,MarkEurophysLett05,YurovskyPRA04,Yurovskycondmat05}. 
Comparisons between predicted and measured molecule production via magnetic 
field sweeps in Bose-Einstein condensates are, therefore, less conclusive than
in the case of a single atom pair confined to an optical lattice site. 

\subsubsection{Limit of high ramp speeds}
In the limit of high ramp speeds the transfer of condensed atoms into 
Fesh\-bach molecules may be estimated using the two-body Landau-Zener approach 
\cite{MiesPRA00,GoralJPhysB04}. To this end, it is instructive to divide the 
dilute gas into regions of virtually constant density. Their volumes, 
$\mathcal{V}$, can be chosen sufficiently large for the thermodynamic limit to 
be applicable. The state of an arbitrary atom pair of a uniform Bose-Einstein 
condensate in each of these periodic boxes is well described by the lowest 
quasi continuum energy level. Accordingly, the associated Landau-Zener 
parameter can be inferred from Eq.~(\ref{deltaLZ}) using the following 
replacement of the initial wave function:
\begin{equation}
  |\phi_0\rangle\to 
  |\phi_0^{(+)}\rangle\sqrt{(2\pi\hbar)^3/\mathcal{V}}.
  \label{boxbarestate}
\end{equation}
Given that the size of a dilute gas is on the order of several $\mu$m, the 
volume $\mathcal{V}$ is sufficiently large for the transition probability of 
Eq.~(\ref{resultp0res}) to reduce just to its first order approximation, 
$p_{0,\mathrm{res}}\approx 2\pi\delta_\mathrm{LZ}$. A typical order of 
magnitude of $p_{0,\mathrm{res}}$ is $10^{-6}$ \cite{MiesPRA00}. 

While the association of a particular pair of condensed atoms is a rare event, 
the fact that each atom has all the others to interact with grossly enhances 
the efficiency of molecule formation. Just in the limit of high ramp speeds 
the Bose-Einstein condensate may be treated as a reservoir whose total atom 
number is hardly affected by an asymptotic magnetic field sweep across $B_0$. 
This assumption implies that the small fraction of lost atoms is well 
described in terms of the pairwise average of the microscopic transition 
probabilities. The number of pairs in a box with $N$ atoms is $N(N-1)/2$ which 
in the limit $|\dot{B}|\to\infty$ yields the following estimate for the 
condensate depletion \cite{GoralJPhysB04}:
\begin{equation}
  N_\mathrm{loss}=
  2\pi (N-1)(N/\mathcal{V})\frac{4\pi\hbar}{m}
  \left|
  \frac{a_\mathrm{bg}\Delta B}{\dot{B}}
  \right|.
  \label{Nlossfast}
\end{equation}
Here $N/\mathcal{V}$ is the uniform density of the gas in the volume 
$\mathcal{V}$. In accordance with the local density approximation, the total 
number of atoms lost from a Bose-Einstein condensate is given by the spatial 
average over the densities of all boxes. 

We note that the Landau-Zener estimate of Eq.~(\ref{Nlossfast}) is applicable 
to both sweep directions of the Fesh\-bach resonance level. In the case of 
downward sweeps the condensed atoms are partly transferred into diatomic 
Fesh\-bach molecules whose final number, $N_\mathrm{d}^\mathrm{f}$, equals one 
half of the atom loss of Eq.~(\ref{Nlossfast}), i.e., 
\begin{equation}
  N_\mathrm{d}^\mathrm{f}=N_\mathrm{loss}/2. 
  \label{Ndfcondensate}
\end{equation}
Upward sweeps lead to the production of correlated pairs with a comparatively 
high relative velocity depending on the ramp speed. In some experiments using 
nonlinear field variations \cite{DonleyNature01,DonleyNature02} such atom 
pairs were detected as a trapped dilute cloud with an average spatial extent 
much larger than the size of the remnant Bose-Einstein condensate.   
 
\subsubsection{Two-level mean field approach}
A description of atom loss from a Bose-Einstein condensate consistent with 
both Eq.~(\ref{Nlossfast}) and the dynamical depletion of pairs during a 
magnetic field sweep across $B_0$ may be based on Eqs.~(\ref{CIdynamicstrap}) 
and (\ref{CIdynamicsres}). Such an extended two-level configuration 
interaction approach to the probability amplitudes, $C_0(t)$ and 
$C_\mathrm{res}(t)$, associated with the zero energy mode of an atom pair and 
its depletion, respectively, is given by the formulae 
\cite{GoralJPhysB04,JulienneJModOpt04}:
\begin{align}
  \label{CIcond}
  i\hbar\dot{C}_0(t)&=E_0 C_0(t)+(N/\mathcal{V})^{1/2}g_\mathrm{res}^*
  C_0^*(t)C_\mathrm{res}(t),\\
  i\hbar\dot{C}_\mathrm{res}(t)&=E_\mathrm{res}(t)C_\mathrm{res}(t)
  +(N/\mathcal{V})^{1/2}g_\mathrm{res}C_0^2(t).
  \label{CIres}
\end{align}
Here $N$ is the total number of atoms of the homogeneous gas in the volume 
$\mathcal{V}$. The inter-channel coupling is determined by the matrix element 
\begin{equation}
  g_\mathrm{res}=(2\pi\hbar)^{3/2}
  \langle\phi_\mathrm{res}|W|\phi_0^{(+)}\rangle,
  \label{gres}
\end{equation}
in accordance with Eqs.~(\ref{CIdynamicstrap}), (\ref{CIdynamicsres}) and 
(\ref{boxbarestate}). Similarly to the two-body configuration interaction 
approach, Eqs.~(\ref{CIcond}) and (\ref{CIres}) lead to a constant of motion, 
$|C_0(t)|^2+|C_\mathrm{res}(t)|^2=1$. This implies an interpretation of the 
quantities $N|C_0(t)|^2$ and $N|C_\mathrm{res}(t)|^2$ in terms of the numbers 
of atoms associated with the remnant Bose-Einstein condensate and correlated 
pairs, respectively. The nonlinear nature of Eqs.~(\ref{CIcond}) and 
(\ref{CIres}) in terms of the Bose enhancement factor, 
$(N/\mathcal{V})^{1/2}C_0(t)$, ensures consistency with Eq.~(\ref{Nlossfast}) 
and, therefore, accounts for the surrounding gas. 

The long time asymptotic populations may be estimated analytically based on a 
linearised version of Eqs.~(\ref{CIcond}) and (\ref{CIres}) using a static 
Bose enhancement factor, $(N/\mathcal{V})^{1/2}$ \cite{MiesPRA00}. This 
treatment of the inter-channel coupling neglects the time dependence of the 
depletion of pairs of condensed atoms and leads to dynamical equations 
formally equivalent to those of the two-body Landau-Zener approach. In 
accordance with Eq.~(\ref{Nlossfast}), the associated Landau-Zener coefficient 
is given by
\begin{equation}
  \delta_\mathrm{LZ}^\mathrm{BEC}=\left(N/\mathcal{V}\right)
  \frac{4\pi\hbar}{m}
  \left|
  \frac{a_\mathrm{bg}\Delta B}{\dot{B}}
  \right|=N\delta_\mathrm{LZ}.
  \label{deltaLZBEC}
\end{equation}
The asymptotic condensate depletion and its remnant uniform density can be 
inferred from Eqs.~(\ref{resultp0res}) and (\ref{resultp00}), respectively, 
using $\delta_\mathrm{LZ}^\mathrm{BEC}$ instead of $\delta_\mathrm{LZ}$.

An approach similar to Eqs.~(\ref{CIcond}) and (\ref{CIres}) but applicable to 
trapped gases beyond a local density treatment was derived on the basis of a 
mean field approximation 
\cite{Tommasini98,Timmermans98,TimmermansPRL99,DrummondPRL98}. The associated 
many-body model Hamiltonian was originally introduced in the context of 
superconductivity \cite{RanningerPhysicaBplusC,FriedbergPRB89}. This procedure 
leads to the following dynamical equations:
\begin{align}
  \label{TimmermansBEC}
  i\hbar\dot{\Psi}(\mathbf{x},t)=&
  H_\mathrm{GP}\Psi(\mathbf{x},t)+g_\mathrm{res}^*\Psi^*(\mathbf{x},t)
  \Psi_\mathrm{res}(\mathbf{x},t),\\
  i\hbar\dot{\Psi}_\mathrm{res}(\mathbf{R},t)=&
  H_\mathrm{res}(t)\Psi_\mathrm{res}(\mathbf{R},t)
  +g_\mathrm{res}\Psi^2(\mathbf{R},t).
  \label{Timmermansres}
\end{align}
Here the background scattering is included in terms of the usual 
Gross-Pitaevskii mean field Hamiltonian, $H_\mathrm{GP}$, in the contact 
pseudo interaction approximation \cite{GrossNuovoCimento61,PitaevskiiJETP61}. 
Given a spherically symmetric harmonic atom trap, $H_\mathrm{GP}$ therefore
consists of the following contributions:
\begin{equation}
  H_\mathrm{GP}=-\frac{\hbar^2\boldsymbol{\nabla}^2}{2m}+
  \frac{m}{2}\omega_\mathrm{ho}^2|\mathbf{x}|^2+
  \frac{4\pi\hbar^2}{m}a_\mathrm{bg}|\Psi(\mathbf{x},t)|^2.
  \label{HGP}
\end{equation}
Typical frequencies $\nu_\mathrm{ho}=\omega_\mathrm{ho}/(2\pi)$ associated 
with such comparatively weakly confining traps are on the order of 100\,Hz. 
The generalised resonance energy, $H_\mathrm{res}(t)$, contains the centre of 
mass kinetic energy of correlated pairs as well as a magnetic field shift from 
$B_\mathrm{res}$ to the measurable position of the singularity of the 
scattering length. Its explicit expression reads:
\begin{equation}
  H_\mathrm{res}(t)=-\frac{\hbar^2\boldsymbol{\nabla}^2}{4m}+
  m\omega_\mathrm{ho}^2|\mathbf{R}|^2+\mu_\mathrm{res}[B(t)-B_0].
\end{equation}
The mean fields $\Psi(\mathbf{x},t)$ and $\Psi_\mathrm{res}(\mathbf{R},t)$ 
refer to the amplitudes of the densities of atoms in the condensate at the 
position $\mathbf{x}$ and of correlated pairs with the centre of mass 
$\mathbf{R}$, respectively. Similarly to the configuration interaction 
approach, Eqs.~(\ref{TimmermansBEC}) and (\ref{Timmermansres}) give rise to a 
constant of motion consistent with the conservation of the total number of 
atoms: 
\begin{equation}
  \int d\mathbf{x}\,|\Psi(\mathbf{x},t)|^2+
  \int d\mathbf{R}\,|\Psi_\mathrm{res}(\mathbf{R},t)|^2=N.
\end{equation}
Since its first applications in the context of Fesh\-bach resonances in the 
physics of cold gases, the concept of this two-level mean field approach has 
been continually extended to a variety of physical situations 
\cite{YurovskyPRA03,AbeelenPRL99,HollandPRL01,GoralPRL01,DuinePRL03}. 

\begin{figure}[htbp]
  \includegraphics[width=\columnwidth,clip]{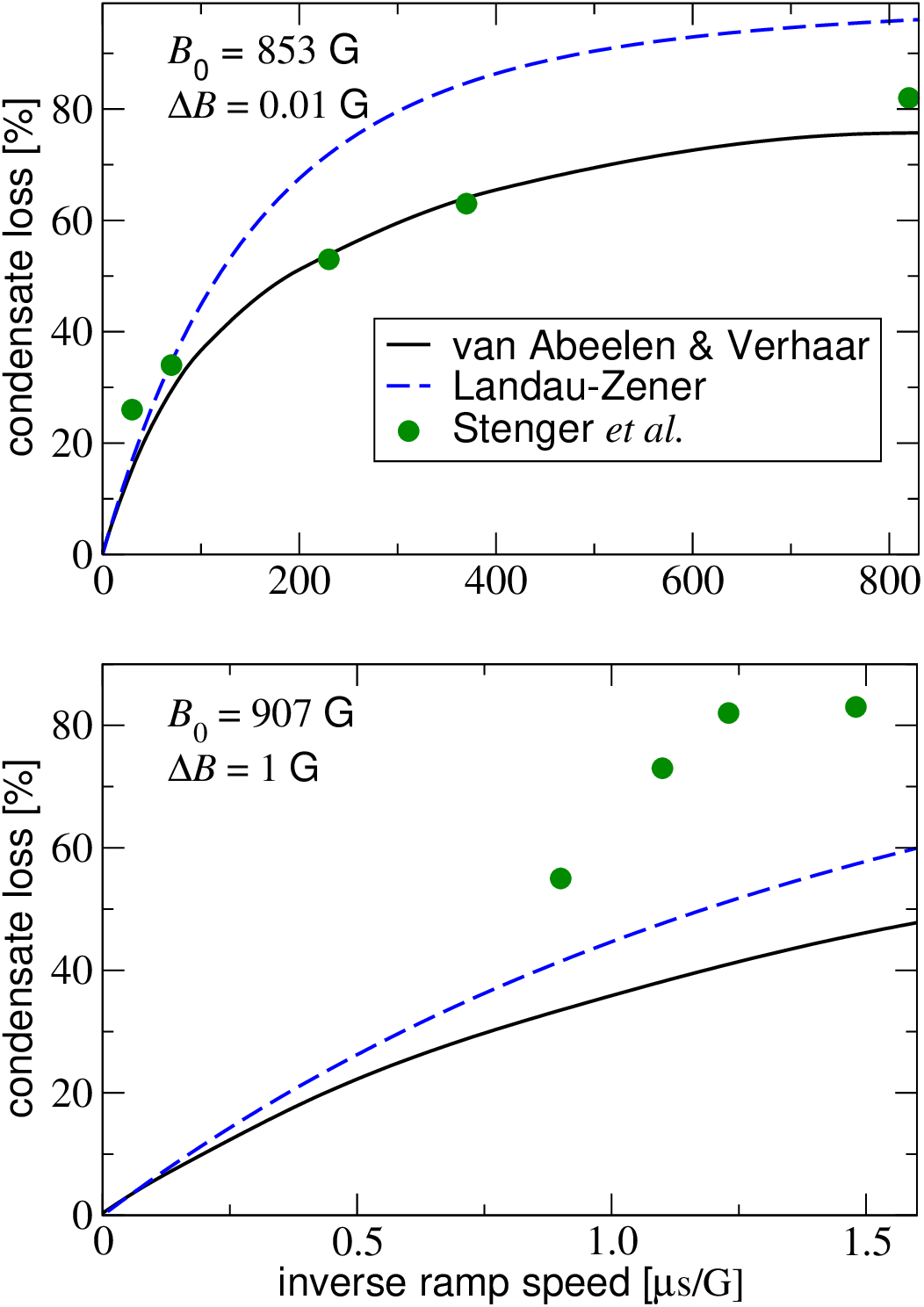}
  \caption{(Colour in online edition)
    Loss of condensate atoms in upward sweeps of the Fesh\-bach resonance 
    level across the 853\,G and 907\,G zero energy resonances of 
    $^{23}$Na versus the inverse ramp speed, $1/|\dot{B}|$ 
    \cite{StengerPRL99}. The experimental data are compared to theoretical 
    predictions using the Landau-Zener \cite{MiesPRA00,GoralJPhysB04} and 
    two-level mean field \cite{AbeelenPRL99} approaches. The theoretical 
    Fesh\-bach resonance parameters employed refer to the values of 
    $a_\mathrm{bg}$ and $\Delta B$ given in Tables~\ref{tab:Vbgparameters} and 
    \ref{tab:couplingparameters}. We note the differences in the ramp speeds 
    of three orders of magnitude between the upper and lower panels which 
    reflect the different widths of the 853\,G and 907\,G zero energy 
    resonances.}
  \label{fig:condensatelossNa}
\end{figure}

Figure~\ref{fig:condensatelossNa} shows that Eqs.~(\ref{TimmermansBEC}) and 
(\ref{Timmermansres}) give reasonable agreement with the loss of condensed 
atoms observed in experiments involving zero energy resonances of $^{23}$Na 
\cite{StengerPRL99}. These measurements refer to asymptotic upward sweeps of 
the Fesh\-bach resonance level across $B_0$ leading to the production of 
unbound correlated pairs of atoms \cite{AbeelenPRL99}. We note that 
Eqs.~(\ref{TimmermansBEC}) and (\ref{Timmermansres}) well account for the 
onset of atom loss over ranges of ramp speeds which differ by three orders of 
magnitude between the upper and lower panels of 
Fig.~\ref{fig:condensatelossNa}. A similarly fair agreement has been reported 
for predictions of several different approaches 
\cite{YurovskyJPhysB03,Mackiephysics02,KoehlerPRA04} on the observed atom loss 
of a Bose-Einstein condensate exposed to magnetic field sweeps across the 
155\,G zero energy resonance of $^{85}$Rb \cite{CornishPRL00}. The dashed 
curves in Fig.~\ref{fig:condensatelossNa} indicate the asymptotic 
Landau-Zener predictions based on the coefficient of Eq.~(\ref{deltaLZBEC}) 
and the local density approximation. Both theoretical approaches recover the 
asymptotic behaviour of the loss of condensed atoms in the case of fast sweeps 
given by Eq.~(\ref{Nlossfast}). Their functional forms differ in the opposite, 
adiabatic limit of low ramp speeds \cite{GoralJPhysB04,IshkhanyanPRA04}, i.e., 
when the parameter $1/|\dot{B}|$ in Fig.~\ref{fig:condensatelossNa} increases. 
We note that both the two-level mean field approach and its associated 
Landau-Zener estimate of asymptotic populations do not distinguish between the 
sweep directions. Consequently, the molecule production in a downward sweep of 
the Fesh\-bach resonance level is treated completely symmetrically to the 
heating of the gas in an upward sweep.

While Eqs.~(\ref{CIcond}) and (\ref{CIres}) refer to a genuinely two-level 
system, the two-level mean field approach accounts for the background 
scattering continuum, in principle, through the parameter $a_\mathrm{bg}$ of 
the mean field Hamiltonian of Eq.~(\ref{HGP}). Such a contact pseudo 
interaction treatment presupposes a separation between the typical time scales 
associated with the evolution of the Bose-Einstein condensate and the diatomic 
collisional duration \cite{ProukakisPRA98,KoehlerPRA02}. Accordingly, the 
two-level mean field approach can be derived formally in terms of the Markov 
limit of microscopic many-body theories of dilute gases \cite{GoralJPhysB04}
outlined in Subsection~\ref{subsec:BECdynamics}. The assumption of a 
separation of time scales between two- and many-body evolutions is violated 
during a magnetic field sweep across a singularity of the scattering length. 
This implies that Eqs.~(\ref{TimmermansBEC}) and (\ref{Timmermansres}) can 
describe the dynamics, at most, in the asymptotic regime where the dilute gas 
parameter, $[N|a(B)|^3/\mathcal{V}]^{1/2}$, is small compared to unity. 
Similarly to the limitations of the two-body Landau-Zener model, the 
intermediate evolution of a Bose-Einstein condensate is influenced by 
phenomena beyond the range of validity of the mean field approximation 
\cite{HollandPRL01,GoralJPhysB04,KoehlerPRA04}. While most of the theoretical 
approaches agree in their predictions on the condensate loss in the fast sweep 
limit of Eq.~(\ref{Nlossfast}), the saturation of molecule production is a 
matter of ongoing research \cite{NaidonPRA03,NaidonPRA06}. 

\subsection{Molecule production in cold Bose and Fermi gases}
\label{subsec:BoseFermi}
Cold Bose and two spin component Fermi gases are subject to a considerable 
momentum spread which generally tends to reduce the efficiency of molecule 
production via linear magnetic field sweeps across a zero energy resonance. 
Its significance is particularly obvious in the case of slow asymptotic 
sweeps, given the spectrum of dressed energies of an atom pair illustrated in 
Fig.~\ref{fig:87Rb39kHz}. Only the $v=0$ mode adiabatically correlates with 
the Fesh\-bach molecular level when the resonance energy is decreased, while 
all the excited states undergo cooling transitions. From this too simplistic 
viewpoint, for instance, two spin components of a dilute vapour of Fermi atoms 
distributed according to the Pauli exclusion principle would produce just a 
single molecule in the adiabatic limit of the ramp speed. Contrary to the case 
of Bose-Einstein condensates, binary physics alone is therefore not even 
sufficient to qualitatively explain the observed substantial molecule 
production in such gases \cite{RegalNature03,StreckerPRL03}. 

\subsubsection{Transitions from continuum to bound states} 
A quantitative analysis of the problems associated with the theoretical 
description of molecule production in the presence of momentum spread may be 
based on the exact treatment of linear magnetic field sweeps of 
Subsection~\ref{subsec:linearsweeps}. Accordingly, the probability for 
transitions from an initial dressed continuum level of a pair of 
distinguishable atoms in a periodic box of volume $\mathcal{V}$ to the 
Fesh\-bach molecular state at the final magnetic field strength is given by
\begin{equation}
  p_\mathrm{ass}(\mathbf{k})=\frac{(2\pi\hbar)^3}{\mathcal{V}}
  \left|
  \langle\phi_\mathrm{b}^\mathrm{f}|
  U_\mathrm{2B}(t_\mathrm{f},t_\mathrm{i})
  |\phi_{\hbar\mathbf{k}}^\mathrm{i}\rangle
  \right|^2.
  \label{pass}
\end{equation}
Here the arguments of the two-body time evolution operator, $t_\mathrm{i}$ and 
$t_\mathrm{f}$, refer to the initial and final times of the sweep, 
respectively, and $\hbar\mathbf{k}$ denotes the initial relative momentum of 
the atoms. Similarly to the derivation of Eq.~(\ref{g2BLZasymp}), the 
asymptotic transition probability including the background scattering 
continuum can be determined analytically using the stationary phase approach 
in Eq.~(\ref{pass}) in the limits $t_\mathrm{i}\to -\infty$ and 
$t_\mathrm{f}\to\infty$. This yields:
\begin{equation}
  p_\mathrm{ass}(\mathbf{k})=\frac{(2\pi\hbar)^3}{\mathcal{V}}
  \frac{2\pi}{\hbar|\dot{E}_\mathrm{res}|}
  \left|
  \langle\phi_\mathrm{res}|W|\phi_{\hbar\mathbf{k}}^{(+)}\rangle 
  \right|^2
  e^{-2\mathrm{Im}\varphi(z_\mathrm{i},t_\mathrm{i})}.
  \label{passasymp}
\end{equation}
Here $z_\mathrm{i}=\hbar^2k^2/m+i0$ denotes the regularised energy argument of 
the initial phase. In accordance with Eqs.~(\ref{definintionphase}), 
(\ref{grefinverse}) and (\ref{principalvalue}), the exponent of 
Eq.~(\ref{passasymp}) is determined by the formula 
\begin{equation}
  \mathrm{Im}\,\varphi(z_\mathrm{i},t_\mathrm{i})=
  \frac{\pi\hbar^2}{|\dot{E}_\mathrm{res}|}
  \int d\mathbf{k}'\,\theta(k-k')
  \left|\langle\phi_\mathrm{res}|W|\phi_{\hbar\mathbf{k}'}^{(+)}\rangle
  \right|^2.
  \label{initialexponent}
\end{equation}
Here the step function of the wave numbers, $\theta(k-k')$, indicates that 
transitions occur just in the intuitive, downward direction of an asymptotic 
sweep across $B_0$ \cite{DemkovJETP68}. According to 
Eq.~(\ref{initialexponent}), the exponential damping of Eq.~(\ref{passasymp}) 
increases in the limit $|\dot{E}_\mathrm{res}|\to0$. This confirms the 
intuitive picture suggested by Fig.~\ref{fig:87Rb39kHz} that adiabatic sweeps 
in a diatomic system with a continuum of modes eventually lead to negligible 
molecule production. While the momentum dependence of the exponent of 
Eq.~(\ref{initialexponent}) recovers experimental dissociation spectra 
\cite{MukaiyamaPRL03}, in the context of molecular association in cold gases, 
Eq.~(\ref{passasymp}) gives rise to exact predictions just in the fast sweep 
limit.

\subsubsection{Fast sweep limit of molecule production}
The onset of molecule production in asymptotic magnetic field sweeps across 
zero energy resonances is sensitive to the statistics associated with 
identical atoms. Subsection~\ref{subsec:howtocount} provides a strict 
approach to the determination of dimer populations on the basis of the 
two-particle correlation function of the gas. Similarly to the prediction of 
the loss of condensed atoms of Eq.~(\ref{Nlossfast}), however, low depletions 
can be inferred intuitively from the fast sweep limit of 
Eq.~(\ref{passasymp}), treating the gas as a reservoir of atom pairs. In this 
limit, $|\dot{B}|\to\infty$, the damping term of Eq.~(\ref{initialexponent}) 
describing cooling transitions into continuum levels below the initial energy 
$\hbar^2k^2/m$ vanishes. In the context of cold collisions in the Wigner 
threshold law regime the matrix element involving the inter-channel coupling 
in Eq.~(\ref{passasymp}) may be evaluated at $\mathbf{k}=0$. Consequently, the 
transition probability of Eq.~(\ref{passasymp}) becomes momentum independent 
and recovers the result of the Landau-Zener approach, 
$p_\mathrm{ass}=2\pi\delta_\mathrm{LZ}$.

For the purpose of studies involving $s$-wave collisions, Fermi gases are 
usually prepared as incoherent mixtures of two different Zeeman states with
occupation numbers $N_1$ and $N_2$. Accordingly, $N=N_1+N_2$ is the total 
number of atoms. Each one of the $N_1$ atoms of the first component has $N_2$ 
atoms of the second component to interact with via $s$-wave collisions. 
Classical probability theory and the incoherent nature of the initial state 
therefore lead to the following estimate for the number of diatomic Fesh\-bach 
molecules produced in the fast sweep limit \cite{ChwedenczukPRL04}:
\begin{equation}
  N_\mathrm{d}^\mathrm{f}=2\pi N_1 N_2\delta_\mathrm{LZ}.
  \label{fastsweepFermi}
\end{equation}
We note that the two-body Landau-Zener coefficient, given explicitly by the 
right hand side of Eq.~(\ref{deltaLZBEC}), is inversely proportional to the 
volume $\mathcal{V}$ of the periodic box. The fraction of molecules, 
$N_\mathrm{d}^\mathrm{f}/N$, is therefore proportional to the density, 
$N/\mathcal{V}$, which allows for an extension of Eq.~(\ref{fastsweepFermi}) 
to inhomogeneous gases via the local density approximation. 

\begin{figure}[htbp]
  \includegraphics[width=\columnwidth,clip]{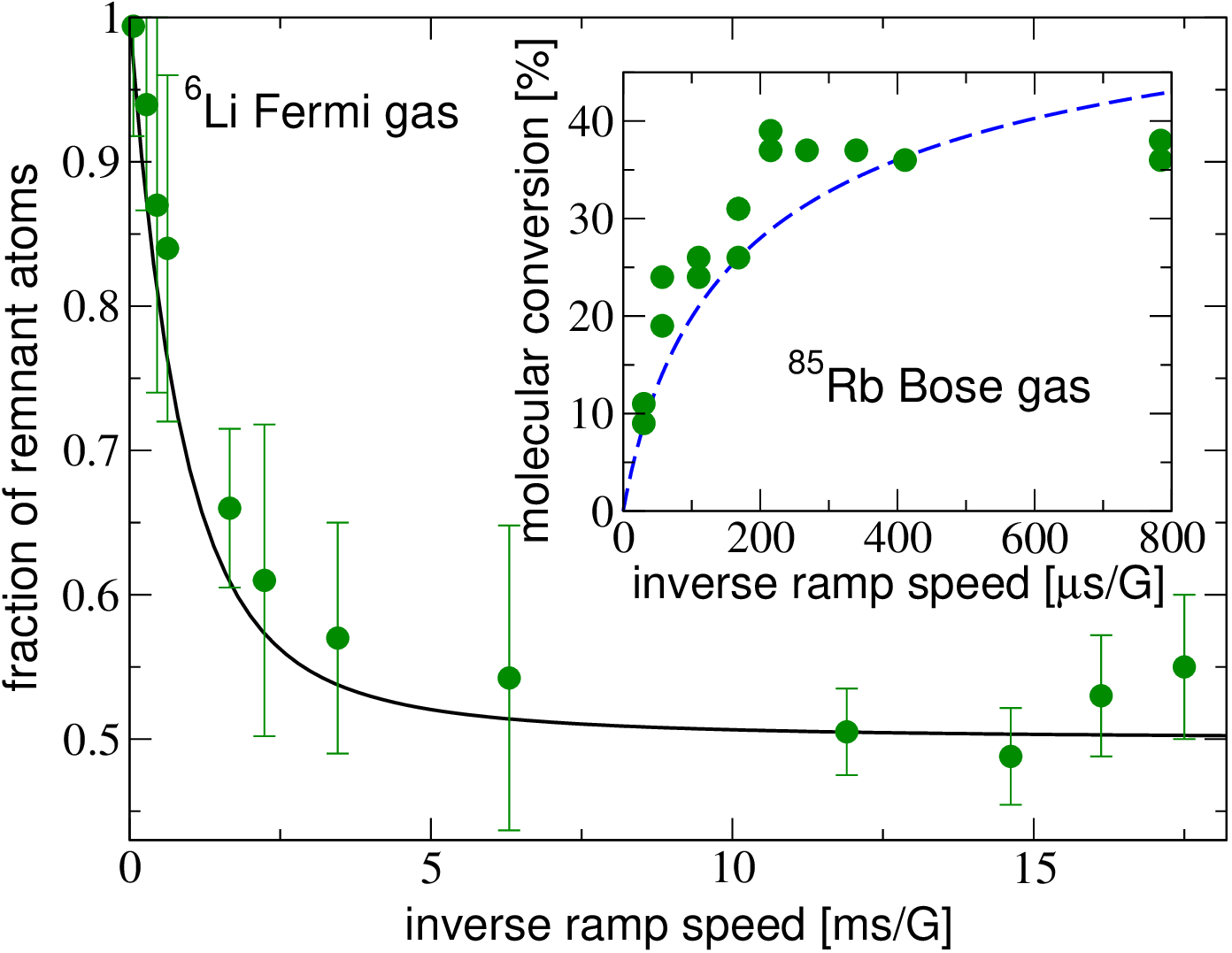}
  \caption{(Colour in online edition)
    The fraction of remnant atoms, $1-2N_\mathrm{d}^\mathrm{f}/N$, at 
    the end of asymptotic magnetic field sweeps across the closed-channel 
    dominated 543\,G zero energy resonance of $^6$Li versus the inverse ramp 
    speed \cite{StreckerPRL03}. The solid curve refers to a Landau-Zener 
    estimate for the onset of molecule production in the fast sweep limit 
    \cite{ChwedenczukPRL04} using the local density approximation. The inset 
    shows the molecular conversion, $2N_\mathrm{d}^\mathrm{f}/N$, observed in 
    a thermal Bose gas with a mean density of 
    $1.3\times 10^{11}\,$atoms/cm$^3$ and a temperature of 40.6\,nK using 
    magnetic field sweeps across the 155\,G zero energy resonance of $^{85}$Rb 
    \cite{HodbyPRL05}. For comparison, the dashed curve indicates the pairwise 
    ensemble average over the transition probabilities, 
    $2p_\mathrm{ass}(\mathbf{k})$, with respect to the Maxwell distribution of 
    relative velocities. This semi-classical estimate for the onset of 
    molecule production in Bose gases is based on Eq.~(\ref{passasymp}) and 
    therefore consistent with the limit of Eq.~(\ref{fastsweepBose}). Both the 
    solid and dashed curves are associated with the parameters $a_\mathrm{bg}$ 
    and $\Delta B$ of Tables~\ref{tab:Vbgparameters} and 
    \ref{tab:couplingparameters}, respectively.}
  \label{fig:6Liloss}
\end{figure}

Given a balanced mixture of Zeeman states, i.e.~$N_1=N_2=N/2$,
Eqs.~(\ref{fastsweepFermi}) and (\ref{Ndfcondensate}) show that the onset 
of molecule production in a Fermi gas is half as large as in a Bose-Einstein 
condensate of identical density and resonance parameters. In the limit of zero 
temperature such a dilute, initially weakly interacting vapour of atoms may be 
approximately described by a pair of filled Fermi seas. As fast magnetic field 
sweeps imply small atomic depletions, it may be argued intuitively that 
two-body cooling transitions leading into occupied modes below the Fermi 
energy are suppressed by the Pauli exclusion principle. This suggests a 
description of the two spin component Fermi sea in terms of a single level 
over a significant range of ramp speeds. The associated Landau-Zener approach 
\cite{ChwedenczukPRL04} consistent with the linear limit of 
Eq.~(\ref{fastsweepFermi}) is illustrated in Fig.~\ref{fig:6Liloss} in 
comparison to experiments on asymptotic magnetic field sweeps across the 
543\,G zero energy resonance of $^6$Li \cite{StreckerPRL03}. Accordingly, a 
suppression of cooling transitions gives an intuitive explanation for the 
onset of a substantial molecule production in two spin component Fermi gases. 

In thermal Bose gases each one of the $N$ constituents can interact with all 
the other $N-1$ atoms to form a dimer Fesh\-bach molecule. As opposed to the 
zero energy mode of condensed pairs, diatomic wave functions in the presence 
of momentum spread need to be explicitly symmetrised which enhances the 
association probability of Eq.~(\ref{pass}) in the thermal average by a factor 
of two \cite{StoofPRA89}. As the number of interacting pairs is 
$N(N-1)/2\approx N^2/2$, the fast sweep limit of the number of Fesh\-bach
molecules produced in a thermal Bose gas is given by the following formula:
\begin{equation}
  N_\mathrm{d}^\mathrm{f}=2\pi N^2\delta_\mathrm{LZ}.
  \label{fastsweepBose}
\end{equation}
This estimate is illustrated in the inset of Fig.~\ref{fig:6Liloss} in 
comparison to experiments associated with thermal clouds of $^{85}$Rb 
\cite{HodbyPRL05}. We note that a similar enhancement of dimer formation in 
thermal Bose gases as compared to condensates was observed in the context of 
inelastic three-body recombination and reflects correlation properties 
\cite{BurtPRL97,KaganJETP85}. 

According to Eqs.~(\ref{fastsweepFermi}), (\ref{Ndfcondensate}) and 
(\ref{fastsweepBose}), the onsets of molecule production increase by factors 
of two between balanced two spin component Fermi gases, Bose-Einstein
condensates, and thermal Bose gases of identical densities and resonance 
parameters. The validity of these statistical estimates based on two-body 
physics depends on the significance of multiple collisions of each atom during 
a magnetic field sweep. This implies that the range of ramp speeds described 
by the fast sweep limits depends on the density of the gas in addition to the 
specific nature of the Fesh\-bach molecules associated with entrance- and  
closed-channel dominated zero energy resonances. Predictions accessing wider 
ranges of ramp speeds require genuinely many-body approaches 
\cite{JavanainenPRL04,PazyPRL04,PazyPRL05,WilliamsJPhysB04,WilliamsNJPhys04}.

\subsubsection{Saturation of molecule production}
Magnetic field sweeps sufficiently slow to convert up to 88\,\% of the atoms 
into Fesh\-bach molecules were employed to produce Bose-Einstein condensates 
of dimers from balanced two spin component mixtures of $^{40}$K Fermi atoms 
\cite{GreinerNature03}. Figure~\ref{fig:Greinercondensate} shows a typical 
density profile of such a final molecular cloud (right image) in comparison to 
a thermal distribution (left image). Motivated by the theory of ideal Bose 
gases, the conditions for condensation in these experiments are likely to 
depend on the phase space density, 
$n_\mathrm{a}^\mathrm{i}\lambda_\mathrm{th}^3$, associated with the initial 
dilute atomic gas of density $n_\mathrm{a}^\mathrm{i}$ and temperature $T$. 
Here $\lambda_\mathrm{th}=[2\pi\hbar^2/(mk_\mathrm{B}T)]^{1/2}$ is the thermal 
de Broglie wavelength. Given the experimental densities on the order of 
$10^{13}$ atoms per cm$^3$ and the 202\,G zero energy resonance parameters, 
magnetic field sweeps with low ramp speeds of typically 160\,G/s 
\cite{GreinerNature03} may be considered to be adiabatic. Such processes 
smoothly alter diatomic wave functions but are not necessarily expected to 
change the occupation of states in phase space. It is therefore plausible that 
not only the condition for molecular condensation but also the saturated
production efficiency, $2N_\mathrm{d}^\mathrm{f}/N$, is determined just by the 
phase space density rather than the number of atoms or their temperature 
individually. 

\begin{figure}[htbp]
  \includegraphics[width=\columnwidth,clip]{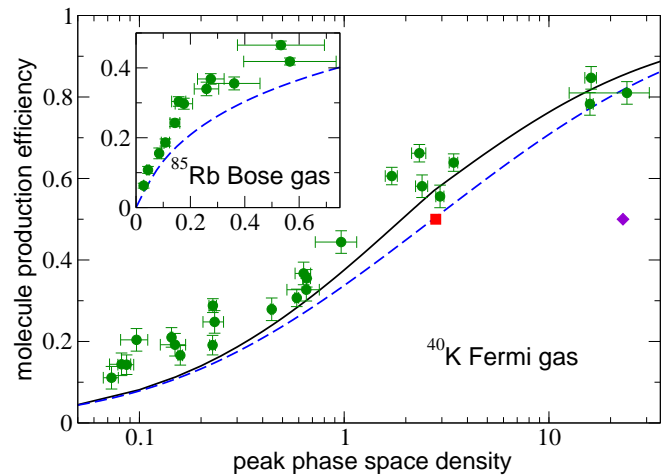}
  \caption{(Colour in online edition)
    The molecule production efficiency, $2N_\mathrm{d}^\mathrm{f}/N$, 
    versus the peak phase space density of cold Bose as well as two spin 
    component Fermi gases. The circles refer to dimer populations measured at 
    the end of adiabatic magnetic field sweeps across the zero energy 
    resonances of $^{85}$Rb at 155\,G (inset) and of $^{40}$K at 202\,G 
    \cite{HodbyPRL05}. The curves indicate predictions 
    \cite{WilliamsNJPhys06} based on coupled Boltzmann equations including 
    quantum statistical effects (solid curve) as well as their associated 
    classical gas limit (dashed curves). For comparison, the square and the 
    diamond refer to the molecular production efficiencies observed for the 
    lowest ramp speeds in earlier sweep experiments using zero energy 
    resonances of $^{40}$K at 224\,G \cite{RegalNature03} and of $^6$Li 
    at 543\,G \cite{StreckerPRL03}, respectively.}
    \label{fig:Jamieplot}
\end{figure}

Subsequent systematic studies of adiabatic magnetic field sweeps in both 
two spin component Fermi gases of $^{40}$K and thermal Bose gases of $^{85}$Rb
have supported this view \cite{HodbyPRL05}. The measured molecule production 
efficiencies of Fig.~\ref{fig:Jamieplot} were analysed in terms of a 
stochastic model relying upon the assumption that the probability for two 
atoms to form a dimer depends solely on their proximity in phase space. Based
on the experimental data, the associated proximity conditions turned out to be 
virtually identical for both atomic species, $^{40}$K and $^{85}$Rb. These 
observations can be understood from first principles using coupled Boltzmann 
equations for the Wigner functions associated with density matrices of
separated atoms and pairs in the resonance state configuration
\cite{WilliamsNJPhys06}. Figure~\ref{fig:Jamieplot} illustrates the accuracy 
of this approach including quantum statistical effects 
\cite{BlochZPhys28,UehlingPR33} as well as its classical gas limit.

Similar experiments using adiabatic magnetic field sweeps across the 834\,G 
zero energy resonance in cold two spin component Fermi gases of $^6$Li yielded 
up to 80\% Fesh\-bach molecular conversion \cite{CubizollesPRL03}. Such 
observations were analysed in terms of theoretical approaches based on the 
assumption of thermal equilibrium throughout the sweep 
\cite{KokkelmansPRA04,ChinPRAthermal04}. In the magnetic field range of 
positive scattering lengths about $B_0$ the Fesh\-bach molecular level gives 
rise to a local minimum of the many-particle action \cite{SzymanskaPRA05} 
which is associated with a meta-stable state. As this state is energetically 
favourable to a gas of separated atoms, it may be argued intuitively that 
production of dimer molecules should occur eventually also at a stationary 
magnetic field strength when the system equilibrates. This principle has been 
employed to convert balanced incoherent two spin component mixtures of $^6$Li 
atoms into dilute vapours of Fesh\-bach molecules \cite{JochimPRL03}. While 
energy conservation inevitably leads to components of comparatively hot atoms 
and dimers, the stability of universal Fesh\-bach molecules discussed in 
Subsection~\ref{subsec:universality} allows such gases to be cooled by 
evaporation. The associated increase in phase space density provides an 
alternative route to the Bose-Einstein condensation of $^6$Li$_2$
\cite{ZwierleinPRL03,JochimScience03} besides the approach of adiabatic 
magnetic field sweeps.

\subsection{Dissociation of Fesh\-bach molecules}
\label{subsec:dissociation}
Asymptotic upward sweeps of the resonance energy lead to the dissociation of 
Fesh\-bach molecules which often serves as a precursor to their detection. To 
this end, a molecular component is usually spatially separated from the 
environment of a remnant atomic gas using, for instance, the Stern-Gerlach 
technique illustrated in Section~\ref{subsec:parameters}. After this 
separation the dissociation allows the fragments to be detected conventionally 
using probe lasers tuned to resonance with an atomic spectral line. The energy 
provided by the time varying homogeneous magnetic field during the sweep is 
transferred to the relative motion of the constituents of a Fesh\-bach 
molecule. Such correlated atom pairs with a relative velocity depending on the 
ramp speed were detected in several experiments 
\cite{MukaiyamaPRL03,DuerrPRA04,VolzPRA05}. Their spectrum of kinetic energies 
of the relative motion is given by the following expression:
\begin{equation}
  n_\mathrm{diss}(\hbar^2 k^2/m)=\frac{m\hbar k}{2}\int d\Omega\,
  \left|\langle\phi_{\hbar\mathbf{k}}^\mathrm{f}|
  U_\mathrm{2B}(t_\mathrm{f},t_\mathrm{i})
  |\phi_\mathrm{b}^\mathrm{i}\rangle\right|^2.
  \label{ndiss}
\end{equation}
Here $d\Omega$ denotes the angular component of $d\mathbf{k}$ describing the 
direction of the momentum $\hbar\mathbf{k}$, while 
$|\phi_\mathrm{b}^\mathrm{i}\rangle$ and 
$|\phi_{\hbar\mathbf{k}}^\mathrm{f}\rangle$ are bound and dressed continuum 
states associated with the initial and final magnetic field strengths, 
respectively. 

Similarly to the derivation of Eqs.~(\ref{passasymp}) and 
(\ref{initialexponent}), the energy spectrum of Eq.~(\ref{ndiss}) can be 
determined analytically in the asymptotic limits $t_\mathrm{i}\to -\infty$ and 
$t_\mathrm{f}\to\infty$. As dissociation and association are related to each 
other just by time reversal, their transition probability densities are 
identical. Typical energies $\hbar^2 k^2/m$ of atomic fragments are on the 
order of $\mu$K in units of the Boltzmann constant which is usually inside 
the Wigner threshold law regime, i.e.~$k|a_\mathrm{bg}|\ll 1$. This implies 
that the matrix elements involving the inter-channel coupling in 
Eqs.~(\ref{passasymp}) and (\ref{initialexponent}) can be evaluated at 
$\mathbf{k}=0$ and $\mathbf{k}'=0$, respectively. Consequently, the asymptotic
dissociation spectrum of Eq.~(\ref{ndiss}) is well approximated by the 
following formula \cite{MukaiyamaPRL03,GoralJPhysB04}:
\begin{equation}
  n_\mathrm{diss}(E)=-\frac{\partial}{\partial E}
  \exp\left(
  -\frac{4}{3}\sqrt{\frac{mE}{\hbar^2}}
    \frac{|a_\mathrm{bg}\Delta B|E}{\hbar|\dot{B}|}
  \right).
  \label{ndissasymp}
\end{equation}
We note that the integral of Eq.~(\ref{ndissasymp}) over all kinetic energies
$E$ of the relative motion gives unity. This implies that Fesh\-bach molecules 
are dissociated with certainty in an asymptotic upward sweep of the resonance 
energy, provided that transitions to the highest excited entrance channel 
vibrational level are negligible. 

\begin{figure}[htbp]
  \includegraphics[width=\columnwidth,clip]{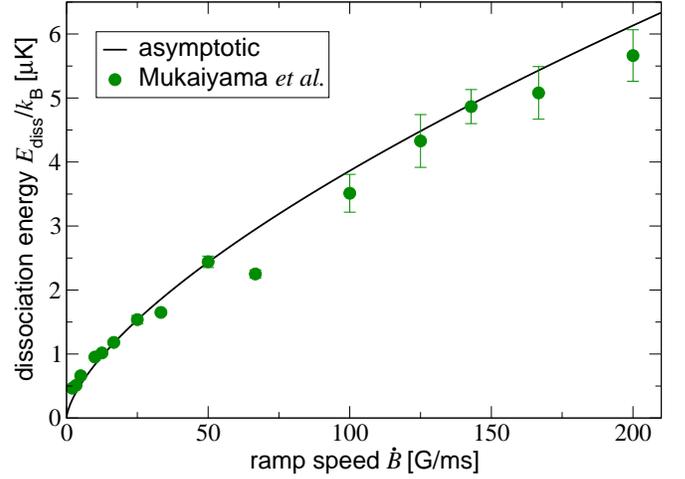}
  \caption{(Colour in online edition)
    The mean energies of the atomic fragments of $^{23}$Na$_2$ 
    Fesh\-bach molecules dissociated by a linear upward sweep of the $907\,$G 
    resonance level versus the ramp speed. Circles indicate 
    experimental data \cite{MukaiyamaPRL03}, while the solid curve refers to 
    the predictions of Eq.~(\ref{Ediss}) using the parameters $a_\mathrm{bg}$ 
    and $\Delta B$ of Tables~\ref{tab:Vbgparameters} and 
    \ref{tab:couplingparameters}, respectively.}
  \label{fig:Na2dissociation}
\end{figure}

According to Eq.~(\ref{ndissasymp}), the width of the dissociation spectrum 
increases with increasing ramp speeds. The associated single particle kinetic 
energies are usually inferred from the velocities of the fragments which 
constitute a radially expanding cloud of atoms. Their average, 
$E_\mathrm{diss}$, amounts to one half of the mean energy of the relative 
motion of all correlated pairs given in terms of Eq.~(\ref{ndiss}) by the 
expression
\begin{equation}
  E_\mathrm{diss}=\frac{1}{2}\int_0^\infty dE\,E\,n_\mathrm{diss}(E).
\end{equation}
Accordingly, the Wigner threshold law approximation of Eq.~(\ref{ndissasymp})
determines $E_\mathrm{diss}$ to be
\begin{equation}
  E_\mathrm{diss}=\frac{1}{3}
  \left(
  \frac{3}{4}\sqrt{\frac{\hbar^2}{ma_\mathrm{bg}^2}}
  \frac{\hbar|\dot{B}|}{|\Delta B|}
  \right)^{2/3}
  \Gamma(2/3).
  \label{Ediss}
\end{equation}
Figure~\ref{fig:Na2dissociation} illustrates the accuracy of the predictions 
of Eq.~(\ref{Ediss}) on average dissociation energies detected at the end of 
asymptotic magnetic field sweeps across the 907\,G zero energy resonance of 
sodium \cite{MukaiyamaPRL03}. While the solid curve refers to  independently 
determined parameters $a_\mathrm{bg}$ and $\Delta B$ \cite{MiesPRA00}, the 
good agreement with the measurements has motivated experiments using molecular 
dissociation to characterise resonances \cite{DuerrPRA04,BrouardPRA05}.

We note that the asymptotic energy spectrum of Eq.~(\ref{ndissasymp}) depends 
just on the product $a_\mathrm{bg}\Delta B$ which, according to 
Eq.~(\ref{aofB}), determines near resonant scattering lengths in the universal 
regime of magnetic field strengths. A similar statement applies to all 
Landau-Zener coefficients as well as the long time asymptotic occupations 
predicted by the two-level mean field approach \cite{GoralJPhysB04}. In the 
limits $t_\mathrm{i}\to -\infty$ and $t_\mathrm{f}\to\infty$ the number of 
correlated atom pairs produced by linear magnetic field sweeps is, therefore, 
insensitive to the parameters $\mu_\mathrm{res}$ and $B_0-B_\mathrm{res}$ 
which directly refer to properties of the resonance level. This suggests that 
such populations can be inferred from any one of the single channel approaches 
outlined in Subsections~\ref{subsec:classification} and 
\ref{subsec:parameters} which all give the correct magnetic field dependence 
of the scattering length \cite{GoralJPhysB04,KoehlerPRA04}. All derivations of 
the exact diatomic dynamics associated with linear magnetic field sweeps of 
Subsection~\ref{subsec:linearsweeps} may be performed, for instance, using a 
single channel Hamiltonian with a time dependent separable potential. Despite 
the fact that such a two-body interaction 
is not suited to describe closed-channel dominated Fesh\-bach molecular 
states, it exactly recovers the asymptotic dissociation spectrum of 
Eq.~(\ref{ndissasymp}).

In accordance with Eq.~(\ref{phaseintegral}), the range of validity of the 
limits $t_\mathrm{i}\to -\infty$ and $t_\mathrm{f}\to\infty$ in applications 
to linear sweeps is determined by the accuracy of the stationary phase 
approach. Given a fixed ramp speed, this approximation is violated when the 
sweep starts or terminates too close to $B_0$ for the variation of the phases 
in Eq.~(\ref{solutiong2Bdownward}) to produce sufficiently many oscillations 
of the complex exponential functions. We note that according to 
Eq.~(\ref{phasederivative}) the phase gradient with respect to the energy 
parameter is proportional to $1/\dot{B}$. This implies that slow sweeps tend 
to require smaller ranges of magnetic field strengths about $B_0$ for the 
asymptotic limits to be applicable. Conversely, the faster a sweep the more it 
resolves details of both the intermediate dynamics and the initial and final 
states. 

In the idealised limit of a jump of the magnetic field strength across $B_0$ 
the Fesh\-bach molecular dissociation spectrum of Eq.~(\ref{ndiss}) just probes
the overlap of the dressed initial bound and final continuum states. Such a 
scenario has been realised using a narrow closed-channel dominated zero energy 
resonance of $^{87}$Rb \cite{DuerrPRA04}. In these experiments the initial 
magnetic field strength was chosen in such a way that the Fesh\-bach molecule 
was virtually identical to the resonance state, 
$|\phi_\mathrm{res},\mathrm{cl}\rangle$. According to Eq.~(\ref{phipcl}), this 
implies that the dissociation spectrum is determined simply by the modulus 
squared of the amplitude of Eq.~(\ref{amplitude}), 
i.e.~$n_\mathrm{diss}(p^2/m)2/(mp)=4\pi |A(B_\mathrm{f},p^2/m)|^2$. Its 
resonance denominator gives rise to a sharp maximum at 
$E_\mathrm{res}(B_\mathrm{f})$ \cite{DuerrPRA04,HaquePRA05} slightly shifted 
by the real part of Eq.~(\ref{Wignerlaw}), while the associated imaginary part 
yields the spectral width. Such dissociation jumps across $B_0$ using a 
resonance state of $d$-wave symmetry can populate several outgoing partial 
waves of the atomic fragments. This approach was employed, for instance, for 
the spectroscopy of a $d$-wave shape resonance \cite{VolzPRA05,DuerrPRA05}.

\section{Atom-molecule coherence}
\label{sec:AMcoherence}
Besides the production of Fesh\-bach molecules via asymptotic linear sweeps 
across $B_0$, several approaches to date rely upon ramp sequences  
\cite{MarkEurophysLett05,Yurovskycondmat05} or resonant oscillating magnetic 
fields \cite{ThompsonRFPRL05}. Such experimental techniques are designed to 
improve the efficiency of dimer formation in Bose gases mainly by avoiding the 
region of large scattering lengths and accordingly strong inter-atomic 
interactions. The associated nonlinear magnetic field variations lead to 
dependences of the molecular population on properties of zero energy 
resonances beyond those included in the Landau-Zener parameter. 

\subsection{Ramsey interferometry with atoms and molecules}
\label{subsec:Ramsey}
Pulses starting and ending on the positive scattering length side of a zero 
energy resonance \cite{ClaussenPRL02} are of particular interest in this 
context. Their repeated application to a $^{85}$Rb Bose-Einstein condensate 
gives rise to the coherent oscillations between the final components of 
separated atoms and Fesh\-bach molecules of Fig.~\ref{fig:DonleyFringe}
\cite{DonleyNature02,ClaussenPRA03}.

\subsubsection{Magnetic field pulse sequence}
\begin{figure}[htbp]
  \includegraphics[width=\columnwidth,clip]{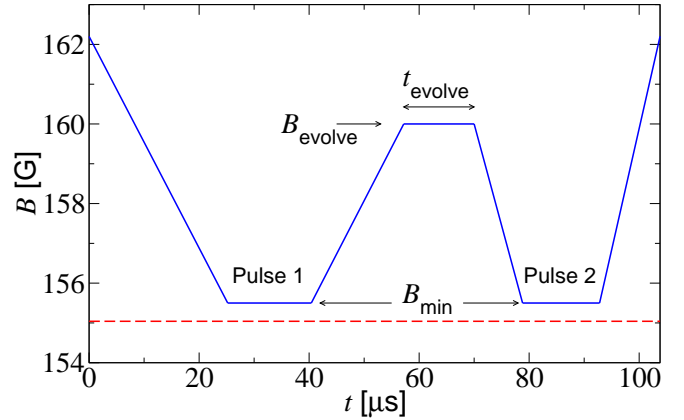}
  \caption{(Colour in online edition)
    Scheme of a typical magnetic field pulse sequence (solid lines) 
    employed in experiments producing coherent oscillations between final 
    atomic and molecular components in a $^{85}$Rb Bose-Einstein condensate 
    \cite{DonleyNature02}. A pair of pulses with a minimum field strength, 
    $B_\mathrm{min}=155.5\,$G, is separated by an evolution period of variable 
    duration, $t_\mathrm{evolve}$, and field strength, $B_\mathrm{evolve}$. 
    The zero energy resonance position, $B_0\approx 155\,$G, is indicated by 
    the dashed line.}
  \label{fig:BpulseClaussen}
\end{figure}

A typical experimental pulse sequence \cite{DonleyNature02} is illustrated in 
Fig.~\ref{fig:BpulseClaussen}. All the associated magnetic field strengths, 
$B$, are on the high field side of the 155\,G zero energy resonance of 
$^{85}$Rb where the scattering length is positive. Consequently, the 
inter-atomic potential supports a loosely bound dimer state whose energy as a 
function of $B$ is shown in Fig.~\ref{fig:85RbEbofB}. The conversion of pairs 
of Bose-Einstein condensed atoms to Fesh\-bach molecules in these experiments 
crucially relies upon intermediate order of magnitude variations of the 
scattering length, $a(B)$, and, accordingly, of the dimer bond length, 
$\langle r\rangle\approx a(B)/2$. The pulse sequence of 
Fig.~\ref{fig:BpulseClaussen} starts at the evaporation field strength of 
about $162\,$G at which a dilute $^{85}$Rb Bose-Einstein condensate is usually 
produced \cite{CornishPRL00} with a scattering length of about 
$200\,a_\mathrm{Bohr}$. Given the experimental mean inter-atomic distance of 
about $12000\,a_\mathrm{Bohr}$, the gas is therefore initially weakly 
interacting. Each pulse tunes the scattering length to  
$a(B_\mathrm{min})\approx 9000\,a_\mathrm{Bohr}$ on time scales too short for 
the Bose-Einstein condensate to adjust its density. Consequently, at the 
magnetic field strength $B_\mathrm{min}$ closest to the zero energy resonance 
the gas is driven into the regime of strong inter-atomic interactions. During 
the course of the experiments \cite{DonleyNature02,ClaussenPRA03}, the 
stationary field, $B_\mathrm{evolve}$, associated with the evolution period 
separating the pulses as well as the duration $t_\mathrm{evolve}$ were varied. 
The variation of $B_\mathrm{evolve}$ between $156\,$G and $162\,$G corresponds 
to the range of Fesh\-bach molecular binding energies displayed in the inset 
of Fig.~\ref{fig:85RbEbofB} with associated scattering lengths between about 
$4000\,a_\mathrm{Bohr}$ and $200\,a_\mathrm{Bohr}$.

All measurements probing the densities of atoms were performed after each 
pulse sequence had terminated and therefore reflect the state of the weakly 
interacting gas at the final magnetic field strength of about 162\,G in
Fig.~\ref{fig:BpulseClaussen}. According to these observations, the fast 
magnetic field variation gives rise to the three components of the atomic 
cloud illustrated in Fig.~\ref{fig:DonleyFringe}. The occupation numbers 
associated with the remnant Bose-Einstein condensate, burst component and 
undetected atoms all oscillate with respect to each other as a function of the 
evolution time, $t_\mathrm{evolve}$. It turned out that their common angular 
frequency, $\omega_\mathrm{e}$, is accurately determined by the Fesh\-bach 
molecular energy in the evolution period, $E_\mathrm{b}^\mathrm{evolve}$, via 
the relation $\omega_\mathrm{e}=|E_\mathrm{b}^\mathrm{evolve}|/\hbar$. 
Donley~{\em et al.} concluded that the undetected atoms were transferred into 
Fesh\-bach molecules whose fast phase evolution as compared to the atomic 
components leads to interference fringes in the final occupations. 

\begin{figure}[htbp]
  \includegraphics[width=\columnwidth,clip]{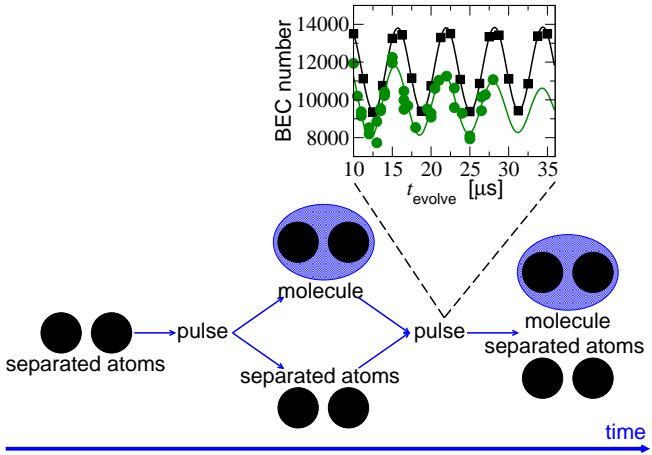}
  \caption{(Colour in online edition)
    Schematic illustration of the Ramsey interferometry with $^{85}$Rb 
    atoms and Fesh\-bach molecules associated with the magnetic field pulse 
    sequence of Fig.~\ref{fig:BpulseClaussen} 
    \cite{DonleyNature02,ZollerNature02}. The first pulse drives the initial 
    Bose-Einstein condensate into a coherent superposition of bound and 
    unbound atom pairs. Between the pulses the individual orthogonal 
    components evolve independently. The second pulse gets the atomic and 
    molecular states to overlap and thereby probes their phase difference. The 
    inset shows a typical interference fringe pattern at the end of the pulse 
    sequence as a function of the evolution time, $t_\mathrm{evolve}$. Circles 
    refer to measurements of the remnant Bose-Einstein condensate 
    \cite{ClaussenPRA03}, while squares indicate theoretical predictions 
    \cite{GoralPRA05}.}
  \label{fig:Ramsey}
\end{figure}

In accordance with such an intuitive explanation, the first pulse provides the 
overlap between the dimer size and the average distance between the condensed 
atoms which is crucial to the molecular association. As this fast magnetic 
field variation is not resonant with the binding energy, it inevitably leads 
to the additional production of excited atom pairs which constitute a burst of 
atoms. During the evolution period all components of the weakly interacting 
gas evolve independently and coherently and therefore accumulate a phase 
difference of $\Delta\varphi=\omega_\mathrm{e}t_\mathrm{evolve}$. Finally, the 
second pulse gets the Fesh\-bach molecules and separated atoms to overlap once
again and thereby probes $\Delta\varphi$ via an interference in their 
occupations. This intuitive scenario is analogous to the principle of a Ramsey 
interferometer using pairs of separated atoms and Fesh\-bach molecules 
\cite{DonleyNature02,ZollerNature02} whose scheme is illustrated in 
Fig.~\ref{fig:Ramsey}. 

\subsubsection{Dynamics of a single atom pair}
In analogy to the analysis of Subsection~\ref{subsec:BoseFermi}, the 
observations of Fig.~\ref{fig:DonleyFringe} can be qualitatively understood in 
terms of the dynamics of just a single pair of atoms 
\cite{BorcaNewJPhys03,GoralPRA05} in a periodic box of volume $\mathcal{V}$. 
To this end, the two-body time evolution operator of Eq.~(\ref{SEU2B}) 
associated with the magnetic field variation of Fig.~\ref{fig:BpulseClaussen} 
may be split into its contributions of the first and second pulse and of the 
evolution period. This yields:
\begin{equation}
  U_\mathrm{2B}(t_\mathrm{f},t_\mathrm{i})=
  U_{2}(t_\mathrm{f},t_2)U_\mathrm{evolve}(t_\mathrm{evolve})
  U_{1}(t_1,t_\mathrm{i}).
  \label{factorisationU2B}
\end{equation}
Here the first pulse starts at the initial time $t_\mathrm{i}$ and terminates 
at $t_1$, while $t_2=t_1+t_\mathrm{evolve}$ indicates the beginning of the 
second pulse which ends at the final time of the sequence, $t_\mathrm{f}$.
Accordingly, $U_{1}(t_1,t_\mathrm{i})$ and $U_{2}(t_\mathrm{f},t_2)$ describe
the two-body dynamics during the first and second pulse, respectively. In 
between the pulses the magnetic field strength is stationary, in accordance 
with Fig.~\ref{fig:BpulseClaussen}. This implies that the associated time 
evolution operator, $U_\mathrm{evolve}(t_\mathrm{evolve})$, depends just on 
the duration $t_\mathrm{evolve}$. Its spectral decomposition in terms of 
dressed states of the relative motion of an atom pair exposed to the magnetic 
field of strength $B_\mathrm{evolve}$ is given by the following formula:
\begin{equation}
  U_\mathrm{evolve}(t_\mathrm{evolve})=\sum_{v=-1}^{\infty}
  |\phi_v^\mathrm{evolve}\rangle 
  e^{-iE_v^\mathrm{evolve}t_\mathrm{evolve}/\hbar}
  \langle\phi_v^\mathrm{evolve}|.
  \label{Uevolvespectral}
\end{equation}
Here the index $v$ labels the vibrational quantum numbers of the box states in
such a way that $|\phi_{-1}^\mathrm{evolve}\rangle$ correlates adiabatically, 
in the limit of infinite volume $\mathcal{V}$, with the Fesh\-bach molecular 
dressed state, $|\phi_\mathrm{b}^\mathrm{evolve}\rangle$. Accordingly, the 
associated vibrational energy levels are denoted by $E_v^\mathrm{evolve}$ 
which implies the asymptotic behaviour 
$E_{-1}^\mathrm{evolve}\sim E_\mathrm{b}^\mathrm{evolve}$ with increasing 
$\mathcal{V}$. The more tightly bound states associated with vibrational 
quantum numbers $v<-1$ contribute negligibly to the fringe pattern of 
Fig.~\ref{fig:Ramsey} and are therefore omitted in 
Eq.~(\ref{Uevolvespectral}).

Similarly to the treatment of Subsection~\ref{subsec:BoseFermi}, the 
interference fringes in the population of the remnant condensate component
shown in the inset of Fig.~\ref{fig:Ramsey} can be described in terms of the 
following transition probability:
\begin{equation}
  p_{0,0}=\left|
  \langle\phi_0^\mathrm{f}|U_\mathrm{2B}(t_\mathrm{f},t_\mathrm{i})
  |\phi_0^\mathrm{i}\rangle\right|^2.
  \label{p00Ramsey}
\end{equation}
To this end, it is instructive to split the associated amplitude, in 
accordance with Eq.~(\ref{Uevolvespectral}), into its contributions of the 
vibrational states with quantum numbers $v\geq 0$ and of the Fesh\-bach 
molecule. This yields:
\begin{equation}
  \langle\phi_0^\mathrm{f}|U_\mathrm{2B}(t_\mathrm{f},t_\mathrm{i})
  |\phi_0^\mathrm{i}\rangle=D_\mathrm{2B}+A_\mathrm{2B}.
  \label{interferenceamplitude}
\end{equation}
Due to the quasi continuum of levels, the first term, $D_\mathrm{2B}$, on the 
right hand side of Eq.~(\ref{interferenceamplitude}) tends to decay via phase 
diffusion as a function of $t_\mathrm{evolve}$. Its explicit expression reads:
\begin{align}
  \nonumber
  D_\mathrm{2B}=&\sum_{v=0}^{\infty}
  \langle\phi_0^\mathrm{f}|U_{2}(t_\mathrm{f},t_2)
  |\phi_v^\mathrm{evolve}\rangle
  e^{-iE_v^\mathrm{evolve}t_\mathrm{evolve}/\hbar}\\
  &\times\langle\phi_v^\mathrm{evolve}|U_{1}(t_1,t_\mathrm{i})
  |\phi_0^\mathrm{i}\rangle.
\end{align}
The magnitude of the second term, $A_\mathrm{2B}$, on the right hand side of 
Eq.~(\ref{interferenceamplitude}) depends on the product of the amplitudes for 
Fesh\-bach molecular association during the first pulse and dissociation into 
the zero mode due to the second pulse. Its evolution as a function of 
$t_\mathrm{evolve}$ stems just from the phase shift associated with the 
energy $E_\mathrm{-1}^\mathrm{evolve}$, in accordance with the following 
formula:  
\begin{align}
  \nonumber
  A_\mathrm{2B}=&\langle\phi_0^\mathrm{f}|U_{2}(t_\mathrm{f},t_2)
  |\phi_{-1}^\mathrm{evolve}\rangle
  e^{-iE_{-1}^\mathrm{evolve}t_\mathrm{evolve}/\hbar}\\
  &\times\langle\phi_{-1}^\mathrm{evolve}|U_{1}(t_1,t_\mathrm{i})
  |\phi_0^\mathrm{i}\rangle.
  \label{A2B}
\end{align}
The modulus squared of the sum of interfering amplitudes, $D_\mathrm{2B}$ and 
$A_\mathrm{2B}$, therefore leads to the anticipated fringe pattern in its 
usual form, 
\begin{equation}
  p_{0,0}=|D_\mathrm{2B}|^2+|A_\mathrm{2B}|^2+
  2|D_\mathrm{2B}||A_\mathrm{2B}|\sin(\omega_\mathrm{e}t_\mathrm{evolve}
  +\varphi),
  \label{p00fringe}
\end{equation}
in agreement with the illustration of Fig.~\ref{fig:Ramsey}. Here the angular  
frequency, $\omega_\mathrm{e}=|E_\mathrm{-1}^\mathrm{evolve}|/\hbar$, is  
determined by the Fesh\-bach molecular binding energy. The absolute phase, 
$\varphi$, is associated with both amplitudes $A_\mathrm{2B}$ and 
$D_\mathrm{2B}$ and depends sensitively on the exact shape of each pulse. A 
similar statement applies to the efficiency of dimer production and 
dissociation which are both crucial to the fringe visibility, according to 
Eqs.~(\ref{A2B}) and (\ref{p00fringe}).

The populations of Fesh\-bach molecules as well as the burst spectrum composed 
of the excited dressed levels at the final magnetic field strength may be 
inferred from the associated transition probabilities, similarly to 
Eq.~(\ref{p00fringe}). Their pairwise averages in a volume $\mathcal{V}$ in 
addition to a subsequent local density approximation give a good account of 
the magnitudes of all components of the gas shown in 
Fig.~\ref{fig:DonleyFringe} \cite{GoralPRA05}. The range of 
validity of such a two-body estimate is set by the requirement that 
$B_\mathrm{evolve}$ is sufficiently far from $B_0$ that the mean distance 
between the atoms is much larger than the scattering length during the 
evolution period. As the 155\,G zero energy resonance of $^{85}$Rb is 
entrance-channel dominated, the two-body dynamics is well described in terms 
of both the effective two- and single-channel approaches of 
Subsection~\ref{subsec:parameters} \cite{KoehlerPRA03,GoralPRA05}. We note, 
however, that the contact pseudo interaction is insufficient. This is 
immediately apparent, for instance, from the inaccuracy of the universal 
estimate of the binding energies in the inset of Fig.~\ref{fig:85RbEbofB} 
whose experimental data were determined using measured fringe frequencies 
\cite{ClaussenPRA03}. The interference experiments are therefore sensitive 
to the van der Waals tail of the inter-atomic potential which is not accounted 
for by the Landau-Zener parameter associated with asymptotic linear magnetic 
field sweeps. 

\subsection{Number of dimers produced in Bose and Fermi gases}
\label{subsec:howtocount}
Another significant difference in the descriptions of the linear sweep and 
magnetic field pulse techniques is associated with the treatment of the 
Fesh\-bach molecular state. The results of 
Subsection~\ref{subsec:linearsweeps} suggest that the number of dimers 
produced via asymptotic magnetic field sweeps may be inferred from two-level 
approaches effectively identifying dressed Fesh\-bach molecules with bare 
resonance states. For the Ramsey interference experiments 
\cite{DonleyNature02,ClaussenPRA03} operating just in the vicinity of the zero 
energy resonance a maximum final molecular conversion of about 16\,\% was 
reported under the conditions referred to in Fig.~\ref{fig:DonleyFringe}. The 
closed channel admixture to the dressed bound state, however, is approximately 
25\,\% at the final magnetic field strength of 162\,G \cite{KoehlerPRA04} and 
even smaller throughout the pulse sequence 
\cite{BraatenComment03,KoehlerPRL03}. In order to predict the observed number 
of dimers, it is therefore necessary to include the long range nature of the 
Fesh\-bach molecules produced in the description of their detection.

An associated generic approach suitable for both magnetic field sweeps and 
pulse sequences may be based on the following observable for the population 
of any diatomic state, $|\phi_\mathrm{d}\rangle$, in a gas consisting of $N$ 
atoms \cite{KoehlerPRA03}:
\begin{equation}
  {\sf N}_\mathrm{d}=\frac{1}{2}\sum_{{i,j=1}\atop{i\neq j}}^N
  |\phi^{\mathrm{d}}_{ij}\rangle\langle\phi^{\mathrm{d}}_{ij}|.
  \label{Ndfirstquantised}
\end{equation}
Here the indices $i$ and $j$ label the atoms, 
$|\phi^{\mathrm{d}}_{ij}\rangle$ refers to the two-body state, 
$|\phi_\mathrm{d}\rangle$, associated with the atoms $i$ and $j$, and the 
factor of $1/2$ on the right hand side of Eq.~(\ref{Ndfirstquantised}) 
prevents double counting of pairs. Given any state of the gas generally 
described by a density matrix, $\rho_{N\mathrm{B}}$, the number of atomic 
pairs in the state $|\phi_\mathrm{d}\rangle$ is consequently determined by the 
following expectation value:
\begin{equation}
  N_\mathrm{d}=\langle{\sf N}_\mathrm{d}\rangle
  =\mathrm{Tr}({\sf N}_\mathrm{d}\rho_{N\mathrm{B}}).
  \label{Ndimergeneral}
\end{equation}
Here the symbol ``$\mathrm{Tr}$'' refers to the trace over the degrees of 
freedom of all atoms constituting the gas. 

Many-body systems consisting of identical atoms are usually described using 
the approach of second quantisation \cite{FetterWalecka71}. Accordingly, any 
$N$ particle state is constructed by repeated application of creation 
operators to the vacuum, $|\Omega\rangle$, which refers to the absence of 
atoms. The associated field operators, $\psi_\alpha(\mathbf{x})$ and 
$\psi_{\beta}^\dagger(\mathbf{y})$, annihilating and creating single atoms at 
the positions $\mathbf{x}$ and $\mathbf{y}$, respectively, fulfil the 
following (anti-)commutation relations:
\begin{equation}
  \psi_\alpha(\mathbf{x})\psi_{\beta}^\dagger(\mathbf{y})\mp
  \psi_{\beta}^\dagger(\mathbf{y})\psi_\alpha(\mathbf{x})
  =\delta_{\alpha\beta}\delta(\mathbf{x}-\mathbf{y}).
  \label{commutation}
\end{equation}
Here the minus sign is associated with bosons while the plus sign refers to
fermions, and the Greek labels $\alpha$ and $\beta$ indicate their Zeeman 
states. Given the diatomic wave function of the relative motion,  
$\phi_{\alpha\beta}^\mathrm{d}(\mathbf{r})=\langle\mathbf{r};\alpha,\beta
|\phi_\mathrm{d}\rangle$, the second quantised representation of the number 
operator of Eq.~(\ref{Ndfirstquantised}) reads: 
\begin{align}
  \nonumber
      {\sf N}_\mathrm{d}=\frac{1}{2}\sum_{\alpha,\beta,\alpha',\beta'}
      &\int d\mathbf{R}\int d\mathbf{r} \int d\mathbf{r}'\,
      \phi_{\alpha\beta}^\mathrm{d}(\mathbf{r})
	  [\phi_{\alpha'\beta'}^\mathrm{d}(\mathbf{r}')]^*\\
	  \times &
	  \psi_{\alpha}^\dagger(\mathbf{x})
	  \psi_{\beta}^\dagger(\mathbf{y})
	  \psi_{\beta'}(\mathbf{y}')
	  \psi_{\alpha'}(\mathbf{x}').
	  \label{Ndsecondquantised}
\end{align}
Here the single-particle coordinates are given in terms of the centre of mass
and relative positions by $\mathbf{x}=\mathbf{R}+\mathbf{r}/2$, 
$\mathbf{y}=\mathbf{R}-\mathbf{r}/2$,
$\mathbf{x}'=\mathbf{R}+\mathbf{r}'/2$, and
$\mathbf{y}'=\mathbf{R}-\mathbf{r}'/2$. 

The approach of Eq.~(\ref{Ndsecondquantised}) was introduced in the context of 
simulations of the Ramsey fringes of Fig.~\ref{fig:DonleyFringe} 
\cite{KoehlerPRA03,KoehlerPRL03} and subsequently applied to the molecule 
production via magnetic field sweeps in Bose \cite{GoralJPhysB04} and Fermi 
gases \cite{PeraliPRL05}. In accordance with Eq.~(\ref{Ndsecondquantised}), 
the expectation value of Eq.~(\ref{Ndimergeneral}) depends on the two-body 
correlation function,
\begin{equation}
  G^{(2)}_{\alpha'\beta',\alpha\beta}
  (\mathbf{x}',\mathbf{y}';\mathbf{x},\mathbf{y})
  =\langle\psi_{\alpha}^\dagger(\mathbf{x})
  \psi_{\beta}^\dagger(\mathbf{y})
  \psi_{\beta'}(\mathbf{y}')
  \psi_{\alpha'}(\mathbf{x}')\rangle.
  \label{G2general}
\end{equation}
Here the average refers to the final state of the gas, 
$\rho_{N\mathrm{B}}(t_\mathrm{f})$. In the context of the dimer production via
fast magnetic field sweeps of Subsection~\ref{subsec:BoseFermi}, for instance, 
the dynamics of $G^{(2)}$ is determined simply by the two-body evolution 
operator, $U_\mathrm{2B}(t_\mathrm{f},t_\mathrm{i})$, in the limit of short 
interaction times. Given an ideal gas initial state, 
$\rho_{N\mathrm{B}}(t_\mathrm{i})$, associated with either Bose or Fermi 
atoms, the exact form of the initial two-body correlation function can be 
inferred from Wick's theorem of statistical mechanics 
\cite{WickPR50,MatsubaraProgTheorPhys55,FetterWalecka71}. In the single 
resonance approach, the diatomic channel wave functions of the Fesh\-bach 
molecular state occurring in Eq.~(\ref{Ndsecondquantised}) are determined in 
terms of the entrance- and closed-channel components by the following formula:
\begin{equation}
  \sum_{\alpha,\beta}|\alpha,\beta\rangle
  \phi_{\alpha\beta}^\mathrm{b}(\mathbf{r})
  =|\mathrm{bg}\rangle\phi_\mathrm{b}^\mathrm{bg}(\mathbf{r})+
  |\mathrm{cl}\rangle\phi_\mathrm{b}^\mathrm{cl}(\mathbf{r}).
  \label{phibmicroscopic}
\end{equation}
Based on these assumptions, the number of Fesh\-bach molecules predicted by 
Eq.~(\ref{Ndsecondquantised}) strictly confirms the fast sweep limits of 
Eqs.~(\ref{fastsweepFermi}), (\ref{Ndfcondensate}) and (\ref{fastsweepBose}), 
in particular, their coefficients associated with the identical nature of the 
atoms. Besides these statistical estimates, the general observable of 
Eq.~(\ref{Ndsecondquantised}) may be applied to a variety of dynamical 
magnetic field variations as well as physical quantities. Given the associated 
two-body correlation function, it recovers, for instance, not only the 
measured maximum conversion into bound dimers of about 16\,\% in 
Fig.~\ref{fig:DonleyFringe} but also the populations of correlated pairs in 
continuum levels which constitute the burst of atoms \cite{KoehlerPRA03}. It 
turns out that due to the nonlinear magnetic field variation of 
Fig.~\ref{fig:BpulseClaussen}, the determination of the time dependence of 
$G^{(2)}$ requires a description via techniques beyond the two-level mean 
field approach of Subsection~\ref{subsec:BECsweeps}.

\subsection{Dynamics of partially condensed Bose gases}
\label{subsec:BECdynamics}
The precise many-body dynamics underlying all observations of 
Fig.~\ref{fig:DonleyFringe} can be derived from the time dependent 
Schr\"odinger equation determined by the general Hamiltonian 
\cite{FetterWalecka71}:
\begin{align}
  \nonumber
  H=&\sum_{\alpha}\int d\mathbf{x}\,
  \psi_\alpha^\dagger(\mathbf{x})
  \left[
  -\frac{\hbar^2}{2m}\boldsymbol{\nabla}^2+E_\alpha^\mathrm{a}
  +V_\alpha^\mathrm{ho}(\mathbf{x})
  \right]
  \psi_\alpha(\mathbf{x})\\
  &+H_\mathrm{int}.
  \label{HMB}
\end{align}
Here $E_\alpha^\mathrm{a}$ refers to an atomic Zeeman energy whose typical 
magnetic field dependence is illustrated in Fig.~\ref{fig:87RbEZeeman} for the 
example of $^{87}$Rb, and $V_\alpha^\mathrm{ho}(\mathbf{x})$ describes the 
harmonic confinement due to an atom trap. In the context of dilute gases, the 
inter-atomic interactions are predominantly pairwise which implies the 
following potential energy contribution:
\begin{align}
  \nonumber
  H_\mathrm{int}=\frac{1}{2}\sum_{\alpha,\beta,\alpha',\beta'}
  \int d\mathbf{x}\int d\mathbf{y}\,
  V_{\alpha\beta,\alpha'\beta'}(\mathbf{x}-\mathbf{y})&\\ 
  \times
  \psi_{\alpha}^\dagger(\mathbf{x})
  \psi_{\beta}^\dagger(\mathbf{y})
  \psi_{\beta'}(\mathbf{y})
  \psi_{\alpha'}(\mathbf{x})&.
  \label{HintMB}
\end{align}
Here $V_{\alpha\beta,\alpha'\beta'}(\mathbf{x}-\mathbf{y})$ denotes the 
two-body potential associated with the incoming and outgoing spin channels 
$|\alpha',\beta'\rangle$ and $|\alpha,\beta\rangle$, respectively. We note 
that Eqs.~(\ref{HMB}) and (\ref{HintMB}) are formulated sufficiently generally 
to treat the inter-atomic interactions on a microscopic level, similarly to 
the coupled channels theory of Subsection~\ref{subsec:coupledchannels}. As a 
consequence, the atomic Zeeman levels, $E_\alpha^\mathrm{a}$, appear 
separately in Eq.~(\ref{HMB}), while in the two-body Hamiltonian of 
Eq.~(\ref{H2B2channel}) the dissociation threshold energy associated with the 
closed channel is included in the potential $V_\mathrm{cl}(B,r)$. Accordingly, 
all interactions, $V_{\alpha\beta,\alpha'\beta'}(\mathbf{x}-\mathbf{y})$, 
vanish in the limit of infinite distances, 
$r=|\mathbf{x}-\mathbf{y}|\to\infty$, similarly to the interaction matrix 
$V_\mathrm{int}(r)$ of Eq.~(\ref{eq:V}).

\subsubsection{Beyond mean field approaches}
Several techniques were employed in treatments of the many-body dynamics of 
dilute gases beyond the Gross-Pitaevskii and two-level mean field approaches 
of Subsection~\ref{subsec:BECsweeps}. These methods involve the 
Schwinger-Keldysh formalism \cite{SchwingerJMP61,KeldyshJETP65} suitable for 
descriptions of phenomena associated with the evolution toward thermal 
equilibrium, such as the dimer production via adiabatic magnetic field sweeps 
of Fig.~\ref{fig:Jamieplot} 
\cite{WilliamsJPhysB04,WilliamsNJPhys04,WilliamsNJPhys06}. Practical 
approaches describing equilibration on even longer time scales, i.e.~beyond 
the range of validity of the generalised Boltzmann equations derived from the 
Schwinger-Keldysh theory, may be based on the two-particle irreducible action
\cite{LuttingerPR60,BaymPR62,CornwallPRD74}. In the context of the dynamics of 
cold gases, such techniques have been implemented, to date, using the contact 
pseudo interaction in a single spatial dimension 
\cite{ReyPRA05,GasenzerPRA05}. While the Ramsey fringes of 
Fig.~\ref{fig:DonleyFringe} are sensitive to parameters of the inter-atomic 
potential besides the scattering length, the associated typical pulse sequence 
of Fig.~\ref{fig:BpulseClaussen} involves time scales sufficiently short for 
equilibration phenomena to be negligible. This implies that the dynamics of 
the gas is captured by extensions of mean field theory which account for the 
two-body time evolution beyond the Markov approximation
\cite{KoehlerPRA02,HollandPRL01,NaidonPRA06,ProukakisPRA98}.

The usual quantities of interest in these short time quantum kinetic 
approaches may be expressed in terms of correlation functions, 
i.e.~expectation values of normal ordered products of field operators, such as 
$G^{(2)}$ of Eq.~(\ref{G2general}). Their general expression reads 
$\langle\psi_{\beta}^\dagger(\mathbf{y})\cdots\psi_{\alpha}(\mathbf{x})
\rangle_t$ where all creation operators appear to the left of all annihilation
operators, and the average refers to the many-body state at time $t$. The 
number of field operators constituting the product shall be referred to in the 
following as the order of the correlation function. Based on the Schr\"odinger 
equation, the associated dynamics is determined by
\begin{equation}
  i\hbar\frac{\partial}{\partial t}
  \langle
  \psi_{\beta}^\dagger(\mathbf{y})\cdots
  \psi_{\alpha}(\mathbf{x})
  \rangle_t
  =\langle
  [\psi_{\beta}^\dagger(\mathbf{y})\cdots
  \psi_{\alpha}(\mathbf{x}),H]
  \rangle_t.
  \label{SEcorrelation}
\end{equation}
Here $H$ is the Hamiltonian of Eq.~(\ref{HMB}), and the symbol $[A,B]=AB-BA$ 
indicates the commutator of the operators $A$ and $B$. Due to the potential 
energy contribution of Eq.~(\ref{HintMB}), the commutator on the right hand 
side of Eq.~(\ref{SEcorrelation}) gives rise to products containing two more 
field operators than the expectation value to the left. This implies that the 
dynamics of any one correlation function is determined by a coupled set of 
equations involving all the others. Approximate solutions to 
Eq.~(\ref{SEcorrelation}), therefore, often rely upon schemes for the
truncation of the associated infinite hierarchy of dynamical equations. The 
range of applicability of such approaches is sensitive to the initial state of 
the many-body system.

In the context of weakly interacting gases close to thermal equilibrium, an 
approximate form of all correlation functions may be determined using Wick's 
theorem of statistical mechanics 
\cite{WickPR50,MatsubaraProgTheorPhys55,FetterWalecka71}. To this end, it is 
instructive to introduce the connected correlation functions 
\cite{WeinbergQFT96}, sometimes referred to as cumulants. These quantities may 
be inferred recursively from a decomposition of each correlation function into 
a sum of all possible products of cumulants which preserve the order of 
appearance of the operators. Given a set of Bose field operators, $A$, $B$ and 
$C$, for instance, the first three cumulants, denoted by 
$\langle A\rangle^\mathrm{c}$, $\langle BA\rangle^\mathrm{c}$ and 
$\langle CBA\rangle^\mathrm{c}$, are determined implicitly by the following 
relations:
\begin{align}
  \label{firstcumulant}
  \langle A\rangle=&\langle A\rangle^\mathrm{c},\\
  \label{secondcumulant}
  \langle BA\rangle=&\langle BA\rangle^\mathrm{c}
  +\langle A\rangle^\mathrm{c}\langle B\rangle^\mathrm{c},\\
  \nonumber
  \langle CBA\rangle=&\langle CBA\rangle^\mathrm{c}+
  \langle BA\rangle^\mathrm{c}\langle C\rangle^\mathrm{c}+
  \langle CA\rangle^\mathrm{c}\langle B\rangle^\mathrm{c}\\
  &+
  \langle CB\rangle^\mathrm{c}\langle A\rangle^\mathrm{c}+
  \langle A\rangle^\mathrm{c}\langle B\rangle^\mathrm{c}
  \langle C\rangle^\mathrm{c}.
  \label{thirdcumulant}
\end{align}
The cumulants associated with Fermi field operators may be inferred from
expressions similar to Eqs.~(\ref{firstcumulant}), (\ref{secondcumulant}) and 
(\ref{thirdcumulant}). In accordance with the anti-commutation relation of 
Eq.~(\ref{commutation}), however, the number of commutations of field 
operators in the decompositions needs to be accounted for by an appropriate 
sign of each contribution. In both cases, Bose and Fermi atoms, Wick's theorem 
reduces to the statement that all cumulants containing more than two field 
operators vanish provided that the gas is ideal and in thermal equilibrium. 
The magnitude of higher order cumulants therefore provides a measure of the 
deviations of the state from the grand canonical density matrix in the absence 
of inter-atomic interactions. 

This suggests that in the context of dilute, weakly interacting gases a 
reasonable quantum kinetic approach may be based on the transformation of 
Eq.~(\ref{SEcorrelation}) into the associated set of dynamical equations for 
cumulants. In accordance with Wick's theorem, this coupled system allows for 
an approximate truncation at any order of correlation usually determined by 
external driving fields as well as the initial state \cite{FrickeAnnPhys96}. 
The dilute gas of $^{85}$Rb atoms in the Ramsey interferometry experiments 
illustrated in Fig.~\ref{fig:Ramsey} was prepared as a Bose-Einstein 
condensate in the $(f=2,m_f=-2)$ Zeeman level. Such a coherent initial state 
gives rise to a mean field, 
$\Psi_\alpha(\mathbf{x},t)=\langle\psi_\alpha(\mathbf{x})\rangle_t$, 
describing the density associated with the macroscopically occupied mode via 
its modulus squared \cite{DalfovoRMP99}. In accordance with 
Fig.~\ref{fig:DonleyFringe}, the condensate is depleted due to the near 
resonant magnetic field variation of Fig.~\ref{fig:BpulseClaussen}. This 
implies that, in addition to the mean field, a minimum set of cumulants 
describing this experiment is given, respectively, by the pair function and 
the one-body density matrix of the non-condensed component:
\begin{align}
  \label{definitionPhi}
  \Phi_{\alpha\beta}(\mathbf{x},\mathbf{y},t)&=
  \langle\psi_{\beta}(\mathbf{y})\psi_{\alpha}(\mathbf{x})\rangle_t
  -\Psi_{\alpha}(\mathbf{x},t)\Psi_{\beta}(\mathbf{y},t),\\
   \Gamma_{\alpha\beta}(\mathbf{x},\mathbf{y},t)&=
  \langle\psi_{\beta}^\dagger(\mathbf{y})\psi_{\alpha}(\mathbf{x})\rangle_t
  -\Psi_{\alpha}(\mathbf{x},t)\Psi_{\beta}^*(\mathbf{y},t).
  \label{definitionGamma}
\end{align}
Accordingly, at any time, $t$, the density of atoms in the Zeeman state with 
the index $\alpha$ is given by $|\Psi_\alpha(\mathbf{x},t)|^2+
\Gamma_{\alpha\alpha}(\mathbf{x},\mathbf{x},t)$. Due to the weak interactions 
of the gas at the beginning of the magnetic field pulse sequence, the second 
order cumulants of Eqs.~(\ref{definitionPhi}) and (\ref{definitionGamma}) are 
negligible initially, 
i.e.~$\Phi_{\alpha\beta}(\mathbf{x},\mathbf{y},t_\mathrm{i})=0
=\Gamma_{\alpha\beta}(\mathbf{x},\mathbf{y},t_\mathrm{i})$.

To a first approximation, the dynamical equations~(\ref{SEcorrelation}) may be 
transformed and truncated in such a way that they just include products of 
normal ordered cumulants containing at most three field operators 
\cite{KoehlerPRA02}. This yields the following relation for the time 
derivative of the condensate mean field:
\begin{align}
  \nonumber
  i\hbar\dot{\Psi}_{\alpha}(\mathbf{x},t)=H_{\alpha}^\mathrm{1B}
  \Psi_{\alpha}(\mathbf{x},t)
  +&\sum_{\alpha,\alpha',\beta'}\int d\mathbf{y}\,
  V_{\alpha\beta,\alpha'\beta'}(\mathbf{r})\\
  &\times\Psi_{\beta}^*(\mathbf{y},t)
  \langle\psi_{\beta'}(\mathbf{y})\psi_{\alpha'}(\mathbf{x})\rangle_t.
  \label{Psidot}
\end{align}
Here $\mathbf{r}=\mathbf{x}-\mathbf{y}$ refers to the relative coordinates  
of a pair of atoms at the positions $\mathbf{x}$ and $\mathbf{y}$, and the 
one-body Hamiltonian, $H_\alpha^\mathrm{1B}$, consists of the kinetic and 
Zeeman energies as well as the trap potential of Eq.~(\ref{HMB}). The 
correlation function, 
$\langle\psi_{\beta'}(\mathbf{y})\psi_{\alpha'}(\mathbf{x})\rangle_t$, on the 
right hand side of Eq.~(\ref{Psidot}) is determined in terms of cumulants by 
Eq.~(\ref{definitionPhi}). In addition to Eq.~(\ref{Psidot}), the dynamical 
equation associated with the pair function reads:
\begin{align}
  \nonumber
  i\hbar\dot{\Phi}_{\alpha\beta}(\mathbf{x},\mathbf{y},t)
  =&\sum_{\alpha',\beta'}
  \big[H_{\alpha\beta,\alpha'\beta'}^\mathrm{2B}
    \Phi_{\alpha'\beta'}(\mathbf{x},\mathbf{y},t)\\
    &+V_{\alpha\beta,\alpha'\beta'}(\mathbf{r})
  \Psi_{\alpha'}(\mathbf{x},t)\Psi_{\beta'}(\mathbf{y},t)\big].
  \label{Phidot}
\end{align}
Here $H_{\alpha\beta,\alpha'\beta'}^\mathrm{2B}$ denotes the two-body
Hamiltonian matrix associated with the incoming and outgoing spin channels
$|\alpha',\beta'\rangle$ and $|\alpha,\beta\rangle$, respectively, describing 
both the centre of mass and relative motions of an atom pair. 

Given this first order truncation scheme, Eqs.~(\ref{Psidot}) and 
(\ref{Phidot}) uniquely determine the condensate mean field as well as the 
pair function. It turns out that the dynamical equation associated with the 
one-body density matrix of the non-condensed component may be solved 
implicitly in terms of the pair function. This yields:
\begin{equation}
  \Gamma_{\alpha\beta}(\mathbf{x},\mathbf{y},t)=
  \sum_{\gamma}\int d\mathbf{z}\,
  \Phi_{\alpha\gamma}(\mathbf{x},\mathbf{z},t)
  [\Phi_{\beta\gamma}(\mathbf{y},\mathbf{z},t)]^*.
  \label{Bogoliubovrelation}
\end{equation}
In accordance with the general expression for the associated observable, 
${\sf N}=\sum_{\alpha}\int d\mathbf{x}\,\psi_{\alpha}^\dagger(\mathbf{x})
\psi_{\alpha}(\mathbf{x})$, the expectation value of the number of atoms, $N$, 
is strictly conserved by Eqs.~(\ref{Psidot}), (\ref{Phidot}) and 
(\ref{Bogoliubovrelation}) at all times, i.e.
\begin{equation}
  \sum_\alpha\int d\mathbf{x}\,
  \left[
    |\Psi_\alpha(\mathbf{x},t)|^2+
    \Gamma_{\alpha\alpha}(\mathbf{x},\mathbf{x},t)
    \right]=N.
    \label{Ntotalcumulant}
\end{equation}
In addition, Eqs.~(\ref{definitionGamma}) and (\ref{Bogoliubovrelation}) 
preserve the positivity of the one-body density matrix 
$\langle\psi_\beta^\dagger(\mathbf{y})\psi_\alpha(\mathbf{x})\rangle_t$,
a consequence of the unitary time evolution, which is not necessarily 
recovered by higher order approximation schemes.

In the context of dimer formation in partially condensed Bose gases, similar 
extensions of mean field theory are based on the Hartree-Fock Bogoliubov 
approach \cite{HollandPRL01} as well as the reduced pair wave function 
approximation \cite{ChernyPRE00,NaidonPRA03,NaidonPRA06}. As these methods 
all originate from Eq.~(\ref{SEcorrelation}), their short time asymptotic 
limits agree with the exact result given by the perturbation expansion of the 
dynamical many-body Schr\"odinger equation. In addition, the correlation 
functions predicted by any one of these approaches 
\cite{HollandPRL01,KoehlerPRA02,NaidonPRA03,NaidonPRA06} are free of 
secular long time asymptotic behaviour, a common artifact of perturbation 
theory associated with a spurious polynomial time dependence. Both the 
cumulant \cite{KoehlerPRA02} and the reduced pair wave function approach 
\cite{NaidonPRA03,NaidonPRA06} were formulated in such a way that they are 
compatible with the use of microscopic potentials beyond contact pseudo 
interactions.

Most implementations of quantum kinetic approaches to the dimer production via 
Fesh\-bach resonances to date are based on single-channel or two-channel 
single resonance binary interactions illustrated in 
Subsection~\ref{subsec:twochannel}. This implies that the Bose-Einstein 
condensate mode is described by a single mean field, $\Psi(\mathbf{x},t)$,
associated with the Zeeman level in which the gas is prepared. In addition, 
the entrance- and closed-channel components of the pair function may be 
inferred, similarly to Eq.~(\ref{phibmicroscopic}), from the following formula:
\begin{equation}
  \sum_{\alpha,\beta}|\alpha,\beta\rangle
  \Phi_{\alpha\beta}(\mathbf{x},\mathbf{y},t)
  =|\mathrm{bg}\rangle\Phi_\mathrm{bg}(\mathbf{x},\mathbf{y},t)
  +|\mathrm{cl}\rangle\Phi_\mathrm{cl}(\mathbf{x},\mathbf{y},t).
  \label{pairfunction2ch} 
\end{equation}
The resonance mean field of Subsection~\ref{subsec:BECsweeps} is determined by 
the relation $\Phi_\mathrm{cl}(\mathbf{x},\mathbf{y},t)=
\Psi_\mathrm{res}(\mathbf{R},t)\phi_\mathrm{res}(r)$. Here 
$\mathbf{R}=(\mathbf{x}+\mathbf{y})/2$ refers to the centre of mass 
coordinates of a pair of atoms and $r=|\mathbf{x}-\mathbf{y}|$ denotes their 
relative distance. Based on a two-channel implementation of 
Eqs.~(\ref{Psidot}) and (\ref{Phidot}), the two-level mean field approach 
\cite{Tommasini98,Timmermans98,TimmermansPRL99,DrummondPRL98} may be recovered 
by formally solving the dynamical equation associated with the 
entrance-channel component of the pair function. A subsequent elimination of 
$\Phi_\mathrm{bg}(\mathbf{x},\mathbf{y},t)$ from the coupled 
equations~(\ref{Psidot}) and (\ref{Phidot}) yields the functional form of the 
right hand sides of Eqs.~(\ref{TimmermansBEC}) and (\ref{Timmermansres}). The 
coupling constants of Eq.~(\ref{gres}) and of the entrance channel contact 
pseudo interaction are determined by the Markov approximation 
\cite{GoralJPhysB04}. 

We note that truncation schemes of higher order than Eqs.~(\ref{Psidot}) and 
(\ref{Phidot}) can involve cross coupling terms that give rise to pair 
functions beyond the two-channel decomposition of Eq.~(\ref{pairfunction2ch}). 
Such a scenario occurs in quantum kinetic approaches associated with both Bose 
and Fermi gases provided that one and the same atomic Zeeman state is shared 
between the two-body entrance and closed channels 
\cite{ParishPRL05,BruunPRA05}. This, in turn, restricts the applicability of 
commonly employed model Hamiltonians 
\cite{RanningerPhysicaBplusC,FriedbergPRB89} which separate out the resonance 
state in the description of Fesh\-bach molecule production.

\subsubsection{The remnant Bose-Einstein condensate}
The dynamics of the condensate mean field associated with the Ramsey 
interferometry experiments \cite{DonleyNature02} was described using the 
Hartee-Fock Bogoliubov method \cite{KokkelmansPRL02} and related techniques 
\cite{MackiePRL02}, as well as the first order cumulant approach of 
Eqs.~(\ref{Psidot}) and (\ref{Phidot}) \cite{KoehlerPRA03}. Such studies 
involving beyond mean field theories, can often be simplified by eliminating 
the pair function from the set of dynamical or associated eigenvalue 
equations, respectively \cite{Burnett99}. In the context of 
Eqs.~(\ref{Psidot}) and (\ref{Phidot}), this procedure yields 
\cite{KoehlerPRA02}:
\begin{align}
  \nonumber
  i\hbar\dot{\Psi}(\mathbf{x},t)=&H_\mathrm{1B}\Psi(\mathbf{x},t)\\
  &-\Psi^*(\mathbf{x},t)\int_{t_\mathrm{i}}^\infty dt'\,
  \Psi^2(\mathbf{x},t')\frac{\partial}{\partial t'}h(t,t').
  \label{NMNLS}
\end{align}
Here $H_\mathrm{1B}$ is the one-body Hamiltonian associated with the initial 
Zeeman state of the condensed atoms. The coupling function on the right hand 
side of Eq.~(\ref{NMNLS}),
\begin{equation}
  h(t,t')=\theta(t-t')(2\pi\hbar)^3
  \langle 0,\mathrm{bg}|VU_\mathrm{2B}(t,t')|0,\mathrm{bg}\rangle,
  \label{couplingfunction}
\end{equation}
is determined by the two-body time evolution operator of 
Eq.~(\ref{factorisationU2B}). Here 
$\langle\mathbf{r}|0\rangle=1/(2\pi\hbar)^{3/2}$ denotes the zero momentum 
plane wave of the relative motion of an atom pair and $V$ is the microscopic 
potential matrix of Eq.~(\ref{HintMB}). 

The representation of Eqs.~(\ref{NMNLS}) and (\ref{couplingfunction}) shows 
that all findings of Subsection~\ref{subsec:Ramsey} about the two-body time 
evolution and physical origin of the Ramsey fringes are included in the 
quantum kinetic approach. Its degree of accuracy in comparison to experimental 
data of the condensate component remaining at the end of a magnetic field 
pulse sequence \cite{ClaussenPRA03} is illustrated in the inset of 
Fig.~\ref{fig:Ramsey}. The associated theory curve is based on an 
implementation of the two-channel singe-resonance approach of 
Subsection~\ref{subsec:parameters} using a separable background scattering 
potential \cite{GoralPRA05}.

\subsubsection{Fesh\-bach molecule and burst components}
In accordance with Eq.~(\ref{Ntotalcumulant}), the total number of atoms can 
be decomposed into a mean field contribution as well as a non-condensed 
component, $N_\mathrm{nc}(t)$, described by the density matrix of 
Eq.~(\ref{definitionGamma}). The physical significance of $N_\mathrm{nc}(t)$
in the Ramsey interferometry experiments \cite{DonleyNature02} may be inferred 
from Eq.~(\ref{Bogoliubovrelation}). To this end, it is instructive to replace 
the spatial average of Eq.~(\ref{Bogoliubovrelation}) using the completeness 
of the set of dressed bound and continuum energy states associated with the 
magnetic field strength at time $t$. This leads to the following relation:
\begin{align}
  \nonumber
  N_\mathrm{nc}(t)=\int d\mathbf{R}\,
  \bigg[
    \int d\mathbf{p}\,
    &\left|\langle\mathbf{R},\phi_\mathbf{p}
    |\Phi(t)\rangle\right|^2\\
    +
    &\left|\langle\mathbf{R},\phi_\mathrm{b}
    |\Phi(t)\rangle\right|^2   
    \bigg].
  \label{decompositionNnc}
\end{align}
Here $\mathbf{R}$ may be interpreted in terms of the centre of mass position 
of an atom pair, and $|\Phi(t)\rangle$ refers to the pair function whose 
channel components in the spatial representation are given by 
Eq.~(\ref{definitionPhi}). For simplicity, the spectral decomposition of 
Eq.~(\ref{decompositionNnc}) includes just the dressed continuum states, 
$|\phi_\mathbf{p}\rangle$, and the Fesh\-bach molecular state, 
$|\phi_\mathrm{b}\rangle$. More deeply bound levels are neglected.  

Equation~(\ref{decompositionNnc}) suggests that the continuum contribution may 
be interpreted in terms of correlated atom pairs associated with relative 
momenta, $\mathbf{p}$, while the bound state part yields the number of atoms 
converted into dimers. An analysis based on the observable of 
Eq.~(\ref{Ndsecondquantised}) confirms this view \cite{KoehlerPRA03}. Its 
practical implementation based on Eqs.~(\ref{Psidot}) and (\ref{Phidot}) 
involves a cumulant expansion of the two-body correlation function in 
accordance with the first order truncation scheme. This yields:
\begin{align}
  \nonumber
  G^{(2)}_{\alpha'\beta',\alpha\beta}
  (\mathbf{x}',\mathbf{y}';\mathbf{x},\mathbf{y};t)=
  &\langle\psi_{\alpha}^\dagger(\mathbf{x})
  \psi_{\beta}^\dagger(\mathbf{y})\rangle_t\\
  &\times
  \langle\psi_{\beta'}(\mathbf{y}')
  \psi_{\alpha'}(\mathbf{x}')\rangle_t.
  \label{G2firstorder}
\end{align}
Here the correlation functions on the right hand side can be expressed in 
terms of pair functions and condensate mean fields via 
Eq.~(\ref{definitionPhi}). Using Eqs.~(\ref{phibmicroscopic}) and
(\ref{G2firstorder}), it turns out that the diatomic number operator of 
Eq.~(\ref{Ndsecondquantised}) gives rise to a dimer mean field,
\begin{align}
  \nonumber
  \Psi_\mathrm{b}(\mathbf{R},t)=&\frac{1}{\sqrt{2}}
  \bigg\{
  \langle\mathbf{R},\phi_\mathrm{b}|\Phi(t)\rangle\\
  &+\int d\mathbf{r}\,\left[\phi_\mathrm{b}^\mathrm{bg}(\mathbf{r})\right]^*
  \Psi(\mathbf{x},t)\Psi(\mathbf{y},t)
  \bigg\},
  \label{Psib}
\end{align}
which determines the number of Fesh\-bach molecules produced via the relation
$N_\mathrm{d}(t)=\int d\mathbf{R}\,|\Psi_\mathrm{b}(\mathbf{R},t)|^2$. 

The first term on the right hand side of Eq.~(\ref{Psib}) recovers the bound 
state contribution of Eq.~(\ref{decompositionNnc}). In accordance with the 
long range of the $^{85}$Rb$_2$ entrance-channel wave functions of 
Fig.~\ref{fig:85Rbboundstates}, the second term may be interpreted in terms of 
the overlap between the dimers produced and the surrounding gas. Its magnitude 
is determined by the dilute gas parameter squared, $n_\mathrm{c}(t)a^3$, where 
$n_\mathrm{c}(t)$ is the average condensate density and $a$ refers to the 
scattering length associated with the magnetic field strength at time $t$ 
\cite{KoehlerPRL03}. Consequently, the existence of this overlap term reflects 
the breakdown of the concept of diatomic molecules in the environment of a 
strongly interacting gas. This breakdown occurs when the bond length 
$\langle r\rangle=a/2$ of the universal dimer wave function of 
Eq.~(\ref{phibuniversal}) is comparable to the mean inter-atomic distance 
$n_\mathrm{c}^{-1/3}$ of the condensate. Similar phenomena are visible in 
Figures~\ref{fig:6LiZofB}, \ref{fig:40Kdecay} and \ref{fig:85Rbdecay}.

At the end of the magnetic field pulse sequence of 
Fig.~\ref{fig:BpulseClaussen}, the gas is weakly interacting and the second,
overlap term on the right hand side of Eq.~(\ref{Psib}) is negligible. Based 
on Eqs.~(\ref{NMNLS}) and (\ref{Psib}), the predicted number of atoms 
converted into dimers recovers the 16\,\% maximum fraction of missing atoms of 
Fig.~\ref{fig:DonleyFringe} as well as its modulation as a function of 
$t_\mathrm{evolve}$ \cite{KoehlerPRA03}. In accordance with 
Fig.~\ref{fig:85Rbdecay}, the fact that these Fesh\-bach molecules were not 
detected reflects their lifetime with respect to spin relaxation on the order 
of only $100\,\mu$s at the final magnetic field strength of about 162\,G 
\cite{ThompsonPRL05,KoehlerPRL05}. An analysis based on 
Eqs.~(\ref{Ndsecondquantised}) and (\ref{G2firstorder}) also shows that the 
continuum contribution on the right hand side of Eq.~(\ref{decompositionNnc}) 
is associated with correlated atom pairs which constitute the measured burst 
component of Fig.~\ref{fig:DonleyFringe} \cite{KoehlerPRA03}.

\subsubsection{Three-component Ramsey fringes}
\begin{figure}[htbp]
  \includegraphics[width=\columnwidth,clip]{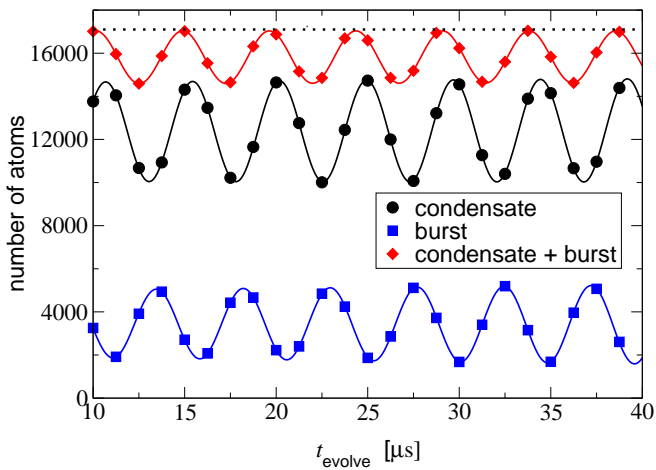}
  \caption{(Colour in online edition)
    Predicted Ramsey fringes between three components of a dilute gas 
    of $^{85}$Rb \cite{GoralPRA05} at the end of the magnetic field pulse 
    sequence of Fig.~\ref{fig:BpulseClaussen} versus the evolution time,
    $t_\mathrm{evolve}$. The circles refer to the remnant Bose-Einstein 
    condensate, while the squares indicate the final population of correlated 
    atom pairs constituting the burst component. The remnant condensate and 
    burst add up to the number of detectable atoms (diamonds). Its difference 
    to the total number of $N=17100$ atoms (dotted line) in the experiments 
    \cite{DonleyNature02} determines the population of Fesh\-bach molecules. 
    The magnetic field strength in the evolution period of 
    $B_\mathrm{evolve}=159.84\,$G associated with these predictions refers to 
    the conditions of the measured fringes of Fig.~\ref{fig:DonleyFringe}.}
  \label{fig:85RbAMcoherence}
\end{figure}

The physical origin of the Ramsey fringes \cite{DonleyNature02} was the 
subject of several theoretical works 
\cite{KokkelmansPRL02,MackiePRL02,KoehlerPRA03,GoralPRA05} using different 
beyond mean field approaches. Figure~\ref{fig:85RbAMcoherence} shows 
such a prediction \cite{GoralPRA05} referring to the magnetic field pulse 
sequence and number of atoms reported for the measurements of 
Fig.~\ref{fig:DonleyFringe}. The overall picture indicates that the quantum 
kinetic approaches support the interpretation of 
Subsection~\ref{subsec:Ramsey}, provided that the gas is weakly interacting 
during the evolution period.

Some measurements were performed also in the regime of near resonant evolution 
fields, $B_\mathrm{evolve}<157\,$G. Under these conditions, the scattering 
length as well as the spatial extent of the Fesh\-bach molecule are comparable 
to the mean inter-atomic distance of the gas. As the atomic and dimer 
components are no longer orthogonal during the evolution period, the classic 
concept of a Ramsey interferometer is expected to break down when 
$B_\mathrm{evolve}$ approaches $B_0$ \cite{GoralPRA05}. Such a phenomenon was
observed in terms of a pronounced damping of the fringe pattern accompanied
by an upward shift of its frequency with respect to the two-body prediction of
$|E_\mathrm{b}^\mathrm{evolve}|/h$ \cite{ClaussenPRA03}. This shift was
subsequently attributed to genuinely many-particle corrections to the 
intuitive viewpoint illustrated in Fig.~\ref{fig:Ramsey} 
\cite{DuinePRL03,GoralPRA05}.

\section{Conclusions and Outlook}
\label{sec:conclusions}
This article has given an overview of a variety of concepts associated with 
the description of properties of Fesh\-bach molecules as well as the 
techniques for their production in the environment of a cold atomic gas. Its 
conclusions can be summarised as follows: Coupled channels theory provides the 
foundation for an accurate understanding of the diatomic molecular and 
collision physics. Its predictions can be recovered, in an experimentally 
relevant energy range about the dissociation threshold, by two-channel 
approaches characterised in terms of just a few measurable quantities. These 
involve the resonance position and width, the background scattering length and 
the van der Waals dispersion coefficient, as well as the difference in the 
magnetic moments associated with the resonance state and a pair of 
asymptotically separated atoms. Universal properties of Fesh\-bach molecules 
and the low energy collision physics are directly related to Wigner's 
threshold law. Such concepts associated with two-body physics treat identical 
bosons and fermions in different Zeeman states as well as distinguishable 
atoms in essentially the same manner. The fundamental transition amplitudes 
for the Fesh\-bach molecular association of an atom pair via linear magnetic 
field sweeps can be represented analytically, in the context of two-channel 
single-resonance approaches. Such an exact treatment provides the foundation 
of the Landau-Zener approach to the dimer production in tight traps, such as 
optical lattice sites, as well as in cold gases in the fast sweep limit. While 
these limits including their quantum statistical phenomena can be understood 
largely in terms of classical probability theory, the saturation of molecule 
production involves the many-body dynamics of thermalisation. Given 
sufficiently short time scales, beyond mean field approaches can provide 
quantitative descriptions of Fesh\-bach molecule production in Bose-Einstein 
condensates also in the context of non-linear magnetic field variations.

Closely related to the enhancement of the $s$-wave scattering cross sections 
discussed in this review are similar magnetically tunable Fesh\-bach resonance 
phenomena involving finite angular momentum quantum numbers, $\ell>0$. 
Associated scattering amplitudes, $f_\ell(p)$, are usually negligible compared 
to the $s$-wave contribution in cold collision physics due to their 
$p^{2\ell}$ scaling with the relative momentum of an atom pair. In gases of 
identical Fermi atoms involving just a single Zeeman state, however, only the 
odd partial waves contribute to diatomic scattering, in accordance with the 
Pauli exclusion principle. In such cases, $p$-wave ($\ell=1$) resonances have 
been observed in gases of both $^{40}$K \cite{RegalPRLpwave03} and $^6$Li
\cite{ZhangPRA04,SchunckPRA05}, providing possibilities of magnetically 
tuning interactions as well as associating dimers. 

Optical Fesh\-bach resonances \cite{FedichevPRL96,BohnPRA97,BohnPRA99} 
provide an alternative approach to tune $s$-wave scattering cross sections in 
cold gases. This technique relies upon coupling the internal states of a pair 
of separated atoms to a closed channel using laser light instead of a 
homogeneous magnetic field. According to experimental studies in cold gases of 
$^{23}$Na \cite{FatemiPRL00} and $^{87}$Rb Bose-Einstein condensates
\cite{TheisPRL04}, significant changes of the scattering length tend to cause 
substantial atom loss due to spontaneous photon emission. Recent predictions 
indicate, however, that such loss may be efficiently suppressed in 
applications to cold gases of alkaline earth metal atoms, such as bosonic 
calcium, as well as ytterbium \cite{CiuryloPRA05}. As these atoms lack 
hyperfine structure due to the absence of nuclear spin, magnetic Fesh\-bach 
resonances are ruled out. Their optical counterpart may therefore provide the 
only possibility of tuning interactions in such gases.

Inter-species magnetic Fesh\-bach resonances were observed in boson-fermion 
mixtures consisting of cold gases of $^6$Li and $^{23}$Na \cite{StanPRL04} as
well as $^{40}$K and $^{87}$Rb \cite{InouyePRL04}. Their binary physics can be 
described using the approaches outlined in this review, whereas the 
collisional stability of inter-species Fesh\-bach molecules requires special 
attention \cite{PetrovJPhysB05}. The production of such diatomic molecules is 
of particular interest because of a possibly polar character of their bonds 
\cite{DoyleEurPhysJD}. Cold polar molecules composed of Bose atoms have been 
produced via photo-association \cite{KermanpolarPRL04,WangPRL04}. According to 
predictions \cite{BaranovPhysScrT02}, the long range, anisotropy and magnitude 
of the dipole-dipole interactions give rise to a host of new features in the 
associated fermionic and bosonic superfuids. Arrays of polar molecules in 
optical lattices may provide an opportunity for studies of super-solid phases 
\cite{GoralPRL02}. It is hoped that heavy polar molecules will enable us to 
improve tests of fundamental physical symmetries, including measurements of 
the electron dipole moment \cite{Sandars75,HudsonPRL02}. Stabilisation with 
respect to collisional relaxation of excited molecular states may be achieved 
via transfer to their vibrational ground states, an approach also demonstrated 
in the context of photo-association \cite{SagePRL05}. Such a de-excitation, in 
turn, tends to enhance the polar character of their bonds.

Few-body scattering phenomena, such as dimer-dimer resonances recently 
observed in cold gases of $^{133}$Cs$_2$ \cite{ChinPRL05}, may provide 
possibilities of extending existing techniques of Fesh\-bach molecule 
production to more complex species. One of the long standing goals in this 
context is an experimental confirmation of predictions associated with 
Efimov's effect in the three-body energy spectrum of identical Bose particles
\cite{EfimovPhysLett70,EfimovSovJNuclPhys}. The Efimov spectrum consists of an
infinite sequence of isotropic three-body bound states accumulating at the 
dissociation threshold, which occurs in the limit of infinite $s$-wave 
scattering length of each two-body subsystem. This scenario is therefore 
directly related to the long range nature of weakly bound two-particle halo 
states \cite{JensenRMP04} and may be realised via magnetic Fesh\-bach 
resonances in cold gases. Signatures of the emergence of such Efimov states 
are predicted to occur in terms of a modulation of three-body recombination 
loss rates as a function of the near resonant magnetic field strength 
\cite{NielsenPRL99,EsryPRL99,BraatenPRL01}. Recently, the first experimental 
evidence for this phenomenon has been reported \cite{KraemerNature06}. Such 
inelastic scattering resonances may be exploited to associate meta-stable 
three-body Efimov molecules from cold Bose gases loaded into optical lattices, 
using the magnetic field sweep technique discussed in this review 
\cite{StollPRA05}.

\section*{Acknowledgements}
We are grateful to Johannes Hecker Denschlag, Stephan D\"urr, Rudi Grimm, 
Tom Hanna, Eleanor Hodby, Randy Hulet, Debbie Jin, Wolfgang Ketterle, 
Servaas Kokkelmans, Tobias Kraemer, Takashi Mukaiyama, Christoph N\"agerl, 
Nicolai Nygaard, Cindy Regal, Gregor Thalhammer, Sarah Thompson, 
Boudewijn Verhaar, Carl Wieman, Jamie Williams and Kaiwen Xu for providing 
their experimental and theoretical data to us. We thank Keith Burnett, 
Cheng Chin, Thomas Gasenzer, Josh Milstein, Bill Stwalley, 
Marzena Szyma{\'n}ska and Eite Tiesinga for valuable discussions. This review 
was supported by the Royal Society (T.K. and K.G.). P.S.J.~gratefully 
acknowledges his support from the Office of Naval Research.

\bibliographystyle{apsrmp}
\bibliography{moleculesrmp}

\end{document}